\documentclass[vecphys]{svmult}

\usepackage{makeidx,amssymb,amsmath}         
\usepackage{graphicx}        
\usepackage{multicol}        
\usepackage[round]{natbib}
\usepackage[bottom]{footmisc}

\makeindex             

\begin{document}

\title*{Special Features of Galactic Dynamics}
\author{C. Efthymiopoulos \and N. Voglis \and
C. Kalapotharakos}
\institute{Research Center for Astronomy and Applied Mathematics,
Academy of Athens Soranou Efessiou 4, 115 27 Athens, Greece\\
\texttt{cefthim@academyofathens.gr}\\
\texttt{nvogl@academyofathens.gr}\\
\texttt{ckalapot@phys.uoa.gr}}
%
%
\maketitle

\section{Introduction}

The present lecture notes are an introduction to selected topics
of {\it Galactic Dynamics}. The focus is on topics that we
consider more relevant to the main theme of this workshop, {\it
Celestial Mechanics}. This is not intended to be a review article.
In fact, any of the topics below could be the subject of a
separate review. Only the main ideas and notions are introduced,
as well as some important currently open problems in each topic.
Some relevant results from our own research are also presented. We
discuss topics related mostly to the so-called {\it ellipsoidal
components} of galaxies. These are \textbf{a)} the dark halos of
both elliptical and disk galaxies, \textbf{b)} the luminous matter
in elliptical galaxies, and \textbf{c)} the bulges of disk
galaxies. We shall only occasionally refer to the dynamics of
disks, bars or spiral structure. These are important chapters of
galactic dynamics which, however, go beyond the limits of the
present article.

The fact that galactic (or stellar) dynamics and celestial
mechanics share many common concepts, tools and methods of study
is nowadays widely recognized in the community of dynamical
astronomers. The connection of the two disciplines is transparent
in recent advanced textbooks such as Contopoulos' {\it Order and
Chaos in Dynamical Astronomy} (2004), or Boccaletti and Pucacco
{\it Theory of Orbits} (1996) (other standard references for
galactic dynamics are Binney and Tremaine 1987, or Bertin 2000).
However, this connection was not always recognized. Until the
sixties, the two fields emphasized rather different aspects of
study, Celestial Mechanics focusing mostly on analytical
expansions of perturbation theory in few body-type problems (e.g.
Szebehely 1967, Hagihara 1970), and Galactic Dynamics focusing on
the properties of the distribution function of stellar systems
composed by a large number of bodies (e.g. Chandrasekhar 1942,
Ogorodnikov 1965). The shift of paradigm in the two fields can be
traced in academic events like a celebrated 1964 Thessaloniki IAU
symposium (Contopoulos 1966, see the description in Contopoulos
2004b).

We would like to point out one more guiding element of the
exposition of ideas followed below. In his talk at the beginning
of this meeting, A. Morbidelli has presented his view of the
division of the problems of Celestial Mechanics into open, i.e.
unresolved, and closed, i.e., resolved problems. In Galactic
Dynamics the very nature of problems does not permit such coarse
classifications. We could claim, instead, that all practically
interesting problems are still largely open. The main obstruction
to closing problems is the lack of sufficient observational data,
which, in many cases, is due to our fundamental inability to
obtain such data. Let us give one trivial example: from the image
of a galaxy in the sky it is impossible to deduce the {\it shape}
of the galaxy without additional dynamical arguments. Such
arguments are to an extent amenable to a posteriori observations,
but the mapping of dynamics to such observations is usually
non-unique. Similarly, the determination of the {\it pattern
speed} of a spiral or barred disk galaxy requires a set of
dynamical assumptions going well beyond the form of the underlying
gravitational potential (the latter can in principle be determined
by the observed rotation curve or distribution of matter in the
galaxy). Since mankind cannot observe galaxies from different
viewpoints, or for times relevant to galactic timescales, these
fundamental constraints will remain with us and require a rather
large effort in dynamical modelling needed to constrain
uncertainties and explain even the simplest available observations
of any particular galaxy. We let apart the fact that large amounts
of matter in a galaxy, with dominant dynamical role, are either
non-detectable by direct observational means (e.g. central black
holes or the dark matter), or subject to non-gravitational
interactions (e.g gas, dust or star formation and evolution), that
seriously complicate the dynamics.

As we shall see in the next section, from the stellar dynamical
point of view the most general information regarding a stellar
system is contained in its phase space density or distribution
function $f(\mathbf{x},\mathbf{v},t)$. This function accounts
for all kinds of photometric or kinematical data that can be
observationally determined. Furthermore,
we can use $f(\mathbf{x},\mathbf{v},t)$ to derive dynamical
properties of the system that cannot be directly observed.
The equilibria of galaxies are described by time-independent
forms of $f$, while evolving galaxies, stellar dynamical
instabilities or density waves are described by time-dependent
forms of $f$. We may thus state that the determination of the
distribution function of galaxies constitutes the {\it central goal}
of galactic dynamics.
The presentation below emphasizes this point of view, by focusing
on {\it dynamical methods} of study of the distribution function.
Other methods, that seek to determine the distribution function
from the observational data via `inversion' algorithms, are not
presented here (see Dejonghe and Bruyne 2003 for a review).

The presentation is organized as follows: section 2 presents some
basic notions of galactic dynamics such as the concept of
relaxation time, Jeans' theorem, third integral of motion etc. In
section 3 we present the statistical mechanical approach to the
study of the distribution function, by dealing mostly with the
theory of violent relaxation and with its modern modifications.
Section 4 deals with the orbital approach. We present the main
types of orbits encountered in spherical, axisymmetric or triaxial
systems, and discuss the methods of `global dynamics' and of
`self-consistent modelling' of galaxies which both occupy an
important place in current research. Section 5 focuses on the
N-Body method. We describe the main techniques to integrate the
N-Body problem when $N$ is large, and discuss recent results from
global dynamical studies of galactic systems from N-Body
simulations.

\section{Basic notions}

\subsection{Time of Relaxation}
The stellar dynamical study of galaxies is simplified by
approximating these systems as {\it collisionless} N-Body systems,
i.e., by assuming that the stars `feel' a mean field gravitational
potential $\Phi(\mathbf{x},t)$, and by ignoring the granularity of
the field due to the point mass distribution of matter. This
approach is justified by the remark that in galaxies the so-called
two body relaxation time $T_R$, i.e., the time needed in order that
close encounters significantly affect an otherwise smooth stellar
orbit, is much larger than the Hubble time of the Universe
(Chandrasekhar 1942, Spitzer and Hart 1971). An order of
magnitude calculation of the two-body relaxation time
can be based on considering deflections of the orbit of a star
that moves in a nearly homogeneous sea of other stars (Fig.1). Let
$\mathbf{v_0}$ be the velocity of the test star at a particular
moment when the impact parameter of its close encounter with a
second star is equal to $b$. Neglecting the attraction by other
stars, the angle of deflection $\psi$ after the encounter is
readily found:
\begin{equation}\label{psi}
\tan\left({\psi\over 2}\right) = {mG\over bv_0^2}
\end{equation}
where $m$ is the mass of the attracting star and $G$ Newton's
constant of gravity. Practically all the angles $\psi$ of
successive scattering events are small, since impact parameters
are in general big. For example, the probability that a second
star passes in the vicinity of the sun at a distance of the order
of 10000AU is about one event in the galaxy's lifetime (see
article by B. Marsden in the same volume). This minimum impact
parameter $b_{min}\approx 10000AU$ is of order $b_{min} \sim
D/N^{1/3}$, where $D$ is the typical length-scale (e.g. diameter)
of the galaxy and $N$ the number of stars in it. We may
also set a maximum impact parameter $b_{max}\sim D$. We may thus
estimate an upper bound for the cumulative deflection angle after
a large number of encounters, within a time interval $T$, by
squaring Eq.(\ref{psi}) (with $\tan(\psi/2)\simeq\psi/2$) and
summing  over the number of stars contained in a differential
cylindrical volume of radius $b$ width $db$, and length $v_0T$:
\begin{equation}
\psi_{cum}^2=\sum_{T}\psi^2 \approx \int_{b_{min}}^{b_{max}}db
2\pi bv_0T\rho {4m^2G^2\over b^2v_0^4}
\end{equation}
where $\rho$ is the mean density. Setting typical values for the
density $\rho\sim mN/D^3$ and stellar velocity $v_0^2\sim GNm/D$,
we find from the above formula that the cumulative deflection will
become of order unity (usually we request $\psi_{cum}=\pi/2$) when
$T=T_R$ becomes equal to
\begin{equation}\label{trelax}
T_R\approx {0.1N\over \ln N}T_D
\end{equation}
where $T_D\sim D/v_0$ is the typical dynamical time or period of a
typical orbit across the galaxy. Setting $T_D\sim 10^8 yr$, and
$N\sim 10^{10} - 10^{13}$, we find $T_R\sim 10^{15} -10^{18}yr$,
i.e., at least five orders of magnitude larger than the Hubble age
of the Universe $T_H\sim 10^{10}yr$. We conclude that close
encounters cannot affect the dynamics in timescales comparable to
the present lifetime of a galaxy.

\begin{figure}[tbp]
\centering{\includegraphics[width=10cm]{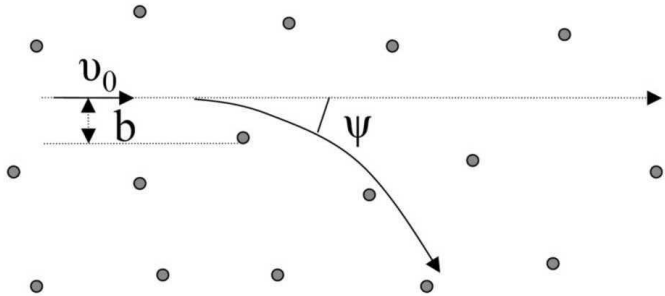}}
\caption{Deflection of a test particle (star) moving in a
homogeneous sea of other particles.} \label{fig01-0}
\end{figure}

Due to Chandrasekhar's calculation of the relaxation time, the
basic paradigm for galaxies is a collisionless stellar
system in which the collisionless Boltzmann equation applies
(subsection 2.3). However, the true nature of relaxation depends
also somewhat on what
region of the galaxy we consider as well as on the properties of
the system's stellar orbits. For example, the above analysis is
not precise at the centers of galaxies, especially when the latter
are occupied by large central mass concentrations. Furthermore, if
a system has a large degree of stochasticity, i.e., many orbits
with Lyapunov times smaller or equal to the Hubble time, then the
two-body relaxation time for such a system is drastically reduced,
perhaps by more than three orders of magnitude (Gurzandyan and
Savvidy 1986, Pfenniger 1986). This is because an initially small
deflection, caused by a two-body encounter, is amplified by the
mechanism of exponential deviations of nearby orbits due to
positive Lyapunov exponents. This may have affected systems that
are `granular', for example galaxies containing a high percentage
of globular clusters (Udry and Pfenniger 1988). The extent to
which such phenomena appear in real galaxies is not yet fully
known.

\subsection{Distribution function}

The most basic quantity in stellar systems is the fine-grained
distribution function:
\begin{equation}\label{df}
f(\mathbf{x},\mathbf{v},t)=\lim_{d^6\mu\rightarrow 0}
\frac{dm(\mathbf{x},\mathbf{v},t)}
{d^3\mathbf{x}d^3\mathbf{v}}
\end{equation}
yielding the mass $dm(\mathbf{x},\mathbf{v},t)$ contained at time $t$ within an
infinitesimal phase-space volume $d^6\mu=d^3\mathbf{x}d^3\mathbf{v}$ centered
around any point $(\mathbf{x},\mathbf{v})$ of the 6D phase space of stellar
motions (called the $\mu-$space in statistical mechanics). In the N-Body
approximation the mass $dm(\mathbf{x},\mathbf{v},t)$ can be considered
proportional to the number of particles, i.e., stars or fluid elements of
the dark matter, within the volume $d^3\mathbf{x}d^3\mathbf{v}$.
Furthermore, it
is often convenient to introduce a coarse-grained distribution function
\begin{equation}\label{coarse}
F(\mathbf{x},\mathbf{v},t)={1\over \Delta^3\mathbf{x}\Delta^3\mathbf{v}}
\int_{\Delta^3\mathbf{x}\Delta^3\mathbf{v}}
f(\mathbf{x},\mathbf{v},t)d^3\mathbf{x}d^3\mathbf{v}
\end{equation}
which gives the average of the fine-grained distribution function
$f$ in small, but not infinitesimal volume elements
$\Delta^3\mathbf{x} \Delta^3\mathbf{v}$ around the phase space
points $(\mathbf{x},\mathbf{v})$. Contrary to the fine-grained
distribution $f$, the value of the coarse-grained distribution $F$
depends on the particular choice of partitioning of the
phase-space by which the volume elements
$\Delta^3\mathbf{x}\Delta^3\mathbf{v}$ are defined. This fact has
some interesting implications in the modelling process of a
galaxy, discussed in section 3 below.

The distribution function can be used to derive several other useful
quantities. For example, the spatial mass
density $\rho(\mathbf{x},t)$ of the system is given by the integral of
the d.f. $f$ over velocities, e.g. (in Cartesian coordinates)
\begin{equation}\label{rho}
\rho(\mathbf{x},t)=
\int_{-\infty}^{\infty}
\int_{-\infty}^{\infty}
\int_{-\infty}^{\infty}
f(\mathbf{x},\mathbf{v},t)dv_xdv_ydv_z
\end{equation}
The latter quantity, $\rho(\mathbf{x},t)$, can be used in turn to calculate
the gravitational potential $\Phi(\mathbf{x},t)$ via Poisson's equation:
\begin{equation}\label{pot}
\nabla^2\Phi(\mathbf{x},t)= 4\pi G \rho(\mathbf{x},t)~~~.
\end{equation}
The orbits of stars are given by the Hamiltonian
\begin{equation}\label{ham}
H(\mathbf{x},\mathbf{p},t)\equiv {\mathbf{p}^2\over 2}+\Phi(\mathbf{x},t)
\end{equation}
setting, for simplicity, $\mathbf{p}=\mathbf{v}$ in Cartesian coordinates
and the average stellar mass equal to unity. We often consider galaxies in
{\it steady state equilibrium} (subsection 2.3), in which case we drop the
explicit dependence of $f$ on the time $t$:
\begin{equation}\label{hamti}
H(\mathbf{x},\mathbf{p})\equiv {\mathbf{p}^2\over 2}+\Phi(\mathbf{x})~~~.
\end{equation}
Assuming a nearly constant mass-to-light ratio, the observable photometric
or kinematic
profiles of a galaxy can be deduced from various moments of $f$.
For example, if the axis $x$ is identified to the direction of the line of
sight,
the surface density at any point $\mathbf{R}\equiv (y,z)$ of the plane of
projection normal to $x$ is given by:
\begin{equation}\label{sig}
\Sigma(\mathbf{R})=
\int_{-\infty}^{\infty}\rho(x,\mathbf{R})dx
\end{equation}
with $\rho$ given by Eq.(\ref{rho}). The quantity
$\Sigma(\mathbf{R})$ can be compared to observed surface
brightness profiles. On the other hand, the line-of-sight
velocity distribution at a particular point $\mathbf{R}$ of the
same plane of projection is given by
\begin{equation}\label{los}
LOSVD(\mathbf{R},v_x)={1\over\Sigma(R)}
\int_{-\infty}^{\infty}
\int_{-\infty}^{\infty}
\int_{-\infty}^{\infty}
dxdv_ydv_z f(x,\mathbf{R},v_x,v_y,v_z)
\end{equation}
and the latter quantity can be compared to the profiles of spectral
lines determined also observationally. Via the line-of-sight
velocity distributions we can determine mean velocity profiles,
\begin{equation}\label{minvel}
\mu(\mathbf{R})=\int_{-\infty}^{\infty} v_x LOSVD(\mathbf{R},v_x) dv_x
\end{equation}
and velocity dispersion profiles
\begin{equation}\label{sigma}
\sigma^2(\mathbf{R})=
\int_{-\infty}^{\infty} (v_x-\mu(\mathbf{R}))^2LOSVD(\mathbf{R},v_x) dv_x~~.
\end{equation}

Also related to observations is the concept of {\it velocity ellipsoid}.
This is an ellipsoid in velocity space assigned to every point
$\mathbf{x}$ of ordinary space. Fixing an orthogonal coordinate
system, say, Cartesian axes $x_1=x,x_2=y,x_3=z$, we calculate the
second moments
\begin{equation}\label{sigmaij}
\sigma_{ij}^2(\mathbf{x}) ={1\over\rho(\mathbf{x})}
\int_{-\infty}^{\infty}
(v_i-V_i)(v_j-V_j)f(\mathbf{x},\mathbf{v})d^3\mathbf{v}
\end{equation}
where the indices $i,j$ run the values $1,2$, or $3$, $V_i$ is the
mean velocity in the i-th direction at the point $\mathbf{x}$ and
the integral denotes a triple integral with respect to the
velocities. The $3\times 3$ matrix $\mathbf{\sigma}$, with
elements $\sigma_{ij}$, is symmetric, thus it has three real
eigenvalues, say $\sigma_1$, $\sigma_2$, and $\sigma_3$ and unit
eigenvectors $\mathbf{e}_{\sigma,k}$, $k=1,2,3$. The velocity
ellipsoid is defined by the equation
\begin{equation}\label{velel}
\sum_{k=1}^3 {(\mathbf{v}\cdot
\mathbf{e}_{\sigma,k})^2\over\sigma_k^2} = 1~~.
\end{equation}
The shape of the velocity ellipsoid at a point $\mathbf{x}$ gives
the dispersion of the distribution of velocities in different
local directions of motion. In particular, a system is called {\it
isotropic} at the point $\mathbf{x}$ if the velocity ellipsoid at
$\mathbf{x}$ is a sphere, otherwise it is called {\it
anisotropic}. In the case of anisotropic systems, we further
distinguish systems with two or three unequal axes of the
velocity ellipsoid. This distinction is important, because it
allows one to link the kinematic observations available for a
particular system to dynamical features of the same system. For
example, the observation that the velocity ellipsoid in the Solar
neighborhood has three unequal axis led to the discovery that the
stellar orbits in the Solar neighborhood are subject to a `third
integral' (Contopoulos 1960), besides the energy and angular
momentum integrals.

\subsection{Stellar dynamical equilibria - Old and new versions of Jeans' Theorem}

The basic equation governing the time evolution of the distribution function
$f$ in collisionless stellar systems is Liouville's equation implemented in the
$\mu-$space of motion of the Hamiltonian (\ref{ham}), otherwise called
Boltzmann's equation (or Vlasov's equation in plasma physics):
\begin{equation}\label{bol}
{df\over dt} = {\partial f\over\partial t}
+ \mathbf{p}{\partial f\over\partial\mathbf{x}}
- {\partial\Phi\over\partial\mathbf{x}}
{\partial f\over\partial\mathbf{p}} = 0
\end{equation}
where we have adopted the notation $\mathbf{p}\equiv\mathbf{v}$
for canonical momenta, i.e., consider stellar masses equal to
unity. Eq.(\ref{bol}) states that the mass contained within any
infinitesimal volume $d^6\mu$ that travels in phase space along
the orbits corresponding to the potential $\Phi$ (determined by
Eq.(\ref{pot})) is preserved. Furthermore, the measure of the
volume $d^6\mu$ is also preserved (Liouville's theorem). Now, the
morphological regularity and the commonly observed characteristics
of most galaxies suggest that the majority of these systems are
close to a state of statistical equilibrium. Thus, we often look
for {\it steady-state} solutions of Eq.(\ref{bol}) that do not
have an explicit dependence of $f$ on time. Setting $\partial
f/\partial t = 0$ in Eq.(\ref{bol}) yields
\begin{equation}\label{jean1}
\mathbf{p}{\partial f\over\partial\mathbf{x}}
- {\partial\Phi\over\partial\mathbf{x}}
{\partial f\over\partial\mathbf{p}} = \{f,H\}=0
\end{equation}
where $\{\cdot,\cdot\}$ denotes the Poisson bracket operator.

Despite its formal simplicity, the physical content of
Eq.(\ref{jean1}) is remarkable. Consider a {\it fixed} phase
volume $d^6\mu$ centered at some phase space point $(\mathbf{x},
\mathbf{p})$ of a galaxy in steady-state equilibrium. The stars
follow orbits determined by the Hamiltonian (\ref{hamti}).
The orbits remain smooth in the course of time, because there
are no short range stochastic force terms affecting the stars,
similar, for example, to collisions in a perfect gas. Nevertheless,
a detailed equilibrium is established in the phase space, i.e.,
if Eq.(\ref{jean1}) is valid the number of stars leaving the volume
$d^6\mu$ at any moment $t$ must be equal to the number of stars
entering the same volume. Furthermore, the gravitational potential
determining the orbits is given by Eq.(\ref{pot}), which involves
also the positions of the stars. This means that the motions of the
stars are combined in such a way so as to reproduce the same macroscopic
distribution of matter continually in time. For this reason, galactic
equilibria are called {\it self-consistent}, i.e., supported solely
by the orbits of stars within the system. It is a great theoretical
challenge to understand the processes by which nature forms such
remarkable systems.

Consider a system in steady-state equilibrium and suppose that the
mathematical form of the function $f(\mathbf{x},\mathbf{p})$ was
given. Then, according to Eq.(\ref{jean1}), the function $f$
constitutes an integral of the motion in involution with the
Hamiltonian. If, on the other hand, we know by independent means
a complete set of functionally independent integrals of motion $I_1,
I_2, ...$ under the Hamiltonian flow of $H$, it follows that $f$
is necessarily a composite function of the phase space
canonical variables $(\mathbf{x}, \mathbf{p})$ through one or
more of the integral functions $I_1, I_2, ...$. That is
\begin{equation}\label{jean2}
f(\mathbf{x},\mathbf{p})\equiv
f(I_1(\mathbf{x},\mathbf{p}), I_2(\mathbf{x},\mathbf{p}),\ldots)~~~.
\end{equation}
The last equation is known as {\it Jeans' theorem} of stellar dynamics
(Jeans 1915).

Although fundamental in theory, Jeans' theorem, in the above
general form, is of limited usefulness, because it specifies
neither \textbf{a)} which integrals out of the set $I_1, I_2, ...$
should actually appear as arguments in the distribution function
of a specific system, nor \textbf{b)} the explicit form of the
dependence of $f$ on these integrals. Regarding point (a), a
`strong' Jeans theorem proved by Lynden-Bell (1962a) asserts that
only {\it isolating} integrals can be arguments of the function
$f$. An integral $I_i$ is called isolating if the constant value
condition $I_i(\mathbf{x}(t),\mathbf{p}(t))=c_i$ defines a
manifold in phase space of dimension lower than the phase space
dimension (equal to six for three dimensional systems). If we have
a set of $M$ isolating integrals $I_1, I_2,...I_M$, any orbit
$(\mathbf{x}(t),\mathbf{p}(t))$ is restricted on a sub-manifold of
phase space which is the intersection of all the manifolds defined
by the constant value conditions
$I_i(\mathbf{x}(t),\mathbf{p}(t))=c_i, i=1,2,...,M$.

A case of particular interest is when the Hamiltonian of motion $H$ is integrable
in the Arnold-Liouville sense. In three degrees of
freedom systems this means that there are three functionally independent
integrals ($H$ itself can be taken as one of them) which are mutually
in involution, namely
\begin{equation}
\{I_i,I_j\}=\sum_{k=1}^3
{\partial I_i\over\partial x_k}{\partial I_j\over\partial p_k}-
{\partial I_i\over\partial p_k}{\partial I_j\over\partial x_k} = 0~~~.
\end{equation}
In that case, the Arnold-Liouville theorem (see e.g. Arnold 1978
or Giorgilli 2002) asserts that if the manifolds defined by the
constant value conditions $I_i(\mathbf{x}(t),\mathbf{p}(t))=c_i,
i=1,2,3$ are compact, then they are topologically equivalent to
3-tori. The integrals $I_i, i=1,2,3$ are isolating and the strong
Jeans theorem takes the following form: {\it if the Hamiltonian of
a collisionless stellar system in steady-state equilibrium is
Arnold-Liouville integrable, the fine-grained distribution
function $f$ has constant value at all the points
$(\mathbf{x},\mathbf{p})$ of an invariant torus of the system.}

We examine below some simple examples of application of the
strong Jeans theorem in stellar dynamics.

\subsubsection{Spherical systems}

A spherical system in equilibrium is the simplest model of a
galactic system. This model is not very realistic, but it
serves \textbf{a)} to introduce some basic concepts, and
\textbf{b)} as a starting point for the analysis of more realistic
systems. In spherical coordinates, the distribution function
depends on $r$ and on the three velocity components $v_r=\dot{r}$,
$v_{\theta}=r\dot{\theta}$, $v_{\phi}= r\sin\theta\dot{\phi}$,
namely $f\equiv f(r,v_r,v_\theta,v_\phi)$. The mass density depends
only on $r,$ $\rho\equiv\rho(r)$. The orbits are determined by a
spherical potential $\Phi(r)$, given by the solution of Eq.(\ref{pot}):
\begin{equation}\label{pot00}
\Phi(r)=-\frac{GM(r)}{r}-\int_r^\infty{Gdm(r')\over r'} = -{G\over
r}\int_0^r 4\pi r'^2\rho(r')dr' -G\int_r^\infty 4\pi r'\rho(r')dr'
\end{equation}
The orbits obey three isolating integrals of motion in involution,
namely the energy $E=H$, and the components of the angular momentum
$p_\theta = r^2\dot{\theta}$ and $p_{\phi}=r^2sin^2\theta\dot{\phi}$.
The angular momentum vector
$\mathbf{L}=\mathbf{r}\times\mathbf{v}$ is constant and an orbit
is restricted on the plane normal to $\mathbf{L}$. The modulus
$L=|\mathbf{L}|$ is an integral in involution with $p_\phi$, and
the triplet $(E,L,p_\phi)$ is the usual choice of integrals in the
study of spherical systems.

According to the strong Jeans' theorem, the general form of the
distribution function $f(r,v_r,v_\theta,v_\phi)$ in equilibrium
can only be a composite function:
\begin{equation}\label{fsphe}
f\equiv f(E(r,v_r,v_\theta,v_\phi),
L(r,v_r,v_\theta,v_\phi),p_\phi(r,v_r,v_\theta,v_\phi))
\end{equation}
Further restrictions in the form of $f$ can be imposed on the basis of
the kinematical properties of the system under study. For example, if
the system has no preferential kinematical axis (e.g. an axis of rotation),
the integral $p_\phi$ cannot appear as an argument in $f$. This implies
that there is equal probability to find a star moving in a plane of any
possible orientation with respect to the galactic frame of reference.
This is applicable, e.g., to the spherical limit of giant elliptical
galaxies, since there is evidence that the these galaxies are
{\it not} rotationally supported against gravity (e.g. Bertola and
Capaccioli 1975, Illingworth 1977, Davies et al. 1983) but they are
`hot systems' with small or no rotation, in which gravity is balanced
by the distribution of velocities in random directions (e.g. Binney 1976,
1978).  In the spherical limit, we use distribution functions of the form
$f(E)$ or $f(E,L)$. If $f\equiv f(E)$ the galaxy is called isotropic.
The expression for the orbital energy $E=v_r^2/2 + v_\theta^2/2 +
v_\phi^2/2 +\Phi(r)$ yields a symmetric dependence of $f$ on either
of the three velocity components. This implies equal axes of the
velocity ellipsoid $\sigma_r^2=\sigma_\theta^2=\sigma_\phi^2$.
On the other hand, if $f\equiv f(E,L)$ the system is called
anisotropic. The appearance of $L=r\sqrt{v_\theta^2+v_\phi^2}$
in $f$ breaks the symmetry of the functional dependence of $f$
on $v_r$ and $v_\theta$ (or $v_\phi$). The velocity ellipsoid has
two equal axes $\sigma_r^2\neq \sigma_\theta^2=\sigma_\phi^2$.
Since every orbit is confined to a plane, we consider the total
velocity in the transverse direction of motion $v_t^2 = v_\theta^2
+v_\phi^2$ and define the {\it anisotropy parameter} $\beta(r)$
(Binney and Tremaine 1987, p.204):
\begin{equation}\label{aniso}
\beta(r) = 1-{\sigma_r^2(r)\over 2\sigma_t^2(r)}
\end{equation}
with
$$
\sigma_r^2(r)={1\over\rho(r)}
\int_0^{\sqrt{-2\Phi(r)}}
\int_0^{\sqrt{-2\Phi(r)-v_r^2}}fv_r^2v_tdv_rdv_t
$$
and
$$
\sigma_t^2(r)={1\over\rho(r)}
\int_0^{\sqrt{-2\Phi(r)}}
\int_0^{\sqrt{-2\Phi(r)-v_r^2}}fv_t^3dv_rdv_t~~~.
$$
The limits of integration in the above equations are imposed by
the consideration of only bound orbits ($E<0$). The parameter
$\beta$ is a function only of $r$. In practice we find that
realistic systems are nearly isotropic in their central parts,
$\beta(r)\rightarrow 0$ as $r\rightarrow 0$, and {\it radially}
anisotropic in their outer parts, i.e., $\beta(r)\rightarrow 1$
for $r$ large. This means that there is a predominance of radial
orbits in the outer part of the galaxy, i.e., orbits with a large
difference between the apocentric and pericentric distances.
This phenomenon is linked to the relaxation process of galaxies
(section 3). In particular, this is the expected final behavior
of systems  subject to a phase of `collapse' (Eggen et al. 1962),
and this behavior is confirmed by N-Body experiments of violent
relaxation (e.g. Aguilar and Merritt 1990, Voglis 1994a).

\subsubsection{Axisymmetric systems and the `third integral' of motion}

The Hamiltonian of motion in an axisymmetric galaxy can be written
in cylindrical canonical variables $(R,\phi,z,p_R,p_\phi,p_z)$:
\begin{equation}\label{hamaxi}
H \equiv {p_R^2\over 2} + {p_\phi^2\over 2R^2} + {p_z^2\over 2} +
\Phi(R,z)
\end{equation}
where $z$ is the axis of symmetry, and $p_R=\dot{R}$, $p_\phi =
R^2\dot{\phi}$, $p_z=\dot{z}$. Since the azimuthal angle $\phi$
is ignorable, the canonical momentum $p_\phi$ is a second integral
of motion, besides the energy $E=H$. This can be identified to
the z-projection of the angular momentum vector $p_\phi=L_z$.
The study of orbits can be simplified
by considering only the motion on the meridional plane $(R,z)$
\begin{equation}\label{eqmomer}
\ddot{R} = -{\partial\Phi\over\partial R} + {L_z^2\over R^3},~~~~
\ddot{z} = -{\partial\Phi\over\partial z}~~.
\end{equation}
The form of these equations implies that Eq.(\ref{hamaxi}) can be
viewed as a two degrees of freedom Hamiltonian, where $p_\phi$,
replaced by $L_z$, is considered as a parameter. The angular
motion is readily found via $\dot{\phi}=L_z/R^2$. The orbits
on the equatorial plane are defined by a central potential
$\Phi(R,0)$ (provided that the system is symmetric with respect
to the equatorial plane, i.e., the function $\Phi(R,z)$
is even with respect to $z$).

If we consider circular orbits in the equatorial plane for a particular
value of $L_z$, the circular radius is given by the root of the equation:
\begin{equation}
-{\partial\Phi(R_c,0)\over\partial R} + {L_z^2\over R_c^3}=0
\end{equation}
The circular orbit appears as a equilibrium point on the meridional
plane, at $R=R_c,\dot{R}=0$. If we expand the Hamiltonian with respect
to this point we get (ignoring a constant term $L_z^2/2R_c^2$):
\begin{equation}\label{hamaxiexp}
H = {1\over 2}\big(p_Y^2 + p_z^2 + \omega_Y^2Y^2 +
\omega_z^2z^2\big) + \sum_{k=3}^\infty H^{(k)}(Y,z;L_z)
\end{equation}
where $Y=R-R_c$, $\omega_Y^2 =
\dfrac{\partial^2\Phi(R_c,0)}{\partial^2R}+
\dfrac{3L_z^2}{R_c^4}$, $\omega_z^2 =
\dfrac{\partial^2\Phi(R_c,0)}{\partial^2z}$, and the functions
$H^{(k)}(Y,z;L_z)$ are polynomials of degree $k$ in the variables
$Y,z$, depending also on $L_z$ as a parameter.

The Hamiltonian (\ref{hamaxiexp}) has a particular place in the
history of both galactic dynamics and dynamical systems theory
because \textbf{a)} it is the first Hamiltonian for which a `third
integral' of motion was calculated (Contopoulos 1960), and
\textbf{b)} its third order truncation yields the H\'{e}non -
Heiles (1964) Hamiltonian that has served as a prototype of many
studies in nonlinear Hamiltonian dynamical systems.

Special forms of the third integral, e.g. quadratic in the velocities,
were considered by various authors (see references in Ogorodnikov
1965). On the other hand, Contopoulos (1960) explored the question
of whether a third integral of motion
$I$, besides $H$ and $L_z$  can be constructed {\it algorithmically}
for the Hamiltonian (\ref{hamaxiexp}). The existence of $I$
implies that all the orbits are regular (no chaos is present).
Furthermore, according to Jeans' theorem's the integral can
possibly appear as an argument in the distribution function.
Recalling arguments similar to the spherical case, we then find
that if $f$ depends on $I$ the velocity ellipsoid at any point
of ordinary space has unequal axes $\sigma_R\neq \sigma_z$,
while if $f$ does not depend on $I$ the dispersions are equal
$\sigma_R=\sigma_z$. The observational data in our own Galaxy,
in the Solar neighborhood, favored the former case to be true.

Contopoulos (1960) combined two earlier methods of Whittaker (1916)
and Cherry (1924a,b) in order to show that, in the so-called
non-resonant case, when the frequencies $\omega_Y$, $\omega_z$ are
incommensurable, an integral can be {\it formally} constructed in
the form of a polynomial series in the canonical variables, by an
algorithm which is significantly simpler than the use of canonical
transformations as in the Birkhoff - von Zeipel method
(Birkhoff 1927), widely used in Celestial Mechanics. Given that
such formal series are, in general, {\it not} convergent (Siegel
1941), the above series does not represent a real third integral
of the system. However, we shall see below that the series has an
{\it asymptotic} behavior. Namely,
if we define a remainder $R^{(n)}$ for the series at the n-th
order of truncation, the remainder initially decreases as $n$
increases, giving the impression that the series is convergent.
However, after an optimal order $n_{opt}$ the remainder becomes
an increasing function of $n$ (Fig.2), implying divergence of the
series. If we truncate the integral series at the order $n_{opt}$,
we obtain a function $I=I^{(2)}+I^{(3)}+...+I^{(n_{opt})}$ which
is an approximate integral of motion, in the sense that the time
variations $dI/dt$ are quite small, of order $R^{(n_{opt})}$.
The apparent improvement of the accuracy of the integral as $n$
increases (below $n_{opt}$) was checked by a computer program
that calculated the series (Contopoulos and Moutsoulas 1965).
This was confirmed later by Gustavson (1966) with a calculation
of the Birkhoff series in the H\'{e}non - Heiles Hamiltonian.

\begin{figure}[tbp]
\centering{\includegraphics[width=\textwidth]{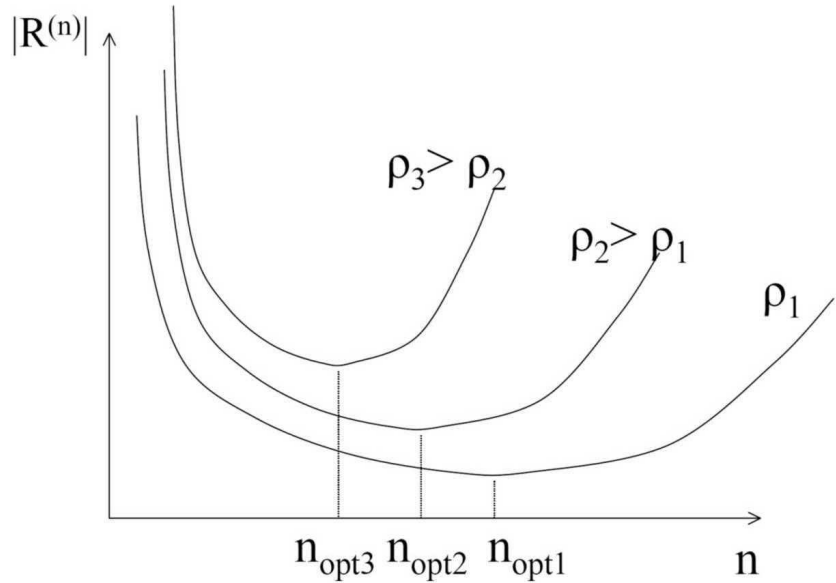}}
\caption{Asymptotic behavior of the `third integral'. The size of
the remainder $|R^{(n)}|$ of the integral series as a function of
the order of truncation $n$ for three different effective
distances from the equilibrium point.} \label{fig02-0}
\end{figure}

While in the non-resonant case the calculation of the third integral by
the direct method of Contopoulos is simpler than by the Birkhoff normal
form, the situation is reversed in the case of resonant integrals, i.e.,
when the frequencies satisfy a commensurability relation $m_1\omega_Y
+m_2\omega_z = 0$, with $m_1,m_2$ integers. A direct method to construct
a resonant integral without use of a normal form was given by Contopoulos
(1963) and exploited in the case of particular resonances by Contopoulos
and Moutsoulas (1965). However, this method involves a `back and forth'
algorithm between successive orders of truncation, which is an essential
complication. The discrete analog of the direct method for symplectic
mappings was given by Bazzani and Marmi (1991), in the non-resonant case,
and by Efthymiopoulos (2005) in the resonant case. However, all these
direct methods are currently superseded by the use of the Birkhoff
method via Lie canonical transformations (Hori 1966, Deprit 1969, Giorgilli
and Galgani 1978, Verhulst 1979), which is the simplest method to
implement in the computer (e.g. Giorgilli 1979).

The Lie method of construction of a third integral was implemented
in axisymmetric galaxies by Gerhard and Saha (1991). These authors
studied various constructive methods of the canonical perturbation
theory. A particular method is to express
the Hamiltonian in the action-angle variables of the
{\it spherical} part of the potential, since analytical
expressions yielding the action-angle variables in terms of the
usual canonical variables are explicitly known in that case. The
Lie method can then be used in order to construct a formal third
integral, besides the energy and $L_z$. The so-obtained
expressions represented a well-preserved integral if the system's
axial ratio was greater than 0.5. Further models of this type were
given by Dehnen and Gerhard (1993), starting from the spherical
isochrone model (subsection 3.4) to represent the unperturbed
system. On the other hand, Matthias and Gerhard (1999) tested
whether the boxy elliptical galaxy NGC 1600 is better fitted by
a two-integral or three-integral model. They found that three-integral
models better reproduce the available kinematic data for the galaxy.
This conclusion was confirmed in subsequent studies
(e.g. Gebhardt et al. 2000, 2001, Cappellari et al. 2002,
Verolme et al. 2002, Dejonghe and Bruyne 2003, sect. 4 and references
therein). We will return
below to the question of the choice between two-integral or
three-integral axisymmetric models, when discussing relevant
results from N-Body simulations.

Besides the harmonic oscillators, or the spherical model, there
are other integrable axisymmetric models that can serve as
starting models for the construction of formal third integrals.
For example, Petrou (1983) constructed a third integral starting
from an axisymmetric model of the form $\Phi(r,\theta) = \Phi_0(r)
+ \Phi_1(\theta)/r^2$ (in polar coordinates) which is known to be
integrable (e.g. Goldstein 1980, p.457). Another possible choice
is an axisymmetric St\"{a}ckel potential (Stiavelli and Bertin
1987, Dejonghe et al. 1996). Other models based on local
St\"{a}ckel fits are reviewed in Dejonghe and Bruyne (2003).

When the potential has a central cusp, a convenient method to calculate
third integrals is the semi-analytical (or semi-numerical) method.
Essentially, this means to start with a plausible model in which
action - angle variables are explicitly constructed, and then to
introduce canonical transformations to new action - angle variables
with generating functions specified through a numerical criterion.
This criterion can be based on either the `theoretical' Hamiltonian
flow (found by the normal form) fitting the true Hamiltonian flow of
the system, or the theoretical tori, viewed as geometrical objects,
fitting the real tori of the system. Such fitting methods were introduced
in galactic dynamics by Ratcliff et al. (1984), McGill and Binney
(1990) and Kent and de Zeeuw (1991).

\subsubsection{(Non-)convergence properties of the `third integral'.
The theory of Nekhoroshev}

The optimal order of truncation $n_{opt}$ of the third integral series, as well
as the size of the optimal remainder $R^{(opt)}$ are questions that can be
examined in the framework of the theory of Nekhoroshev (Nekhoroshev 1977,
Benettin et al. 1985, Lochak 1992, P\"{o}shel 1993), implemented, in particular,
in the case of elliptic equilibria (Giorgilli 1988, Fass\`{o} et al. 1998,
Guzzo et al. 1998, Niederman 1998).
This theory states that as the parameter $\epsilon$ that quantifies the
perturbation of the system from an integrable system decreases, the size
of the optimal remainder becomes {\it exponentially small} in $1/\epsilon$,
that is:
\begin{equation}\label{nekho}
R_{opt} = O\left(\exp\left(-{1\over\epsilon^p}\right)\right)
\end{equation}
where the exponent $p$ depends on the number of degrees of freedom
of the system under study. Conversely, approximate integrals of
the type of the `third integral' retain almost constant values for
times exponentially long in $1/\epsilon$, that is, $T_{Nek} =
O(\exp(1/\epsilon^p))$. In galactic Hamiltonians such as
(\ref{hamaxiexp}), the effective perturbation $\epsilon$ is
identified to the average distance $\rho$ of an
orbit from the elliptic equilibrium. Thus, without being able to
prove the existence of an exact third integral for the orbits
$R(t),z(t)$ on the meridional plane, we can assert that, even if
such orbits are chaotic, an orbit will behave effectively like
regular for a time exponentially long in $1/\rho$, where $\rho$ is
the distance of the orbit from the equilibrium $R=R_c,z=0$.

A heuristic derivation of the formula $R_{opt} = O\big(\exp(-1/\rho)\big)$,
based on the use of integrals calculated by the direct method, can be given
following a theorem by Giorgilli (1988). We make the derivation in
action - angle variables $(J,\phi)$. We set
$Y=\sqrt{2J_1}\sin\phi_1$,
$p_Y=\sqrt{2J_1}\cos\phi_1$, $z=\sqrt{2J_2}\sin\phi_2$,
$p_z=\sqrt{2J_2}\cos\phi_2$, and $\omega_1\equiv\omega_Y$, $\omega_2\equiv
\omega_z$. The Hamiltonian (\ref{hamaxiexp}) takes the form
\begin{equation}\label{hamaxiacan}
H=\omega_1J_1+\omega_2J_2 + \sum_{k=3}^\infty
H^{(k)}(J_1,J_2,\phi_1,\phi_2)
\end{equation}
where the functions $H^{(k)}$ are of degree $k/2$ in the actions and
contain trigonometric terms of the form
$e^{i(k_1\phi_1+k_2\phi_2)}$ with $k_1,k_2$ integers,
$|k_1|+|k_2|\leq k$ and of the same parity as $k$. We look
for a third integral as a series yielding a correction to the
action $J_1$, or $J_2$ (each of the actions is an exact integral
in the harmonic oscillator limit of Eq.(\ref{hamaxiexp})). We thus
set
$$
I=J_1 + \sum_{k=3}^\infty I^{(k)}(J_1,J_2,\phi_1,\phi_2)~~,
$$
the functions $I^{(k)}$ satisfying the same properties as the functions
$H^{(k)}$ (we set $I^{(2)}=J_1$).
The integral is calculated by splitting the integral condition $\{I,H\}=0$
to terms of equal order. This yields the relation
\begin{equation}\label{intrec}
\{I^{(k)},H^{(2)}\} = -\sum_{s=2}^{k-1}\{I^{(s)},H^{(k+2-s)}\}
\end{equation}
with $H^{(2)} = \omega_1J_1+\omega_2J_2$. Eq.(\ref{intrec}) can be
solved recursively to yield $I^{(k)}$ in the k-th step from the
terms $I^{(s)}$, $s=2,...,k-1$ determined in the previous steps.
If we express the terms $I^{(s)}$ in sums of Fourier terms of the
form $J_1^{s_1\over 2}J_2^{s_2\over 2}
e^{i(k_1\phi_1+k_2\phi_2)}$, we readily see that the algebraic
nature of the direct scheme (\ref{intrec}) is quite similar to
that of the Birkhoff-von Zeipel normal form scheme: Each Fourier
term is an {\it eigenfunction} of the linear differential operator
$\{\cdot,H^{(2)}\}$ with eigenvalue equal to
$-i(k_1\omega_1+k_2\omega_2)$, that is
$$
\{J^{s\over 2}e^{ik\cdot\phi},H^{(2)}\} =
-i(k\cdot\omega)J^{s\over 2}e^{ik\cdot\phi}
$$
where we use the abbreviations $J^{s\over 2}\equiv J_1^{s_1\over
2} J_2^{s_2\over 2}$, $k\equiv(k_1,k_2)$,
$\omega\equiv(\omega_1,\omega_2)$. This implies that the solution
of Eq.(\ref{intrec}) for $I^{(k)}$ yields precisely a sum of the same
Fourier terms as in the r.h.s. of the same equation, each term
being divided by the divisor $k\cdot\omega$. The presence of
divisors is important because, for generic incommensurable
frequency vectors $\omega$ there are integer vectors $k$ that can
be found, which render the product $k\cdot\omega$ a {\it small
divisor}. For example, from number theory it is known (e.g. Berry
1978) that most irrationals satisfy diophantine conditions of the
form
\begin{equation}\label{dioph}
|k\cdot\omega|\geq {\gamma\over |k|^\tau}
\end{equation}
with $\gamma$ an $O(1)$ constant and $\tau$, the diophantine
exponent, depending on the number of degrees of freedom ($\tau\geq
n-1$). This means that, as $|k|$ increases, the minimum size of
divisors appearing in the recurrent solution of Eq.(\ref{intrec})
decreases, i.e., the divisors become smaller and smaller.
Furthermore, as one repeatedly implements the recurrence relation,
such small divisors {\it accumulate} in the form of products in
the denominators of the various integral terms. That is, there are
Fourier terms $f^{(k)}$ in $I^{(k)}$ with an accumulation of divisors
yielding a size
\begin{equation}\label{fk}
||f^{(k)}||\sim{F^{(k)}\over a_3a_4...a_k}
\end{equation}
with divisors $a_s, s=3,4,\ldots,k$ satisfying $a_s\sim 1/s^\tau$
according to Eq.(\ref{dioph}). The numerator $F^{(k)}$ in
Eq.(\ref{fk}) can be estimated by the remark that, for any
term $I^{(s)}$, the Poisson bracket in the r.h.s. of Eq.(\ref{intrec})
means to take the derivatives $\partial I^{(s)}/\partial J$, or
$\partial I^{(s)}/\partial\phi$, which both cause the appearance of
a factor $O(s)$ in front of the corresponding Fourier terms of
$I^{(s)}$. Hence, the repeated action of Poisson brackets, up to order
$k$ creates a factor $F^{(k)}\sim O(3)O(4)...O(k)\sim k!$ in the
numerator of the Fourier terms of $I^{(k)}$ (see Efthymiopoulos et al.
2004 for a more detailed analysis). Putting these remarks
together, the size of Fourier terms (\ref{fk}) can be estimated as
$||f^{(k)}||\sim k!^{\tau+1}$. If we now consider an orbit of
effective distance $\rho$ from the equilibrium, we have
$J^{s/2}\sim\rho^s$ for this orbit, so that the value of the
remainder of the formal series at the k-th order of truncation can
be estimated as:
\begin{equation}\label{rk}
R^{(k)}\sim k!^{\tau+1}\rho^k
\end{equation}
The estimate (\ref{rk}) contains the essential result regarding the
asymptotic character of formal series: using Stirling's formula $k!\sim
(k/e)^k$, for large k, we have $R^{(k)}\sim (k^{\tau+1}\rho/e^{\tau+1})^k$.
We then want to check whether the remainder decreases or increases as
the order $k$ of calculation of the formal integral increases. We see
immediately that as long as $k<<e/\rho^{1/(\tau+1)}$, the remainder
decreases with $k$, while if $k>>e/\rho^{1/(\tau+1)}$ the remainder
increases with $k$. Thus the optimal order is at an order $n_{opt}$
were the remainder is minimum, which can be estimated as $n_{opt}\sim
e/\rho^{1/(\tau+1)}$. Inserting this in Eq.(\ref{rk}) we find the
value of the remainder at the optimal order of truncation
$R_{opt}\sim \exp(-n_{opt})\sim \exp(-1/\rho^{(1/(\tau+1)})$, which
leads to Nekhoroshev's formula of exponentially small time variations
of the truncated integral $I=I^{(2)}+I^{(3)}+...+I^{(n_{opt})}$.

In generic nearly-integrable
Hamiltonian systems of the form $H(J,\phi) = H_0(J)+
\epsilon H_1(J,\phi)$ the Nekhoroshev theory is much more complicated
than in the simple case of elliptic equilibria. The main complication
is that the frequencies $\omega(J)=\partial H/\partial J$ depend on
the actions, a fact that renders necessary the separate treatment
of several non-resonant or resonant domains that coexist in the
space of actions. This treatment is the so-called geometric part
of Nekhoroshev theorem (see Morbidelli and Guzzo 1997 for an
instructive introduction and Giorgilli 2002 for a rigorous but still
pedagogical proof). On the other hand, the analytic part
of the theorem is treated more easily if we avoid dealing with repeated
Poisson brackets, as above, acting on the series terms of successive orders
of normalization. This is done by setting from the start a number of
assumptions regarding the analyticity properties of the Hamiltonian
under consideration in a complexified space of actions
and angles and by using various forms of Cauchy theorem for analytic
functions. This simplifies considerably the proof of the analytical
part of the theorem.

Nevertheless, it seems that when one wants to find realistic estimates
as regards the optimal order of truncation and the optimal value
of the remainder, one has to rely on the classical methods of
analysis of series convergence. The first systematic exploitation
of these questions, referring to the method of Birkhoff series,
was made by Servizi et al. (1983), who calculated
`pseudoradii of convergence' for the Birkhoff normal form in
symplectic mappings representing the Poincar\'{e} surface of
section of 2D Hamiltonian systems. Kaluza and Robnik (1992)
found that there was no indication of divergence of the formal
series below the order $n=15$ in the H\'{e}non - Heiles model. A
particular application in the problem of stability of the Trojan
asteroids (Giorgilli and Skokos 1997) showed that the optimal
order of truncation of the integrals in this case is beyond $n=32$
(in some cases we find $n_{opt}>60$, Efthymiopoulos and S\'{a}ndor
2005). But a precise treatment of the problem was made only very
recently (Contopoulos et al. 2003, Efthymiopoulos et al. 2004). In
these works scaling formulae are given yielding the optimal order
of truncation as a function of the distance from the elliptic
equilibrium and of the number of degrees of freedom. These
formulae are derived theoretically and verified by computer
algebraic calculations. A recent application in the case
of galactic potentials was given by Belmonte et al. (2006).

The estimate $n_{opt}\sim e/\rho^{1/(\tau+1)}$ implies that the
optimal order of truncation is smaller, and the value of the
optimal remainder is larger, for larger $\rho$. This behavior is
shown schematically in Fig.2. On the other hand, when $\rho$
surpasses a threshold value $\rho_c$, at which $n_{opt}$ approaches
the lowest possible value $n_{opt}=3$,
there is no more meaning in calculating a third
integral $I$, since the series will be divergent from the start.
This situation corresponds physically to the fact that for
$\rho>\rho_c$, or energy $E>E_c\sim\rho_c^2$, the majority of
orbits in phase-space are {\it chaotic}. In fact, in generic
Hamiltonian systems of the form (\ref{hamaxiexp}), some degree of
chaos exists in the phase space of motions for arbitrarily small
values of the energy. When regular and chaotic orbits co-exist,
the system is said to have a {\it divided} phase-space (e.g.
Contopoulos 2004a, pp.17-19). However, for values
$E<E_c\sim\rho_c^2$, the largest measure in phase space is
occupied by regular orbits, laying on invariant tori, while for
$E>E_c$ it is occupied by chaotic orbits. In the H\'{e}non -
Heiles system, for example, $E_c=1/6$.

The occurrence of a divided phase space, which is a generic
phenomenon, renders problematic the implementation of Jeans'
theorem in realistic stellar systems because there is no uniform
answer regarding the number and the form of integrals (or approximate
integrals) which are preserved in different regions of the phase
space. We shall come back to this question in subsection (2.5).

\subsubsection{Triaxial systems}

The paradigm of integrable triaxial galactic potential models are
ellipsoidal St\"{a}ckel potentials (St\"{a}ckel 1890, 1893, Eddington
1915, Kuzmin 1956, Lynden-Bell 1962b, de Zeeuw and Lynden-Bell 1985):
\begin{equation}\label{stackel}
\Phi(\lambda,\mu,\nu) =
-{F_1(\lambda)\over(\lambda-\mu)(\lambda-\nu)}
-{F_2(\mu)\over(\mu-\nu)(\mu-\lambda)}
-{F_3(\nu)\over(\nu-\lambda)(\nu-\mu)}
\end{equation}
where $(\lambda,\mu,\nu)$ are the so-called ellipsoidal coordinates.
These can be related to Cartesian coordinates $(x,y,z)$ via the three
solutions for $u$ of the equation
\begin{equation}
{x^2\over u-a^2}+{y^2\over u-b^2}+{z^2\over u-c^2}=1
\end{equation}
where the constants $a^2\geq b^2\geq c^2$ represent the axes of concentric
ellipsoids. The form of the two integrals, besides the Hamiltonian, is given
e.g. in Contopoulos (1994) where the main types of orbits are also analyzed.
A case of particular interest for galactic dynamics is the
{\it perfect ellipsoid}. The density is given by
\begin{equation}\label{perfell}
\rho = {\rho_0\over (1+m^2)^2}
\end{equation}
where $m$ is the ellipsoidal radius defined as:
\begin{equation}\label{muel}
m^2 = {x^2\over a^2}+{y^2\over b^2}+{z^2\over c^2}~~.
\end{equation}
The form of the integrals in that case is given e.g. in de Zeeuw
and Lynden- Bell (1985). The density function (\ref{perfell})
belongs to a more general class of density functions that can
serve as models of triaxial galaxies
\begin{equation}
\rho = {\rho_0\over (1+m^2)^q}~~~.
\end{equation}
However, a numerical study (Udry and Pfenniger 1988) indicated that,
for $q>0$, only the value $q=2$ of the perfect ellipsoid yields an
integrable system, since other values yield systems containing stochastic
orbits with positive Lyapunov exponents.

The use of an ellipsoidal radius $m$ is an easy method to
`produce' triaxial systems from known spherical systems, namely if
one has a given potential or density function for the spherical
system $\Phi(r),\rho(r)$, one obtains a triaxial system by
replacing $r$ with $m$ in {\it either} the potential $\Phi$ {\it
or} the density $\rho$. One then has to solve again Poisson's
equation for the missing function. Examples of this type of models
are reviewed in Merritt (1999, sect.1).

If the potential near the center of a triaxial galaxy is close to
harmonic, one may try to calculate approximate integrals of motion
of the type of the `third integral'. Namely, expanding the
potential as:
\begin{equation}\label{pot3dexp}
\Phi(x,y,z) = {1\over 2}(\omega_x^2x^2+\omega_y^2y^2+\omega_z^2z^2) +
\sum_{k=3}^\infty P_k(x,y,z)
\end{equation}
where the functions $P_k$ are polynomial of degree $k$ in the cartesian
coordinates $x,y,z$, one looks for approximate integrals of the form
\begin{equation}\label{ix}
I_x = {1\over 2}\omega_x^2x^2 +...
\end{equation}
and similarly for $I_y$, $I_z$. Such integrals can be constructed
either by the direct method or by the Birkhoff normal form. If
resonances are present among the frequencies, one may look for
resonant integrals that are given as linear combinations of the
actions $J_x$, $J_y$ and $J_z$ with integer coefficients (Verhulst
1979, de Zeeuw and Merritt 1983, Belmonte et al. 2006).  As in
the two degrees of freedom case, the validity of the approximation
of such integrals is determined by the theory of Nekhoroshev. A
particular example was studied in Contopoulos et al. (1978). It
was found that there are cases where \textbf{a)} {\it two}
integrals, \textbf{b)} {\it no} integrals, and \textbf{c)} {\it
only one} integral besides the Hamiltonian, appear to be
well-preserved along the orbits' flow. Case (c) is the most
interesting, because it contradicts a claim by Froeschl\'{e} and
Scheidecker (1973) that the number of preserved integrals besides
the energy is either two or zero, that is, the orbits either lie
on 3D invariant tori of the phase space or they are completely
chaotic. There was a recent revival of interest in this issue
after the remark (Varvoglis et al. 2003) that the case of
preservation of one more integral besides the energy (or the
Jacobi constant in rotating systems) may be
associated with the phenomenon of `stable chaos' (Milani and
Nobili 1985, 1992) that is well known in Celestial Mechanics.
Besides the differences in the form of local velocity ellipsoids,
that were discussed above, the question of other consequences of
the number and form of preserved integrals on the dynamical
structure of galactic systems is still open.

\subsubsection{Jeans' theorem in systems with divided phase-space}

As already mentioned, the occurrence of a divided phase space,
which is a generic phenomenon in stellar systems apart from the
idealized spherical or St\"{a}ckel cases, renders problematic the
implementation of Jeans' theorem in realistic stellar systems.
This is because \textbf{a)} it is not clear how to incorporate
approximate integrals of the form of the `third integral' in the
arguments of the distribution function, \textbf{b)} such integrals
have different expressions when {\it resonances} are present, each
resonance being characterized by its own form of resonant
integral, and \textbf{c)} the integrals are not valid for chaotic
orbits, which, however, co-exist with the regular orbits within
any hypersurface of the phase space defined by a constant energy
condition.

As regards the form of the distribution function in the chaotic
sub-domain of the phase-space, the theorem of Arnold (1964) suggests
that in generic Hamiltonian systems of more than two degrees of freedom
there is an a priori topological possibility for $O(1)$
excursions of chaotic orbits in phase space, even if the system
differs from an integrable system by an arbitrarily small perturbation
$O(\epsilon)$. Such excursions are possible through heteroclinic
chains that span the whole interconnected chaotic subset of the
phase space, i.e., the {\it Arnold web}. Furthermore, if
$\epsilon$ is large enough, there are large chaotic domains formed
by the `resonance overlap' mechanism (Contopoulos 1966, Rosenbluth
et al. 1966, Chirikov 1979). In that case, the results of numerical
integrations (e.g. Contopoulos et al. 1995)
indicate that the transport of chaotic orbits is efficient enough
so as to create a uniform measure throughout any connected chaotic
domain. On the other hand, as $\epsilon\rightarrow 0$, the
resonance overlap mechanism almost disappears and the transport of
chaotic orbits through the Arnold web occurs in a timescale
characteristic of Arnold diffusion. The latter is much slower than
any timescale of relevance to galactic dynamics, as exemplified in
a number of studies (e.g. Laskar 1993a, Giordano and Cincotta 2004,
Guzzo et al. 2005). The slowness of Arnold diffusion has the
consequence that there may be considerable deviations of the
phase-space density from a uniform measure in the chaotic
subdomain of the phase space. Such deviations are opposed to the
validity of Jeans' theorem, i.e., that the distribution function
$f(E)$ is constant within the chaotic subdomain of any
hypersurface of constant energy. In that sense, the latter
statement should be true only in integrable isotropic systems such
as the spherical systems considered in subsection 2.3.

This is precisely the claim made by Binney (1982a) in a paper that
initiated a fruitful discussion on the interconnection between
{\it global phase space dynamics}, on the one hand, and the form
of the distribution function, on the other hand. In particular, an
important line of research in galactic dynamics since the 80s has
been the detailed exploration of the various types of regular or
chaotic orbits that
co-exist in a galaxy, as well as their relative statistical
importance in creating building blocks of the self-consistent
distribution function of the system. This research on {\it
self-consistent models} of galaxies, discussed in section 4 below,
is today a very active area of research. Furthermore, an even more
powerful line of research on the same problem has appeared in recent
years: exploring the orbital content of systems
resulting from {\it N-Body simulations}. This was made possible
after the use of appropriate `smooth potential' techniques
of simulation of the N-Body problem that yield smooth solutions of
the equations of motion and of the variational equations for stellar
orbits. In section 5, we refer to the main results of this
approach which yields the closest approximations to the study of
realistic stellar systems, since the equilibria reached in N-Body
simulations are by definition \textbf{a)} self-consistent and
\textbf{b)} stable. The above methods are quite powerful
and have yielded some important results towards understanding
the equilibria of systems with divided phase space.

Our basic understanding today is that there are two types of
orbits that play a major role in the equilibria of galaxies. These
are \textbf{a)} the regular orbits, and \textbf{b)} chaotic orbits
exhibiting significant chaotic diffusion over times comparable to
the Hubble time. In particular, the orbits in the chaotic subdomain
are important if they can spread and produce an almost uniform
measure in this domain at times comparable to the age of the
system. On the other hand, weakly chaotic orbits that exhibit
`stickiness' phenomena (e.g. Contopoulos 1971, Karney 1983,
Efthymiopoulos et al. 1997) play a role similar to the role of
regular orbits. Such differences can be quantified by the
introduction of appropriate measures of the inverse Lyapunov
number, i.e., the Lyapunon time of orbits (e.g. Voglis et al.
2002).

In the context of the above discussion, we can mention a proposal
of a new form of Jeans' theorem by Merritt (see for example
Merritt and Fridman 1996, Merritt 1999), that is applicable to
systems with a divided phase space: ``{\it The phase-space density
of a stationary stellar system must be constant within every
well-connected region}''. The definition of `well-connected' is
``...one that cannot be decomposed into two finite regions such
that all trajectories lie on either one or the other (what the
mathematicians call `metric transitivity')'' (Merritt 1999).

\begin{figure}[tbp]
\centering{\includegraphics[width=10cm]{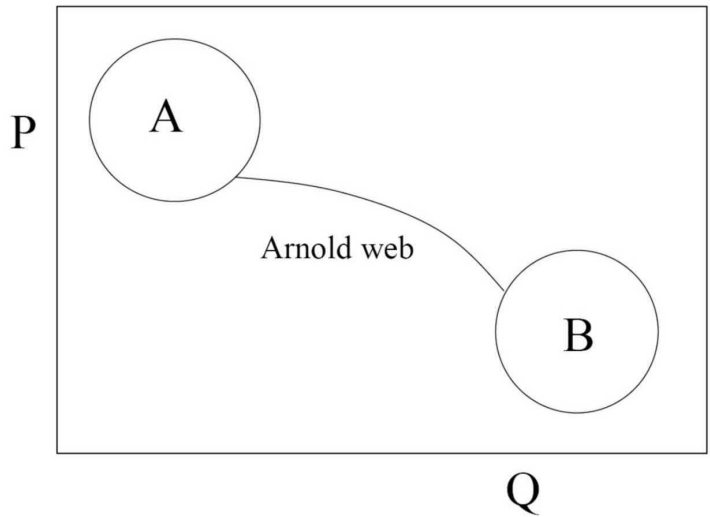}}
\caption{A schematic representation of the phase-space. Regions A
and B communicate via the Arnold web.} \label{fig03-0}
\end{figure}

In the idealized case of a phase fluid set from the start to
satisfy the above condition, the above version of Jeans' theorem
corresponds essentially to the preservation of the phase space
density under the system's Hamiltonian flow. In practice, however,
a definition of `well-connected' region such as the above one,
i.e., based only on topological arguments, may not be so
convenient in describing galactic equilibria. We can give the
following qualitative argument: Suppose the 6D phase space of a
galactic system is represented schematically as the $(Q,P)$ space
of Fig.3. Suppose also that the system's Hamiltonian $H$ differs
from an integrable Hamiltonian, with exact integrals
$I_1,I_2,I_3$, by an arbitrarily small perturbation, of order
$O(\epsilon)$. According to Nekhoroshev theorem, if $\epsilon$ is
below a threshold, there are approximate integrals $\tilde{I}_1,
\tilde{I}_2,\tilde{I}_3$ that have variations of order
$O(\exp(-1/\epsilon))$ over timescales of order
$O(\exp(1/\epsilon))$, i.e., much longer than the age of the
galaxy. Thus, for all practical purposes, we may describe the
system by a distribution function depending on these approximate
integrals $f(\tilde{I}_1,\tilde{I}_2,\tilde{I}_3)$. Consider now
two different regions of $(Q,P)$, region A and region B (Fig.3),
with an $O(1)$ separation in phase-space (in units normalized to
the overall extent of the phase space in $P$ and $Q$). As a consequence,
the values of $\tilde{I}_i$, which are functions of the variables $(Q,P)$,
will in general also have an $O(1)$ difference in the two regions,
that is $|\tilde{I}_i(A)-\tilde{I}_i(B)|=O(1)$ . Since these integrals
are arguments of the distribution function, it follows that the
same order of the difference will also appear in $f$, that is
\begin{equation}\label{fafb}
|f(A)-f(B)| = O(1)~~.
\end{equation}

Given, now, that the Nekhoroshev theorem for approximate integrals
is valid in open domains of the phase space, it follows that
Eq.(\ref{fafb}) is valid for $f(A),f(B)$ standing for the value of
the distribution function at {\it any} pair of points inside the
regions A and B respectively, provided that the approximate
integrals $\tilde{I}_i$ are well preserved in both regions. On the
other hand, according to the KAM theorem (Kolmogorov 1954, Arnold 1963,
Moser 1962), there is a chaotic subset of
measure $O(\epsilon)$ in region A, which is the compement of the
invariant tori of A, and a similar subset in region B. Suppose that
the two subsets communicate via the Arnold web. Then, according to
the previous definitions, the two subsets belong to one 'well-connected'
chaotic region and we should have $f(A)-f(B) = 0$ for any pair of
points in A and B belonging to this region. Thus we see that if we
use the approximate integrals $\tilde{I}_i$ as arguments in the
distribution function we reach a different conclusion (Eq.(\ref{fafb}))
than if we use the concept of well-connectedness. This is
because the integrals $\tilde{I}_i$ are not exact, but they are
preserved for times of the order of the Nekhoroshev time $t_{Nek}$.
Thus the equalization of  $f(A)$ and $f(B)$ in the chaotic subset
can happen only after a time $t>t_{Nek}$, which is much larger than
the age of the system.

The above example shows that a more pragmatic definition of what
`well-connected' means is required in the case of galaxies, in
order to take into account the fact that the topological
well-connectedness may not have always practical dynamical
implications for the equilibria of galaxies. This is because the
lifetime of galaxies is much smaller than the typical Nekhoroshev
time.

A numerical example of the form of the distribution function in
systems with divided phase space was given by analyzing the orbits
and approximate integrals in the phase space of systems produced
by N-Body simulations (Efthymiopoulos 1999, Contopoulos et al.
2000, Efthymiopoulos and Voglis 2001, Contopoulos et al. 2002).
Fig.4 (Contopoulos et al. 2000) shows one example of a nearly
prolate system. This system resulted from a collapse simulation
with cosmological initial conditions (Efthymiopoulos and Voglis
2001). The self-consistent gravitational potential is calculated
by the self-consistent field code of Allen et al. (1990). If we
ignore triaxial terms, the potential can be expanded in a
polynomial series in the $(R,z)$ variables, namely:
\begin{equation}\label{nbpotaxi}
\Phi(R,z) = \sum_{k=0}^8\sum_{l=0}^8g_{kl}R^{2k}z^{2l}
\end{equation}
where $z$ is the long axis of the system and the coefficients
$g_{kl}$ are specified numerically, via the code potential. The
form of the potential (\ref{nbpotaxi}) is such that a third
integral can be calculated in the form of series. We calculate a
different integral for box orbits (non-resonant integral) or for
higher-order resonant orbits (e.g. 1:1 resonance for tube orbits).
\begin{figure}[tbp]
\centering{\includegraphics[width=\textwidth]{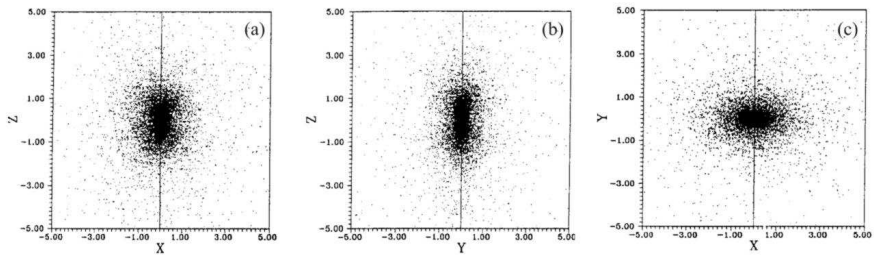}}
\caption{Projection of the final state of an N-Body collapse
experiment in the three planes (a) X-Z, (b) Y-Z, (c) X-Y of
ordinary space. The system is nearly prolate, with one long axis
(Z) and two short axes (X,Y) (after Contopoulos et al. 2000).}
\label{fig04-0}
\end{figure}

The question, now, is whether such integrals should appear as
arguments in the distribution function of the system. The answer
is affirmative, as indicated by Fig.5. Panel (a) shows a
Poincar\'{e} surface of section $(R,\dot{R})$ for an energy
$E=-1.6\times 10^6$ (in the N-Body units) which is close to the
central value of the potential well ($-2\times 10^6$), and angular
momentum $L_z$ close to zero. We then integrate the orbits of the
{\it real} particles of the N-Body system with energies in a bin
centered at the above value of $E$, until each orbit intersects
the Poincar\'{e} section for the first time. By this numerical
process, a particle located on an invariant torus of the system,
that corresponds to a particular value of the third integral $I$,
is transferred to a point on an {\it invariant curve} of the
section $(R,\dot{R})$ where the section is intersected by the
torus. This also means that if the phase-space density
(distribution function $f$) depends on $I$, the {\it surface
density} of points in the section $(R,\dot{R})$ will also be
stratified in such a way that the {\it equidensities} should
coincide with the invariant curves corresponding to different
label values of $I$. Precisely, this is what we see in Fig.5b.
Namely, the equidensity contours of the distribution of the real
particles in the surface of section have a good coincidence with
the invariant curves (shown together in Fig.5c). This is a
numerical indication that the integral $I$ should, indeed, be
included as an argument in $f$ (see also Contopoulos et al. 2002).
We have calculated numerically the dependence of the surface
density $f_S$ on $I$ and found it to be exponential (Fig.5d).

\begin{figure}[tbp]
\centering{\includegraphics[width=\textwidth]{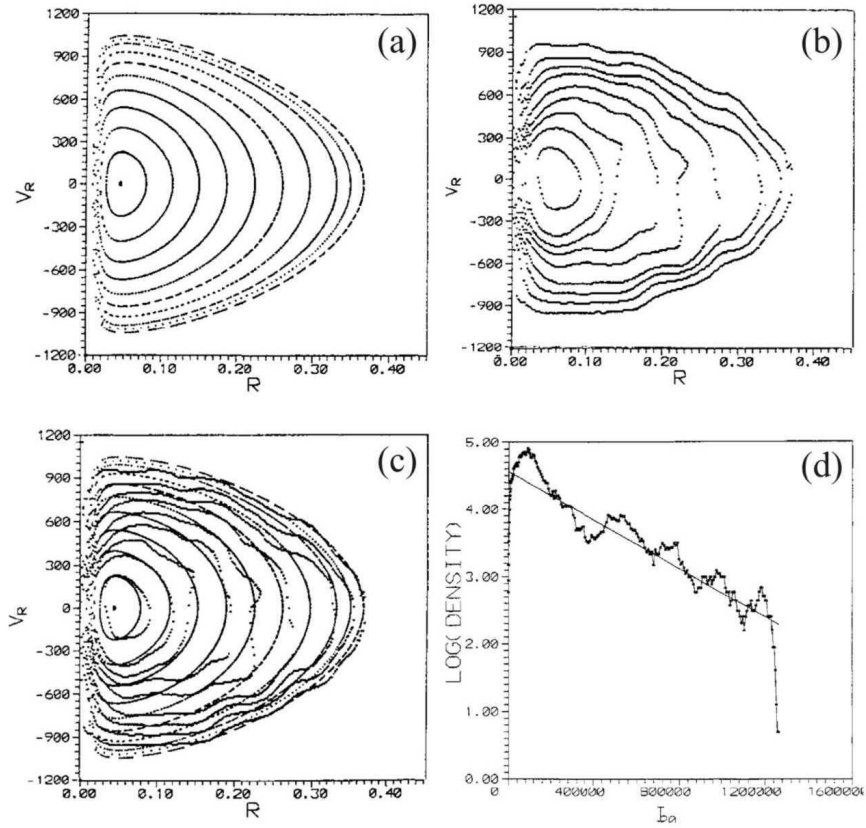}}
\caption{(a) The Poincar\'{e} surface of section of the system of
Fig.4 for energy $E=-1.6\times 10^6$ (in the N-Body units) and
$L_z$ very close to zero $L_z = 0.045$. (b) The equidensity
contours of the distribution of the real particles in the
Poincar\'{e} section, for energies within a bin $\Delta E =
2\times 10^4$ around the value $E$ of (a) and angular momentum
$|L_z|\leq 0.09$. (c) the plots (a) and (b) together. (d)
Exponential dependence of the distribution function on the value
of the third integral along the invariant curves of (a) (after
Contopoulos et al. 2000).} \label{fig05-0}
\end{figure}

For larger energies, the divided nature of the phase-space is
clearly manifested (Fig.6a). In particular, besides the region of
invariant curves corresponding to box orbits (A), we distinguish a
second island around the 1:1 resonance (B) as well as a connected
chaotic domain (C) separating the two regular domains. Some finer
details, e.g., secondary resonances (D) are distinguished but they
are not dynamically so important. If, now, we compare this figure
to the equidensity plot of the distribution of particles
(Figs.6b,c), the tendency to have a distribution stratified
according to the underlying phase-space structure is again visible
to a large extent. This indicates that both the non-resonant third
integral, yielding the tori of region (A), and the resonant
integral yielding the tori of (B), should appear locally as
arguments of the distribution function $f$ (the dependence of $f$
on $I$ in region A is again exponential, Fig.6d).

\begin{figure}[tbp]
\centering{\includegraphics[width=\textwidth]{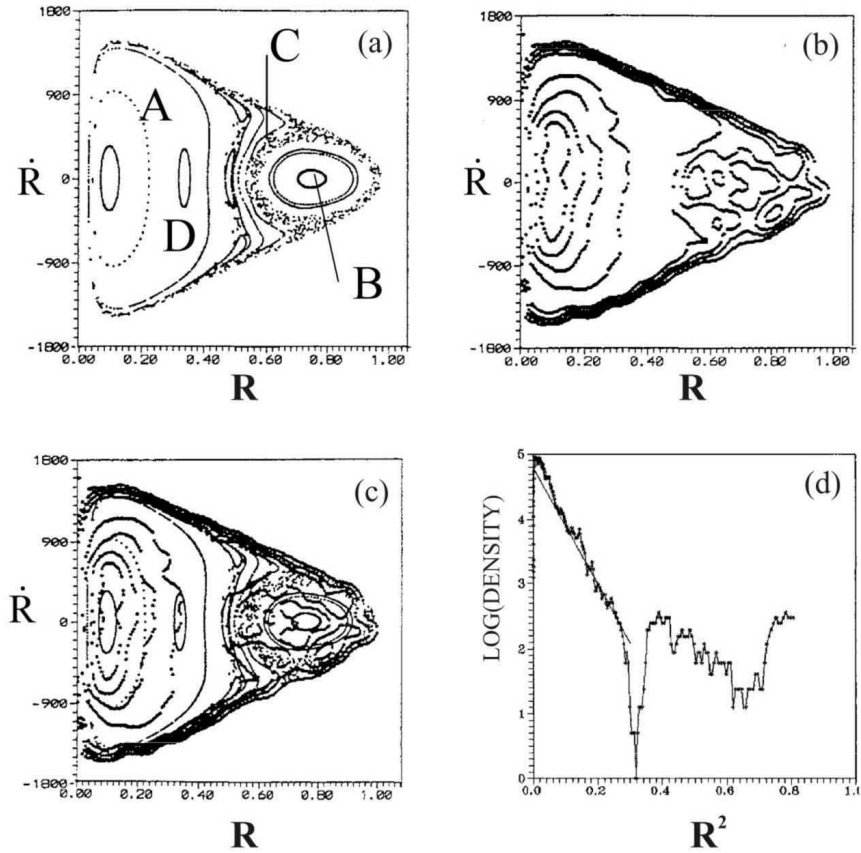}}
\caption{As in Fig.5, for a different value of the energy
($E=-9\times 10^5$). The region (A) corresponds to box orbits, (B)
to loop orbits (1:1 resonance), (C) chaotic orbits, and (D) a
secondary resonance inside the domain of box orbits. In (d) we use
the approximation $I\propto R^2$ (after Contopoulos et al. 2000).}
\label{fig06-0}
\end{figure}

For still larger energies, the chaotic domain occupies a large
volume of the phase space (Fig.7a). We find that the phase-space
density of the real particles in the connected chaotic domain is
nearly constant. Fig.7b shows the density on the Poincar\'{e}
section along a constant line $R=0.65$, as a function of $\dot{R}$.
The variations shown in Fig.7b are {\it not} completely due to the
sampling noise, but, in general, they are small enough so as to
allow us to characterize the density as nearly constant in the
connected chaotic domain (C). In this case Merritt's version of
Jeans' theorem is applicable.

\begin{figure}[tbp]
\centering{\includegraphics[width=\textwidth]{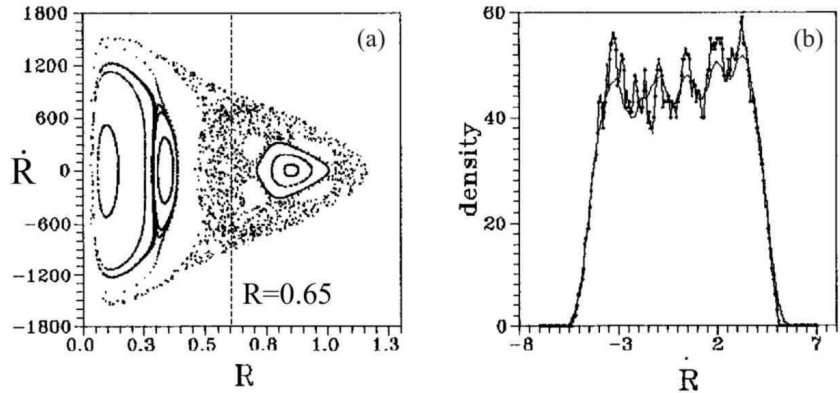}}
\caption{(a) As in Fig.5a for energy $E=-8\times 10^5$. (b) the
density of real particles of the N-Body system in a slice around
the line $R=0.65$ is almost constant (after Contopoulos et al.
2000).} \label{fig07-0}
\end{figure}

The nature of the above questions prevents one from making clearcut
statements as per what phenomena introduced by regular or
chaotic orbits should be considered as dynamically important. Let
us notice, however, that galaxies are quite complex systems and
such questions have not yet been fully explored even in simple toy
models of basic research in Hamiltonian dynamical systems. We
mention one example which is of particular importance in the
study of global dynamics of galaxies: the distinction between
Arnold diffusion (Arnold 1964) and resonance overlap diffusion
(Contopoulos 1966(7), Rosenbluth et al. 1966, Chirikov 1979) and the
role of these two types of diffusion in galaxies. The difference
between these two types of diffusion is topological, but it is
also a difference in the diffusion rate. As regards the rate of
Arnold diffusion, there is a general belief that this should be
connected to Nekhoroshev theorem and that, hence, it is very slow
to be of any importance in galaxies. This was partly
verified recently by the interesting numerical experiments of
Froeschl\'{e} and his collaborators (Froeschl\'{e} et al. 2000,
Guzzo et al. 2002, 2005, Lega et al. 2003). More work is requested
in order that such simulations help us clarify questions such as
what is a pragmatic definition of `well-connected' domains of
phase space and how to implement such ideas in galactic dynamics.

\section{The Statistical Mechanical Approach - Violent Relaxation}

\subsection{Observational evidence of the equilibrium state assumption}

\begin{figure}[tbp]
\centering{\includegraphics[width=\textwidth]{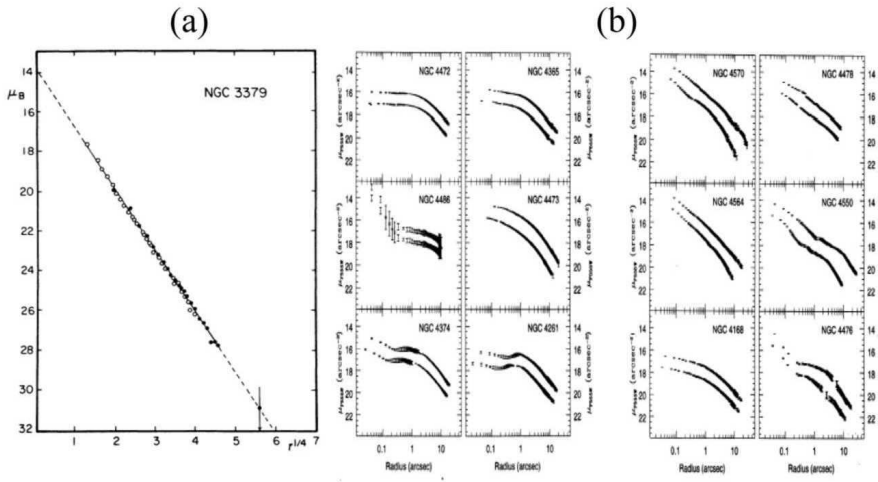}}
\caption{(a) Fitting of the elliptical galaxy NGC3379 by de
Vaucouleurs' law (after de Vaucouleurs and Capaccioli 1979). (b)
Shallow (left) and cuspy (right) profiles of the cores of
elliptical galaxies (after Ferrarese et al. 1994).}
\label{fig08-0}
\end{figure}

The smoothness of observed photometric profiles suggests that at
least the spheroidal components of galaxies are in a form of
statistical equilibrium. The surface brightness profiles of many
elliptical galaxies are well-fitted by the de Vaucouleurs' (1948)
$R^{1/4}$ law (Fig.8a):
\begin{equation}\label{devauc}
I(R)=I_e\exp\big(-7.67[(R/R_e)^{1/4}-1]\big)
\end{equation}
where $I_e$ is the value of the surface brightness (in
$mag/arcsec^2$) at the radius $R_e$ of a disk in the plane of
projection containing half of the total light. In a number of
galaxies this relation is verified in a range up to ten magnitudes
(e.g. de Vaucouleurs and Capaccioli 1979). The profiles of bulges
and of some ellipticals follow a similar law, namely the Sersic
$R^{1/n}$ law (Sersic 1963, 1968). On the other hand, the
central profiles of elliptical galaxies were reliably observed by
the Hubble space telescope. It was found that the profiles have
central cusps, i.e., the surface brightness grows as a power-law
in the center $I(R)\propto R^{-\gamma}$, $\gamma>0$ (Crane et al.
1993, Ferrarese et al. 1994, Lauer et al. 1995). There are two
groups of observed central profiles (Fig.8b), namely \textbf{a)}
shallow profiles ($\gamma \leq 0.2$), and \textbf{b)} abrupt
profiles ($\gamma \sim 1$). However, as emphasized by Merritt
(1996), even shallow profiles in the surface brightness correspond
to power-law cusps in the 3D density profile $\rho(r)\propto
r^{-a}$ with power exponents $a>1$. This means that at least the
centers of galaxies deviate considerably from simple isothermal
models with a Boltzmann - type distribution function such as the
King models (King 1962):
\begin{equation}\label{king}
f(\mathbf{x},\mathbf{p})=A\exp\left[-\beta
\left(E(\mathbf{x},\mathbf{p})-E_0\right)\right] =
A\exp\left[-\beta \left({\mathbf{p}^2\over
2}+\Phi(\mathbf{x})-E_0\right)\right]
\end{equation}
with $A,\beta,E_0$ constants, or their non-isotropic generalizations
(Michie 1963). The latter models are characterized by flat density
profiles at the center (Binney and Tremaine, 1987, p.234). Thus,
the nature
of statistical equilibrium of galaxies should be quite different to
the isothermal equilibrium. Furthermore, since the time of two-body
relaxation is much larger than the age of the Universe, galaxies had
no time to approach such an equilibrium. The very fact that galaxies
are statistically relaxed systems seems, at first, to be a paradox
(the so-called `Zwicky's paradox').

\subsection{The theory of violent relaxation}

A way out of the paradox developed gradually in the sixties, after
a systematic study of the hypothesis that in the early phase of
galaxy formation, galaxies were subject to a sort of `violent
relaxation' (Lynden-Bell 1967) caused by the collapse and ultimate
merger of clumps of matter produced by the nonlinear evolution of
initially small density inhomogeneities in the early Universe. We
can mention in this context an influential paper by Eggen,
Lynden-Bell and Sandage (1962) under the characteristic title
``Evidence from the Motions of Old Stars that the Galaxy
Collapsed'', as well as one of the first numerical simulations of
a spherical gravitational system in the computer by M. H\'{e}non
(1964).

The theoretical foundations of the statistical mechanics of violent
relaxation were set by Lynden-Bell (1967), using a continuum approach
for the distribution
function, and re-derived by Shu (1978) with a particle approach to the
same distribution. These analytical studies are now considered classical,
despite the fact that the so-derived equilibrium distribution functions are
far from able to account for the properties of systems produced by realistic
N-Body simulations or for the data of observed galaxies in the sky.

\begin{figure}[h]
\centering{\includegraphics[width=11cm]{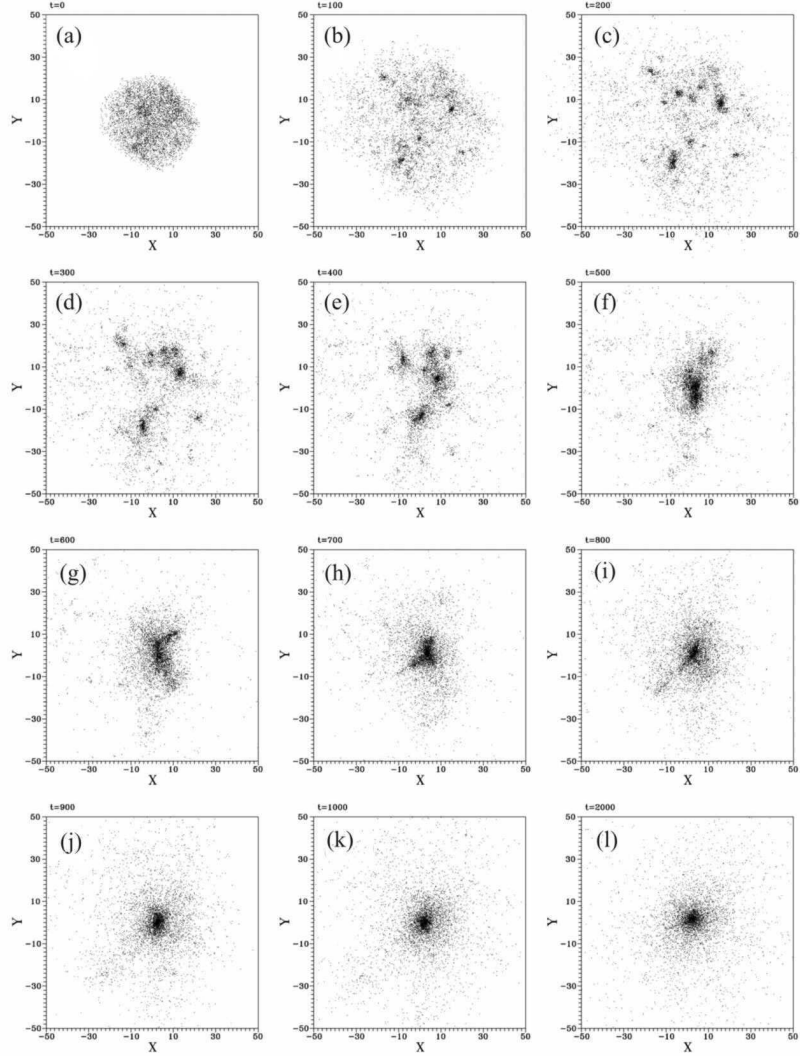}}
\caption{N-Body simulation of the collapse and violent relaxation
of a nearly spherical mass (protogalaxy). The initial conditions
correspond to a hierarchical clustering scenario (clumpy initial
conditions) in a $\Lambda-$CDM expanding Universe (after
Efthymiopoulos and Voglis 2001).} \label{fig09-0}
\end{figure}

An example of such a collapse process is shown in Fig.9
(Efthymiopoulos and Voglis 2001). This is an N-Body simulation of
an isolated system containing one galactic mass represented with
5616 particles. The mass is initially contained in a nearly
spherical subvolume of the Universe. The particles are assigned
positions and velocities in agreement with the general Hubble
expansion of the Universe in a $\Lambda-$CDM scenario, but the
position and velocity vectors of each particle are perturbed
according to a prescription for the spectrum of density
perturbations in the Universe at the moment of decoupling, and
following a well-known technique of translating these
perturbations into perturbations of positions and velocities
introduced by Zel'dovich (1970). As seen in Fig.9, the system
initially expands following the general expansion of the Universe
(Figs.9a,b), but the extra gravity due to local overdensities
results in a gradual detachment of the system from the average
Hubble flow, so that the system reaches a maximum expansion radius
(Fig.9c) and then begins to collapse. At the initial phase of
collapse, small subclumps are formed within the spherical volume
which collapse to local centers forming larger bound objects
(Figs.9d,e). However, these clumps also collapse towards a common
center of gravity (Fig.9f), until the overall system relaxes,
after a phase of rebound, to a final equilibrium (Figs.9g-l).

There is a wide variety of initial conditions that lead to the
above type of relaxation process. For example, a currently popular
scenario of formation of elliptical galaxies via the merger of
spiral galaxies (Toomre and Toomre 1972, Gerhard 1981,
Negroponte and White 1983, Barnes 1988, 1992, Hernquist 1992,
Naab et al. 1999, Burkert and Naab 2003)
corresponds to a case where, instead of many clumps, as in Fig.9,
we have only two major clumps corresponding to the dark halos of
the spiral galaxies. In that case the presence of gas dynamical
processes must be taken into account. Nevertheless, the main process
driving the system towards a final equilibrium state is again a violent
relaxation process, although the initial
conditions and the detailed time evolution of the system is
different than in the case of a simple collapse or a multiple
merger event.

The statistical mechanical theory of violent relaxation aims,
precisely, at justifying theoretically the tendency of such systems
to settle down to an equilibrium, and to find the form of the
distribution function $f$ at this equilibrium.

A simplified version
of the main steps in the derivation of Lynden-Bell's statistics
is the following:

\begin{figure}[tbp]
\centering{\includegraphics[width=10cm]{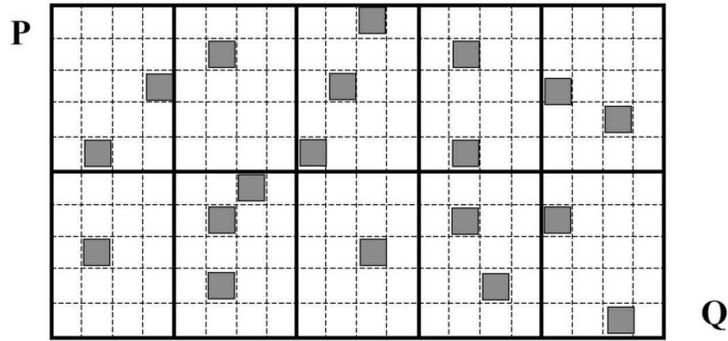}}
\caption{A simple schematic representation of the phase-space
(Q,P) of a violently relaxing system. The gray squares represent
elements of the phase fluid. The phase-space is partitioned in
micro-cells (fine grid of dashed lines) and macro-cells (coarse
grid of bold lines).} \label{fig10-0}
\end{figure}

1) We consider a compact $\mu-$space (i.e. we consider that
escapes are negligible), and implement a coarse - graining process
by dividing the $\mu-$space in a number of, say, $N$ {\it
macrocells} of equal volume (Fig.10) labelled by an index
$i=1,2,...,N$. We further divide each macrocell into a number of
{\it microcells} that may or may not be occupied by {\it elements
of the Liouville phase flow} of the stars moving in $\mu-$space.
In Fig.10 these phase elements are shown by dark squares that
occupy some microcells within each macrocell.

2) We adopt the {\it equal a priori probability} assumption,
namely we assume that each element of phase flow has equal a
priori probability to be found in any of the macrocells of Fig.10.
As the system evolves in time, each phase element travels in phase
space by respecting this assumption. We should note that, because
of phase mixing, the form of the phase elements also changes in
time. However, this deformation does not change the volume of an
element. We can thus proceed in counting the number of phase
elements  in each macrocell by keeping the simple schematic
picture of Fig.10.

3) We denote by $n_i$ the occupation number of the i-th macrocell, i.e., the
number of fluid elements inside this macrocell at any fixed time $t$. The set
of numbers $(n_1,n_2,...,n_N)$, called a {\it macrostate}, can thus be viewed
as a discretized realization of the coarse-grained distribution function of
the system at the time $t$.

4) For any given macrostate, the mutual exchange of any two phase elements,
or the shift of an element in a different cell within the same macrocell
leaves the macrostate unaltered. Thus, we can calculate the number
$\Omega(n_1,n_2,...,n_N)$ of all possible microscopic configurations that
correspond to a given macrostate, and define a Boltzmann entropy $S=\ln\Omega$
for this particular macrostate. If we denote by $n=\sum_{i=1}^N n_i$ the total
number of phase elements and by $\nu$ the (constant) number of
microcells within
each macrocell, the combinatorial calculation of $\Omega$ readily yields:
\begin{equation}\label{omega}
\Omega(n_1,n_2,...,n_N)={n!\over n_1! n_2!...n_N!} \prod_{i=1}^N
{\nu!\over (\nu-n_i)!}
\end{equation}

5) We finally seek to
determine a statistical equilibrium state as the most probable macrostate,
i.e., the one maximizing $S$ under the constraints imposed by all preserved
quantities of the phase flow. Besides mass conservation $n=\sum_{i=1,N}n_i$,
we can assume conservation of the total energy of the system
$E=\sum_{i=1}^N n_i\epsilon_i$ (where $\epsilon_i$ is the average energy
of particles in the macrocell $i$), and perhaps of other quantities such
as the total angular momentum (if spherical symmetry is preserved during
the collapse) or any other `third integral' of motion.
In the simplest case of mass and energy conservation, we maximize $S$ by
including the mass and energy constraints as Lagrange multipliers
$\lambda_1$, $\lambda_2$ in the maximization process, namely:
\begin{equation}\label{deltaome}
\delta\ln\Omega -\lambda_1\delta n - \lambda_2\delta E = 0
\end{equation}
We furthermore apply Stirling's formula for large numbers $\ln
N!\approx N\ln N - N$. In view of Eq.(\ref{omega}),
Eq.(\ref{deltaome}) then yields
\begin{equation}\label{lb1}
F_i = {\eta n_i\over\nu}|_{S=max} =
{\eta\over \exp(\lambda_1+\lambda_2\epsilon_i) +1}
\end{equation}
where $\eta$ is the (constant) value of the phase-space density
inside each moving phase-space element. Eq.(\ref{lb1}) is
Lynden-Bell's formula for the value $F_i$ of the coarse-grained
distribution function within the i-th macrocell at statistical
equilibrium. Following the conventions of thermodynamics, we
interpret $\lambda_2$ as an inverse temperature constant
$\lambda_2\equiv\beta \propto 1/T$ and $\lambda_1$ in terms of an
effective `chemical potential' $\epsilon_0=-\lambda_1/ \beta$ (or
`Fermi energy'). We
thus rewrite Eq.(\ref{lb1}) in a familiar form reminiscent of
Fermi-Dirac statistics
\begin{equation}\label{lb2}
F_i =  {\eta\over \exp[\beta(\epsilon_i-\epsilon_0)] +1} =
{\eta\exp[-\beta(\epsilon_i-\epsilon_0)]\over 1+\exp[-\beta(\epsilon_i-\epsilon_0)]}
\end{equation}
by recalling, however, that the energy and effective chemical potential in
Eq.(\ref{lb2}) have in fact dimensions of energy per unit mass, in accordance
to our general treatment of orbits in $\mu-$space (subsection 2.2). Therefore,
contrary to two-body relaxation, the process of violent relaxation cannot lead
to mass segregation at the equilibrium state. At any rate, in the so-called
non-degenerate limit $F_i<<\eta$, Eq.(\ref{lb2}) tends to the form of a
Boltzmann distribution $F_i \simeq A\exp(-\beta\epsilon_i)$, that is, the final
state approaches the isothermal model.

The above exposition of Lynden-Bell's theory is simplified in many
aspects. In particular: \textbf{a)} The expression given for the
constraint in the total energy is not precise. One should calculate
the energy self-consistently by the gravitational interaction of
the masses contained in each phase element. However, the final
result turns out to be the same with this more precise
calculation.  \textbf{b)} All phase elements in the above
derivation are assumed to have the same value of the phase space
density $\eta$, i.e., the same `darkness' in Fig.10.

A more general distribution function was derived by Lynden-Bell
when the phase elements of Fig.10 can be grouped into $K$ groups
of distinct darkness $\eta_J$, $J=1,\ldots K$. The final formula,
derived also by the standard combinatorial calculation, reads:
\begin{equation}\label{lb3}
F_i = \sum_{J=1}^K{\eta_J\exp\big(-\beta_J(\epsilon_i-\epsilon_{0J})\big)
\over
1+\sum_{J=1}^K\exp\big(-\beta_J(\epsilon_i-\epsilon_{0J})\big) }
\end{equation}
that is, it depends on a set of $K$ pairs of Lagrange multipliers
$\beta_J, \epsilon_{0J}$, $J=1,\ldots,K$. This more realistic
formula links the initial conditions of formation of a system,
parameterized by the values of $\eta_J$ which are conserved during
the relaxation, to the final distribution function. In the
non-degenerate limit, the latter is a superposition of nearly
Boltzmann distributions, meaning that each group of phase elements
is characterized by its own Maxwellian distribution of velocities
which yields a different velocity dispersion in each group,
depending on the value of $\eta_J$ This poses a problem as regards
the possibility to express the overall distribution of velocities
in the galaxy by a single Maxwellian function. (see for example
Shu 1978 and the debate Shu 1987 - Madsen 1987). We return to this
question in subsection (3.5) where we discuss alternative
formulations of the statistical mechanics of violently relaxing
systems.

\subsection{Incomplete relaxation}

The basic prediction of Lynden-Bell's theory, namely the
possibility for a stellar system to settle down to a statistical
equilibrium within a time comparable to a few system's mean
dynamical times, has been completely verified in a series of
numerical simulations over subsequent years (e.g. Gott 1973, 1975,
White 1976, 1978, Aarseth and Binney 1978, Hoffman et al. 1979,
van Albada 1982, 1987, May and Van Albada 1984, McGlynn 1984,
Villumsen 1984, Aguilar and Merritt 1990, Burkert 1990, Mineau et
al. 1990, Katz 1991, Dubinski and Calberg 1991, Londrillo et al.
1991, Cannizzo and Holister 1992, Curir et al. 1993, Voglis 1994a,
Voglis et al. 1995, Carpintero and Muzzio 1995, Henriksen and
Widrow 1997, 1999, Efthymiopoulos and Voglis 2001, Merrall and
Henriksen 2003, Trenti et al. 2005). However, the data of these
experiments, as well as other considerations converge to the
conclusion that Lynden-Bell's formula (\ref{lb2}) is not
applicable even in the simplest cases of realistic galactic
systems (Cuperman et al. 1969, Goldstein et al. 1969, Lecar and
Cohen 1972, White 1976, Binney 1982b, May and Van Albada 1984,
Severne and Luwel 1986, Madsen 1987, Hjorth and Madsen 1991,
Voglis et al. 1991, Voglis 1994a, Takizawa and Inagaki 1997,
Efthymiopoulos and Voglis 2001, Trenti et al. 2005).

There are many phenomena which act as factors of obstruction to a
convergence of $F$ towards Lynden-Bell's prediction. We refer
below to one of the most important factors considered in the
literature: incomplete relaxation. This is a phenomenon that may
happen even in the simplest case of systems relaxing via a
monolithic collapse. The term `incomplete' means that the process
of mixing of phase elements in $\mu$ space, during the relaxation
process, is not efficient enough so as to justify the assignment
of equal a priori probability on a phase element to be in any of the
macrocells of $\mu$. This also implies that some memory of initial
conditions survives in the final equilibrium state. This
phenomenon is commonly verified by N-Body experiments (May and Van
Albada 1984, Stiavelli and Bertin 1987, Voglis et al. 1991, 1995,
Efthymiopoulos and Voglis 2001, Trenti et al. 2005). An example is
shown in Fig.11 (Voglis et al. 1995) which shows a plot of the
final versus initial energies (Fig.11a) or angular momenta
(Fig.11b) for each particle in a N-Body collapse experiment. The
correlation between initial and final values of the angular
momentum is obvious from the concentration of points towards the
diagonal.

\begin{figure}[tbp]
\centering{\includegraphics[width=\textwidth]{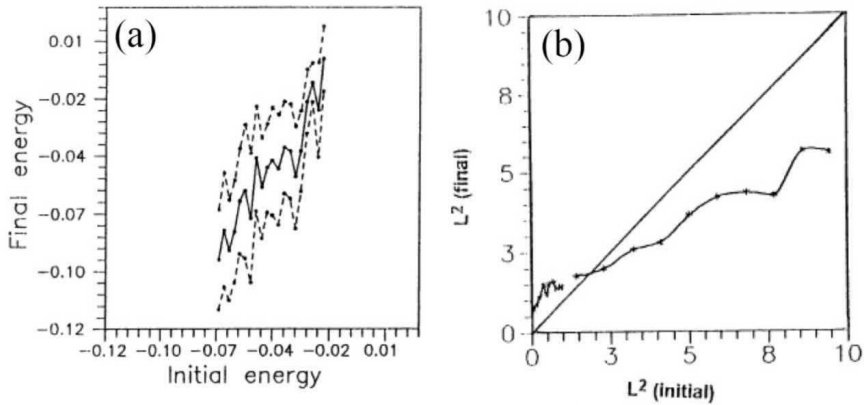}}
\caption{(a) Initial versus final energies (solid line=mean,
dashed lines = lower and upper limits) of the particles before and
after the collapse. (b) Initial versus final angular momenta for
the same particles (after Voglis et al. 1995).} \label{fig11-0}
\end{figure}

We may quantify this correlation by calculating, in N-Body collapse
experiments, the time-dependence of the correlation coefficient defined by
\begin{equation}\label{corel}
CR(t)={\sum_{i=1}^N(E_{0i}-\bar{E}_0)(E_{ti}-\bar{E}_t)\over
\sqrt{\sum_{i=1}^N(E_{0i}-\bar{E}_0)^2\sum_{i=1}^N(E_{ti}-\bar{E}_t)^2}}
\end{equation}
where $E_{0i}$ , $i=1,...,N$ are the energies of the N particles
at the initial snapshot of the experiment, $E_{ti}$ the energies
of the same particles (each labelled by $i$) at a time $t$, and
$\bar{E}_0$, $\bar{E}_t$ the mean energies respectively (Fig.12).
In this and in subsequent plots we refer to a series of collapse
experiments corresponding to the time evolution of the matter
distributed in a spherical volume in the Universe containing one
galactic mass, in which, at the moment of decoupling, we impose
a field of density perturbations consistent with a standard
$\Lambda-$~CDM cosmological scenario. We furthermore distinguish
between \textbf{a)} experiments with a spherically symmetric field
of initial density perturbations, and \textbf{b)} clumpy initial
density perturbations (S and C experiments, see Efthymiopoulos and
Voglis 2001 for a detailed description of the initial conditions
of the experiments). Finally, we examine various exponents $n$ of
the power spectrum of density perturbations, that is, the r.m.s.
dependence of a density perturbation on scale $r$ is given by
\begin{equation}\label{pwsp}
{\delta\rho(r)\over\rho}\propto {1\over r^{n+3\over 2}}
\end{equation}

\begin{figure}[h]
\centering{\includegraphics[width=10cm]{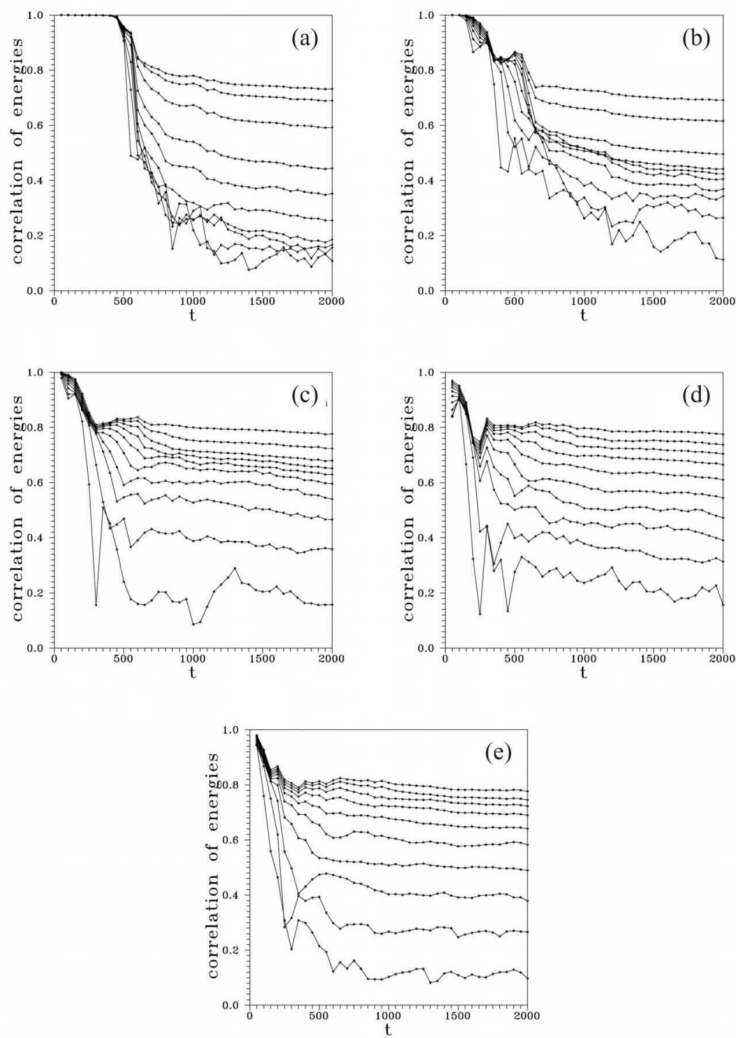}}
\caption{The time evolution of the initial - final energy
correlation coefficient (Eq.(\ref{corel})) for five collapse experiments
differentiated by the value of the power exponent $n$ of initial
density perturbations, namely (a)$n=-2.9$, (b)$n=-2$, (c)$n=-1$,
(d)$n=0$, (e)$n=1$. The curves in each panel from down to the top
correspond to the values of the correlation coefficient for the
innermost 10\%,20\%,...90\% of the bound matter.}
\label{fig12-0}
\end{figure}

\noindent according to standard cosmological considerations (see
Voglis 1994b). In the case of S-experiments, Eq.(\ref{pwsp}) is
viewed as the radial profile of a spherically symmetric density
perturbation, while in the C-experiments the perturbation field
inside the spherical volume is determined by a superposition of
plane waves with power spectrum
$P(k)\propto k^n$ and random phases. The resulting perturbation
field is translated to perturbation of the particles'
positions and velocities with respect to an ideal Hubble flow by
means of Zel'dovich approximation (Zel'dovich 1970). We choose
different values of the exponent $n$ in the range $-3<n\leq 1$,
consistent with the hierarchical clustering scenario.

\begin{figure}[tbp]
\centering{\includegraphics[width=10cm]{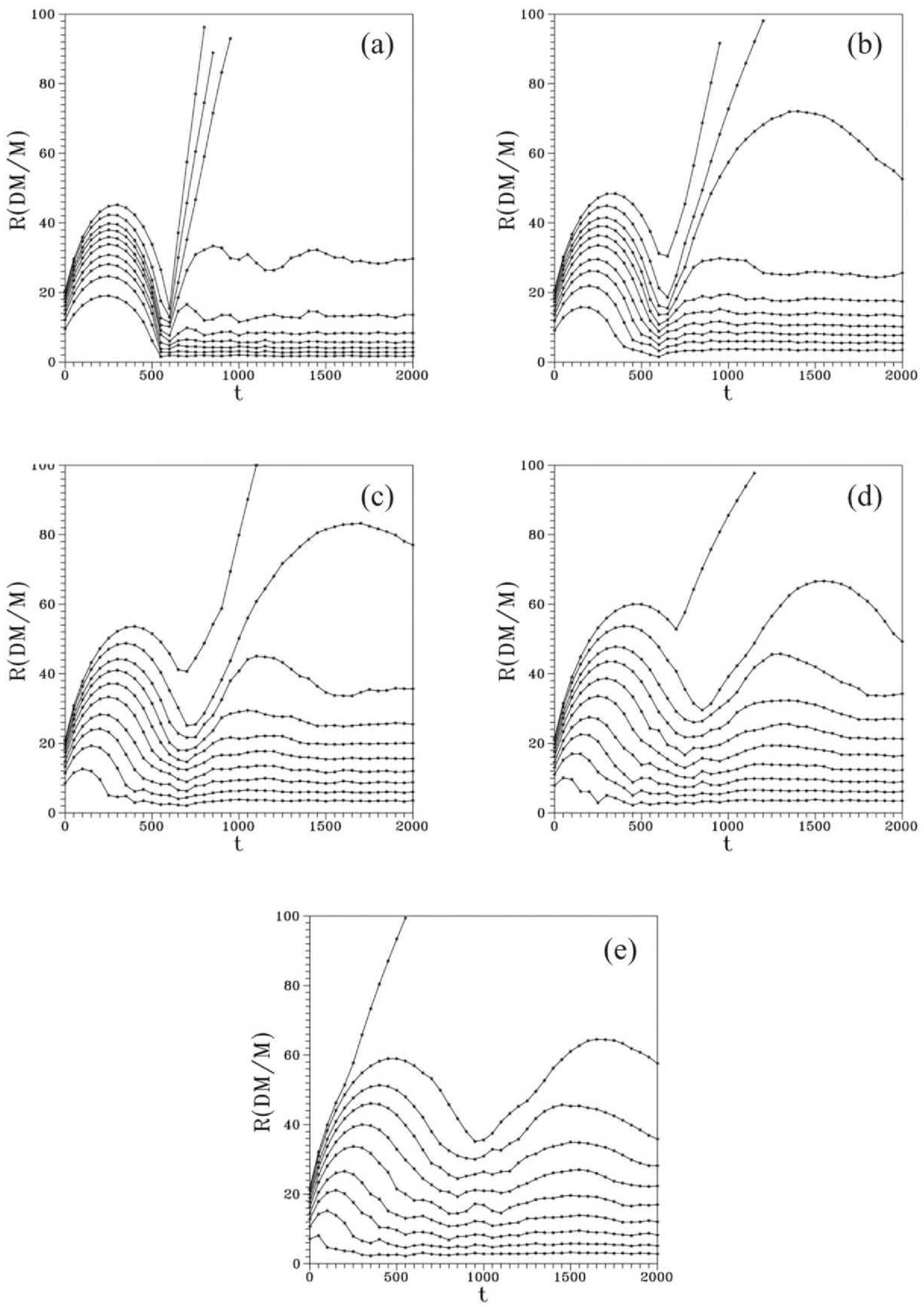}}
\caption{The time evolution of the radii of spherical shells
containing 10\%,20\%,...90\% of the matter in the same experiments
as in Fig.12.}
\label{fig13-0}
\end{figure}

The value of $n$ is a parameter that regulates the violence of the
collapse phase by affecting the distribution of power between
perturbations of small and large scale. This can be seen by the
following analysis, due to Palmer and Voglis (1983): if the r.m.s
profile of mass perturbation in a structure of scale $h$ at the
moment of cosmological decoupling is, according to Eq.(\ref{pwsp})
taken to be $\mu^2(h) = \mu_0h^{-(n+3)}$, then the total mass
contained in the interior of a sphere of radius $h$, given by
$M(h)={4\pi \over 3}\rho_0h^3[1+\mu(h)]$, where $\rho_0$ is the
average density of the Universe at decoupling, will cause a
gravitational attraction of the spherical shell at radius $h$ so
that the expansion of the shell will gradually detach from the
average Hubble expansion of the Universe. If $r(t)$ denotes the
radius of the shell at the moment $t$, the solutions of the
equations of motion in a $\Omega=1$ Universe can be given
parametrically in the form of cycloid motion
$$
r=h{1+\mu(h)\over 2\mu(h)}(1-\cos u),~~~
t=t_0{3[1+\mu(h)]\over 4\mu(h)^{3/2}}(u-\sin u)
$$
where $t_0$ is the time of decoupling, $r(t_0)=h$, and we use
units in which $G=1$ and the Hubble constant at decoupling is
$H_0=\sqrt{2}$. From these equations we find that a shell of
radius $h$ will reach its maximum expansion at $t_{max}\approx
3t_0\pi/4\mu(h)^{3/2}$, and from there on the shell will begin to
collapse, the collapse time being almost equal to the expansion
time. We may now use the form of the profile $\mu(h)\propto
h^{-(n+3)/2}$ and find that the collapse time for a spherical
shell including in its interior spherical volume a percentage
$\Delta M\over M$ of the total mass of the system is given by:
\begin{equation}\label{tcollapse}
t_{collapse} \propto \left({\Delta M\over M}\right)^{n+3\over 4}~~.
\end{equation}
This power-law is well verified in N-Body experiments. In Fig.13
we show the evolution of the radii $r(t)$ of spherical shells
containing a percentage $10\%,20\%,...,90\%$ of the total mass of
the collapsing object for different values of $n$, namely
\textbf{a)} $n=-2.9$, \textbf{b)} $n=-2$, \textbf{c)} $n=-1$,
 \textbf{d)} $n=0$, and \textbf{e)} $n=1$. It is immediately seen
that in the limit $n\rightarrow -3$ (Fig.13a), meaning a homogeneous
profile of the initial density perturbation (Eq.(\ref{pwsp})),
all shells collapse at about
the same time. This is the well known spherical `top-hat' model.
On the other hand, as $n$ increases, the collapse becomes more
gradual, and in the other limit $n\rightarrow 1$ (Fig.13e) the
outer shells collapse at a time which is an order of magnitude
larger than the collapse time of the inner shells.

Fig.14 shows the value of the correlation coefficient
(\ref{corel}) of the particles' energies at the initial and final
snapshot of the experiment, as a function of the exponent $n$.
There are nine curves in this diagram, corresponding to the value
of the correlation coefficient for the innermost 10\%, 20\%, ...,
90\% of the matter. We see that, independently of the value of
$n$, the innermost 20\% of the matter yields low correlation
coefficients (0.2 to 0.3), meaning that we can speak about almost
complete relaxation. In the case of the $n=-2.9$ experiment, this
percentage raises to 40\%. However, for the rest of matter
the correlation coefficient has values that can be as high as 0.7.
This means that the mixing of energies is incomplete. Such high
values of the correlation coefficient are observed in all the
experiments, including the limit of the `top-hat' model ($n=-2.9$).

\begin{figure}[tbp]
\centering{\includegraphics[width=7cm]{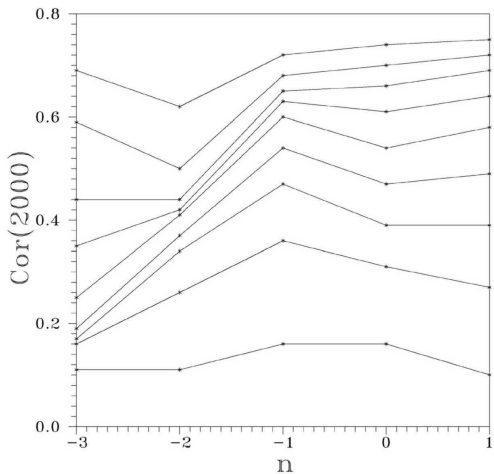}}
\caption{The final value of the correlation coefficient (at
$t=2000$) for the 10\%, 20\%,...90\% of matter (from down to top)
as a function of the exponent $n$ of the initial density
perturbations. The correlation coefficient has relatively high
values for an important fraction of the matter of all the systems,
indicating that the relaxation is incomplete.} \label{fig14-0}
\end{figure}

This fact is remarkable, and
requires some further explanation. This is related to a problem
regarding the very nature of violent relaxation that was posed by
Miller (private correspondence with Lynden-Bell, see Merritt
2005). In the original approach of Lynden-Bell, the energies of
stars are subject to stochastic changes caused by the time
fluctuations of the self-gravitational potential of the system,
since the rate of energy change of each star is given by:
\begin{equation}\label{dedt}
{dE\over dt} = {\partial\Phi\over\partial t}
\end{equation}
The rate of relaxation is thus linked to the mean
timescale of the time-dependent variations in the r.h.s. of
Eq.(\ref{dedt}), that is $T_{rel}\sim<(\Phi/\dot{\Phi})^2>^{1/2}$.
Lynden-Bell established that this timescale is of the order of the
mean dynamical period of the system, hence the term `violent'
relaxation. Nevertheless, Miller notices that if we have an
isolated galaxy and a mass $m$ which is uniformly distributed
in a spherical shell surrounding the galaxy,
then, if we let the mass $m$ vary in time $m\equiv m(t)$, the
total gravitational potential $\Phi = \Phi_{galaxy}+\Phi_{shell}$
becomes time-dependent. As a result, the energy of each star in
the galaxy changes, according to Eq.(\ref{dedt}), but these
changes are only due to the addition of a time-dependent uniform
term to the energies of all stars and, in reality, they have {\it
no} effect in the stars' orbits, since the shell does not excert
any force to particles in its interior. Miller concludes
that Eq.(\ref{dedt}) cannot characterize the effectiveness, or
timescale, of mixing of the energies in a violent relaxation
process, but other criteria must be established in order to
distinguish when and how fast such a mixing actually occurs.

The results for the `top-hat' case $n=-2.9$ are in certain aspects
similar to Miller's example. Since the shells collapse all at the
same time, the variations of the energies of all the stars are
{\it in-phase}, that is, all the stars gain or lose energy during
collapse and rebound of the system, so that the mixing of energies
is not very effective despite the fact that the rate of change of
energies is very fast. On the other hand, in the limit
$n\rightarrow 1$, the variations of energies of the stars are to a
large extent out-of-phase, since the inner shells are at the
rebound phase when the outer shells are still in the collapse
phase (Fig.13). This is caused by the decreasing profile of mass
perturbation $\mu(h)\propto h^{-2}$. At the same time, this
mechanism implies that the overall time fluctuations of the
potential are less violent than in the `top-hat' model. As a
conclusion, in both limits $n\rightarrow -3$ and $n\rightarrow 1$
the relaxation cannot be complete, although the reasons for that
are different in each limit.

The question of more refined criteria characterizing the violence
or effectiveness of the relaxation process is still unanswered to
a large extent. A recent proposal in this direction was made by
Kandrup (Kandrup 2003, Kandrup et al. 2003).
When the potential has strong time fluctuations, these
fluctuations introduce chaos to the relaxing system through
time-dependent terms of the Hamiltonian. This happens even in a
spherically symmetric, but pulsating, or collapsing, system. For
example, such chaos is found in models of spherical galaxies
in which the galaxy undergoes stable periodic oscillations
(e.g. Louis and Gerhard 1988, Miller and Smith 1994, Smith and
Contopoulos 1995). Now, in regions of phase space where chaos is
prominent, the rate of mixing is determined by the Lyapunov times
of the orbits of stars that move as ensembles within the phase
space (Kandrup and Mahon 1994). This so-called {\it chaotic
mixing} process is much faster than the phase mixing process
discussed already in Lynden-Bell (1967). In Kandrup's view
the rate of chaotic mixing determines essentially the rate
of approach of the system to equilibrium.

There is no direct experimental test so far, e.g. by N-Body
collapse simulations, of the validity of Kandrup's suggestion. One
way to produce such tests is by a detailed exploration of plots
from N-Body collapse experiments showing in detail the spreading
of particles in phase space during the relaxation phase. A
schematic example is given in Fig.15. If we consider a `frozen'
spherical potential corresponding to one snapshot of the collapse
experiment, the invariant tori of the Hamiltonian of this momentary
potential have the form shown schematically in Fig.15. As long as
the system is not in equilibrium, the phase flow is transverse to
the direction determined by the foliation of these tori (arrows in
Fig.15). However, as the system approaches
closer and closer to the equilibrium, the flow becomes more
and more tangent to the directions defined by the foliation of
the tori.  This simple picture is
not precise when chaotic mixing takes place. This causes
irregularities of the flow both in the transverse and tangent
directions that may be the dominant source of mixing. Such
irregularities are distinguishable in some real plots of the
phase flow in collapse experiments (e.g. Burkert 1990, Henriksen
and Widrow 1997, 1999), but, to our knowledge, there has been no
systematic qualitative or quantitative study of the time evolution
of this flow so far.

\begin{figure}[tbp]
\centering{\includegraphics[width=10cm]{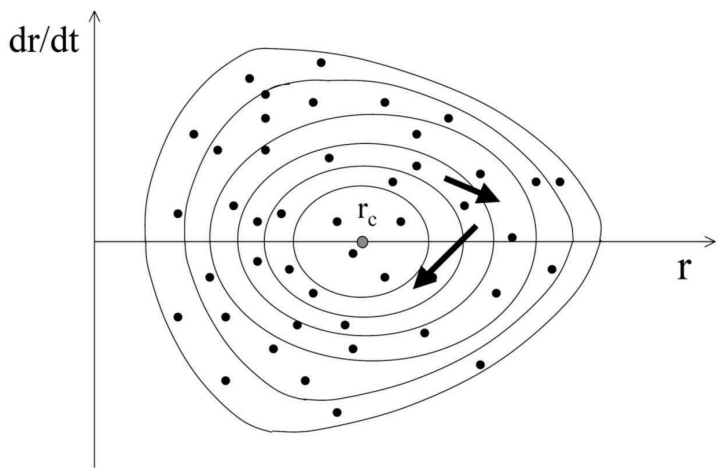}}
\caption{Schematic representation of the theoretical invariant
tori in the space $r,\dot{r}$ of a spherical system for a constant
pair of energy - angular momentum values. If the system is
collapsing, these tori correspond to a `frozen' snapshot of the
time-dependent spherical potential. The relaxation process
continues for as long as the phase flow of the real particles
(bold arrows) is transverse to the tori. The points represent the
n-body sampling of the distribution function.} \label{fig15-0}
\end{figure}

\subsection{Collective instabilities}

Stellar systems relaxing from different initial conditions cannot
in general be expected to relax to the {\it same} equilibrium
endstate, since the properties of the latter are determined, to a
large extent, by dynamical instabilities affecting the system in
the course of or after the relaxation process. The topic of
instabilities in collisionless stellar systems is a whole chapter
of galactic dynamics (see Fridman and Polyachenko 1984 and Palmer
1995 for a review).

Collective instabilities in the simplest case of a spherical
system where first discussed by Antonov (1960). Such instabilities
may lead to interesting phenomena such as the `gravothermal catastrophe'
(Lynden-Bell and Wood 1968) that is believed to have played some role
in dense systems such as the cores of spherical clusters.
The main result of Antonov's studies is that a spherical isotropic
system is stable against radial or non-radial instabilities if its
distribution  function is a monotonically decreasing function of
the energy (see Binney and Tremaine 1987, p.307).
Subsequent studies (H\'{e}non 1973, Dejonghe and
Merritt 1988) gave criteria for the stability of anisotropic
systems under various types of radial perturbations. The analog
of such instabilities in the case of disks are axisymmetric
instabilities (e.g. Toomre 1964).

A type of instability relevant to elongated galaxies is the
`radial orbit instability' (Polyachenko 1981, Polyachenko and Shukman
1984, Palmer and Papaloizou 1987). If a galaxy contains initially
many radial orbits, i.e., $\sigma_r>>\sigma_t$ (subsection 2.2),
a small deviation of the angular distribution of these
orbits from spherically symmetric creates a collective
collaboration of the orbits, based on their mutual torques, that results
in a large departure of the system from the spherical
symmetry. The final states can be either axisymmetric (usually prolate)
or triaxial. In the case of disks, Lynden-Bell (1979) examined a
similar collaboration of elongated orbits that can lead to the formation
of a rotating bar inside the inner Lindblad resonance.

The general theory of the radial orbit instability is based on perturbative
solutions to the collisionless Boltzmann equation. The final result
can be cast in the form of Polyachenko's criterion: a system is
stable against the radial orbit instability when
\begin{equation}\label{polya}
{2T_r\over T_t}\leq 1.7\pm 0.7
\end{equation}
where $T_r=<v_r^2/2>$ and $T_t=<v_t^2/2>$ (in non-rotating
galaxies), $v_r$, $v_t$ being the velocities of stars in the
radial and transverse direction respectively. The $\pm 0.7$ error
in Eq.(\ref{polya}) is produced by a compilation of different
values of the ratio $2T_r/T_t$ reported in the literature by use
of different basic models used to study the instability (see
Merritt 1999, subsection 6.2 and references there in). For
example, we may consider a spherical distribution function which
is initially in steady state and find a somewhat different ratio
$2T_r/T_t$ depending on what is the model chosen for the initial
distribution function. Other, similar in spirit, criteria were
proposed by different authors. For example, Merritt and Aguilar
(1985) proposed the criterion $2T/|U|\leq 0.1$, where $T$ is the
initial kinetic energy $U$ the initial potential energy of the
system. Such criteria are verified in N-Body studies of the radial
orbit instability when we start with initial conditions which are
perturbations to a spherical equilibrium (e.g. Merritt and Aguilar
1985, Barnes et al. 1986, Aguilar and Merritt 1990, Canizzo and
Holister 1992).  If, on the other hand, we consider initial
conditions corresponding to a cosmological collapse scenario
(Carpintero and Muzzio 1995, Efthymiopoulos and Voglis 2001), we
find that when we start with a spherically symmetric collapsing
object, which has an overpopulation of radial orbits, then the
system relaxes to its final equilibrium only after the ratio
$2T_r/T_t$, which is initially very large, settles down to a value
near Polyachenko's value 1.7 (Fig.16a). The resulting endstates
are triaxial systems with axial ratios of long to short axis
corresponding to E5 - E6 galaxies. On the contrary, if we start
with clumpy initial conditions (Fig.16b), which are characterized
by a more random initial distribution of the directions of the
particles' velocities, the ratio $2T_r/T_t$ goes below the value
$1.7$ very quickly (at $t\simeq 150$ in Fig.16b), at times smaller
than the collapse time ($t_{collapse}\approx 1000$). Then the
systems do not exhibit a strong radial orbit instability, and the
resulting endstates resemble to E2 - E3 galaxies.

\begin{figure}[tbp]
\centering{\includegraphics[width=\textwidth]{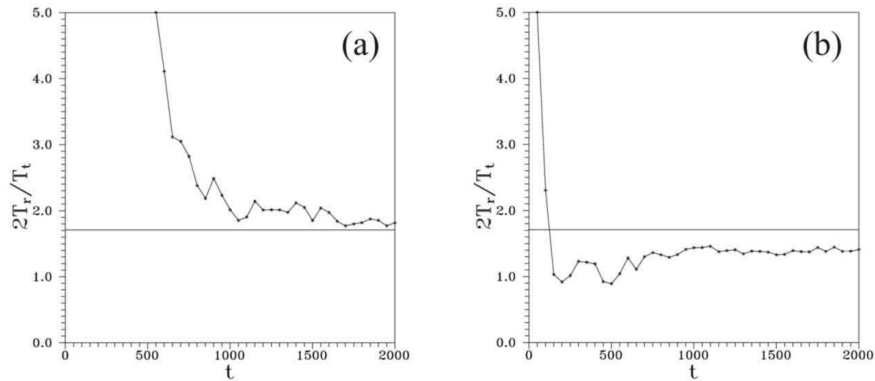}}
\caption{The time evolution of the ratio $2T_r/T_t$ in two
experiments of violent relaxation from (a) quiet (spherically
symmetric) initial conditions, and (b) clumpy initial conditions.
For both systems, the final value stabilizes near Polyachenko's
criterion $2T_r/T_t = 1.7$. Only for the system (a) the radial
orbit instability is prominent.} \label{fig16-0}
\end{figure}

Other types of instabilities (see Palmer 1995) are the `bending'
(Toomre 1966, Fridman and Polyachenko 1984, Merritt and Sellwood
1994),  `tumbling bar' (Allen et al. 1992) instabilities, and the
bar instability in disks (Hohl 1971, Ostriker and Peebles
1973, Athanassoula and Sellwood 1986).

\subsection{Alternative formulations of the statistical mechanics of
violent relaxation}

Both the violent relaxation process and collective instabilities
processes can be described by solutions of Boltzman's equation
(see e.g. Hoffman et al. 1979, Henriksen and Widrow 1997, 1999,
Merrall and Henriksen 2003 for the violent relaxation case and
references in subsection (3.4) for the case of a collective
instability). However, it is not clear how to distinguish
between these two types of solutions, which essentially both
describe excursions of a system in Liouville space, until the
system settles down to a stable equilibrium state. We may say that
when collective phenomena are present, these phenomena constitute
the main factor determining the system's excursion in Liouville
space. The extent that this happens determines also the limits of
applicability of statistical mechanical considerations such as
those forming the basis of the violent relaxation theory.

On the other hand, we can always say that a stable equilibrium state of a
system of many particles should correspond to a local, or global, maximum of
a kind of entropy functional $S[F]$ (where $F$ is the coarse-grained
distribution function). The use of such functionals to describe the
endstate of a system subject to dynamical instabilities was pioneered
by Ipser (1974) and Ipser and Horwitz (1979). In the case of violent
relaxation, a debate was caused by a paper of Tremaine et al. (1986),
supporting the view that other functionals than the Boltzman functional
$S[F]=-\int F\log Fd^6\mu $, or its generalization by Lynden-Bell (1967)
can be used in the description of statistical equilibria. In particular,
{\it any} functional $S[F]$ that is convex in $F$ will be an increasing
function of time that reaches a maximum at the equilibrium state, that is,
it can play the role of `entropy' of a stellar system. This approach
to equilibrium can be measured by quantities alternative to the
entropy functionals of Tremaine et al. (Mathur 1988). This point of
view was immediately criticized by Kandrup (1987), Sridhar (1987) and
Dejonghe (1987) on the basis of the remark that Tolman's proof of
H-theorem does not apply in the case of an arbitrary convex functional
$S[F]$ and that the monotonic increase of such an `entropy' cannot be
established by elementary arguments. In order to resolve this issue,
Soker (1996) studied in detail the
time evolution of a particular choice of functional $S[F]$ which is
a variant of a functional proposed by Spergel and Hernquist (1992).
He found that the relaxation process can be divided in two phases:
During the first phase, which includes the first collapse and rebound,
the Spergel - Hernquist entropy functional may increase or decrease with
time. During the second phase (called the `calm' phase), it
is an increasing function of time. The calm phase can perhaps be
identified with the so-called `secondary infall' of matter (Filmore
and Goldreich 1984) that characterizes the formation of dark halos,
or with the process of progressive mixing in finer and finer scales
that takes place in the phase space during the late phase of
relaxation.

Another class of modifications of Lynden-Bell's statistics aims
at curing the problem of superpositions of Maxwelian velocity
distributions with different dispersions when the phase-space
elements are divided in groups of different phase densities
(subsection 3.2). Kull et al. (1997) suggested a statistical
mechanics based on phase elements of unequal volumes but equal
masses. They show that the resulting velocity distribution is
again a superposition of Maxwellians, but this time they all have
the same velocity dispersion. Nakamura (2000) made a completely
different proposal in order to address the same problem. He
suggested to use a particle approach for collisionless systems and
defined an entropy $S = -\sum P_{i,j}\log P_{i,j}$, where
$P_{i,j}$ is the probability that if a particle is at the i-th
cell of the phase space at the initial time $t_0$,  it will be at
the j-th cell in the end. Numerical simulations by Merrall and
Henriksen (2003) yielded the result that in collapse simulations
the final velocity distribution appears to be a unique Maxwellian
in the center, but in merger simulations there were considerable
deviations from such a unique distribution if the centers were
initially well separated. On the other hand, Arad and Johansson
(2005) and Arad and Lynden-Bell (2005) made a detailed comparison
of Lynden-Bell's and Nakamura's theories both by numerical and
analytical means. The final conclusion is somewhat disappointing,
since the authors support that both theories yield results not
compatible with numerical experiments. Arad and Lynden Bell (2005)
conclude that a proper description of the violent relaxation
process should rely on dynamical arguments for the evolution of
the coarse-grained distribution function rather than on the
classical statistical mechanical approach.

A yet different approach is based on the search for criteria that
can characterize an equilibrium of Boltzmann's equation without
reference to the concept of entropy, classical or modified. The
basic proposal in this direction was made by Wiechen et al. (1988)
and Ziegler and Wiechen (1989). We notice first that Boltzmann's
equation can be deduced from a Hamiltonian density function
$H[f]$. Furthermore, an equilibrium state $f_0$ is a fixed point
of $H[f]$. Ziegler and Wiechen (1989) then define a `dynamical
energy function' $W[f]$ such that $f_0$ is a minimum of $W[f]$.
The difference $W[f]-W[f_0]$ defines a kind of energy `dissipated'
during the relaxation process. The same authors proposed an
algorithm of calculation of the dynamical energy functional and of
the state $f_0$. In a similar spirit, Kandrup (1998) proposed to
consider the stability character of `orbits' in the so-called
$\Gamma$ space, the space of all states $f$, by giving a suitable
definition of Lyapunov Characteristic number that is applicable
to the case of the Hamiltonian density $H[f]$.

A final proposal has been to use the well known Tsallis entropy (Tsallis 1988)
\begin{equation}\label{tsallis}
S_q[f]= -{1\over q-1}\int (f^q-f)d^6\mu
\end{equation}
as more relevant to the description of the relaxation process,
since gravitational systems are, in general, non-extensive
(Plastino and Plastino 1993, Taruya and Sakagami 2002, 2003). If
the functional (\ref{tsallis}) is maximized under the usual
constraints of mass and energy conservation, the resulting
distribution function has the form of a {\it polytropic}
distribution $f\propto |E|^p$, where the power index $p$ is
related to the q-index of Tsallis' entropy. This approach was
criticized by Chavanis (2006), who points out the fact that the
equilibria of galaxies are far from polytropic. Chavanis
emphasizes that the use of the Tsallis entropy in stellar dynamics
is somewhat ad hoc, because the Tsallis entropy applies when the
phase space of a system is a fractal, or multifractal, while the
fractal properties of the phase-space structure of stellar systems
are not known. This is an open problem that requests more work to
be clarified. On the other hand, Chavanis (1998, 2002) proposed a
method to study the approach to equilibrium that is close in
spirit to Arad and Lynden-Bell's call upon a dynamical description
of violent relaxation. The proposal is to consider a
Boltzmann-type equation that describes the time evolution of
either the coarse-grained distribution function
$F(\mathbf{x},\mathbf{v},t)$ (Chavanis 1998), or a distribution
function $\rho(\mathbf{x},\mathbf{v},\eta,t)$ that is different
for each subset of phase elements with initial phase-space density
equal to $\eta$ (subsection 3.2). In this studies, the analog of
the partial derivative $\partial f/\partial t$ in Boltzmann's
equation (\ref{bol}), namely $\partial F/\partial t$ or $\partial
\rho/\partial t$, is replaced by a diffusion-like term the form of
which is chosen on the basis of dynamical considerations.

\subsection{The number density function in the space of integrals of motion.
Stiavelli - Bertin statistical mechanics}

The distribution function $f$ is a density function in the six-dimensional
phase space, i.e., it gives the mass of stars per unit values of the phase
space coordinates. If, however, the orbits obey integrals of motion, the
distribution function depends on these integrals
$f\equiv f(I_1,I_2,...,I_K)$,
thus it can be expressed in terms of a different function,
$N(I_1,I_2,...,I_K)$
which yields the mass of stars $dm$ per unit value of each of the integrals
$I_i$. The latter function, $N$, is called the {\it number density} function
\begin{equation}\label{nunden}
N(\mathbf{I})={dm\over d\mathbf{I}}
\end{equation}
were $\mathbf{I}$ is the K-dimensional vector of integrals considered and
$d\mathbf{I}$ is an infinitesimal volume in the space of integrals. The
relation between $f$ and $N$ is specified by providing the {\it density of
states function}
\begin{equation}\label{densta}
W(\mathbf{I}) = {d\Omega(\mathbf{I})\over d\mathbf{I}}
\end{equation}
where $d\Omega(\mathbf{I})$ is the elementary volume of phase space that
comprises all phase-space points $(\mathbf{x},\mathbf{p})$ yielding values
of the integrals in the range $\mathbf{I},\mathbf{I}+d\mathbf{I}$.

There are indications that the number density function $N$ may be
more fundamental than the distribution function $f$ in the
characterization of particular properties of stellar dynamical
systems.  A first such suggestion was made by Binney (1982b) who
found that in spherical isotropic galaxies obeying de Vaucouleurs
law, the number density function $N(E)$ depends exponentially on
the energy $N(E)\propto \exp(-\beta E)$, a fact that allows one
to characterize these systems as ``isothermal after all'' (Binney
1982b). In order to find the isotropic spherical equilibrium associated
with de Vaucouleurs' law (Eq.(\ref{devauc})), we recall that in
isotropic systems only the energy $E$ appears as an argument of
the distribution function $f$. We can then make use of a the
well-known Eddington's inversion formula (Eddington 1916):
\begin{equation}\label{eddi}
f(E)={1\over\sqrt{8}\pi^2}{d\over dE}\int_E^0{d\rho\over d\Phi}
{d\Phi\over\sqrt{E-\Phi}}
\end{equation}
which allows one to find the unique isotropic distribution
function $f(E)$ consistent with a given density - potential
profile $\rho(r), \Phi(r)$ (by eliminating $r$ we use the function
$\rho(\Phi)$ in the actual calculation). Since $\Phi$ is derived
from Eq.(\ref{pot00}), the only unknown of the problem is the
density profile $\rho(r)$. However, we may also invert
Eq.(\ref{sig}) and obtain $\rho(r)$ from a known surface
brightness profile $\Sigma(R)$.

In the case of de Vaucouleurs' surface brightness profile (\ref{devauc})
we find, numerically, a particular distribution function $f(E)$. The value
of $f$ is the same at all the points of phase space which lie on the same
hypersurface of constant energy $E$. We next consider an elementary phase
space volume $\Delta\Omega(E) = \Delta^3\mathbf{x}\Delta^3\mathbf{p}$ by
taking all the points of phase space corresponding to energies in a small
interval $E,E+\Delta E$. The density of states function $W(E) =
\Delta\Omega(E)/\Delta E$ is then calculated. Finally, we define the
number density function
\begin{equation}\label{nd}
N(E)={\Delta N\over \Delta E} =
{\Delta N\over \Delta\Omega}
{\Delta\Omega\over\Delta E} = f(E)W(E)
\end{equation}
yielding the number of particles per unit energy of the system. We stress
again that $N(E)$ represents a density in energy space, while $f(E)$ represents
a density in phase space. The two functions
can be linked only because the distribution
of velocities is isotropic. Binney's numerical calculation showed that
the number density function $N(E)$ for a system with de Vaucouleurs'
profile is, to a good approximation, an exponential function $N(E)\simeq
N_0\exp(-\beta E)$. This suggests that a kind of statistical mechanics
is applicable in these systems, which, however, should introduce a
{\it non-uniform partition} of the phase-space in terms of elementary
volumes $\Delta\Omega$ corresponding to the energies in intervals
$E,E+\Delta E$.

This approach can be generalized in the case of anisotropic systems.
In that case we consider distribution functions of the form $f(E,L^2)$, and
look for a number density distribution $N(E,L^2)$ in the space $(E,L^2)$,
called the `Lindblad space' (Merritt 1985). The calculation of the elementary
volume $\Delta\Omega(E,L^2)=\Delta^3\mathbf{x}\Delta^3\mathbf{p}$, corresponding
to the volume of the union of invariant tori with energy and angular momentum
values in the range $E,E+\Delta E$, $L^2,L^2+\Delta L^2$ can be done as
follows (Ogorodnikov 1965): a phase-space element corresponding to values
of the phase-space variables in the range $r,r+dr$, $\theta,\theta+d\theta$,
$\phi,\phi+d\phi$, $v_r,v_r+dv_r$, $v_t,v_t+dv_t$ (where $v_r$, $v_t$
denote the modulus of the radial and transverse velocity respectively)
is given by
\begin{equation}\label{omededl1}
d^3\mathbf{r}d^3\mathbf{v} = r^2\sin\theta drd\theta d\phi
4\pi v_t dv_tdv_r
\end{equation}
Considering the transformation $r\rightarrow r$,
$\theta\rightarrow\theta$, $\phi\rightarrow\phi$,
$v_t^2\rightarrow L^2/r^2$, $v_r^2\rightarrow
2[E-\Phi(r)]-L^2/r^2$, we can take the determinant of the
transformation's Jacobian matrix and write Eq.(\ref{omededl1}) in
the form
\begin{equation}\label{omededl2}
d^3\mathbf{r}d^3\mathbf{v} = 2\pi\sin\theta d\theta d\phi {dr\over v_r}dEdL^2
\end{equation}
Then, the total phase space volume occupied by tori with energies in the
interval $E,E+dE$ and angular momenta $L^2,L^2+dL^2$ is given by
\begin{equation}\label{domel1}
d\Omega(E,L^2) = dEdL^2\pi\int_{r_p(E,L^2)}^{r_a(E,L^2)}
{2dr\over v_r}
\int_0^\pi\sin\theta d\theta
\int_0^{2\pi}d\phi = dEdL^24\pi^2T_r(E,L^2)
\end{equation}
where $r_p(E,L^2)$, $r_a(E,L^2)$ are the radii of pericenter and apocenter
respectively, for given $(E,L^2)$, that is, the roots of the equation
\begin{equation}\label{apside}
E-\Phi(r)-{L^2\over 2r^2}=0
\end{equation}
and $T_r(E,L^2)$ is the radial period of orbits, i.e., the time needed to
go from pericenter to apocenter and back to pericenter, given by
\begin{equation}\label{tr}
T_r(E,L^2)=2\int_{r_p(E,L^2)}^{r_a(E,L^2)}
{dr\over\sqrt{2(E-\Phi(r))-{L^2\over r^2}}}
\end{equation}
We thus have that:
\begin{equation}\label{omeel}
W(E,L^2) = {d\Omega\over dEdL^2} = 4\pi^2T_r(E,L^2)~~.
\end{equation}
It is remarkable that for a wide class of galactic potentials the behavior
of the function $T_r(E,L^2)$ is, to a very good approximation, independent
of $L^2$, and very close to the Keplerian limit $T_r\propto |E|^{-3/2}$.
For example, H\'{e}non (1959) proved that the most general class of
spherical potential functions for which the integral (\ref{tr}) is
strictly independent of $L^2$ is the isochrone model:
\begin{equation}\label{iso}
\Phi(r) = -{GM\over b+\sqrt{r^2+b^2}}
\end{equation}
and for this model Eq.(\ref{tr}) yields precisely the same result
$T_r\propto |E|^{-3/2}$ as in the Keplerian case. This is also
verified in the monopole terms of the potential of N-Body
experiments (Voglis 1994a, Efthymiopoulos and Voglis 2001), and in
the polytropic model (Palmer 1995).

In order, now, to generalize Binney's result $N(E)\propto\exp(-\beta E)$
in the anisotropic case, we request that the number density function
$N(E,L^2)$ has exponential dependence on its arguments, that is
\begin{equation}\label{nelsb}
N(E,L^2)\propto \exp\big(-\beta(E+b'L^2)\big)
\end{equation}
On the other hand, the generalization of Eq.(\ref{nd}) reads:
\begin{equation}\label{felnel}
f(E,L^2)={N(E,L^2)\over W(E,L^2)}
\end{equation}
Thus, substituting the ansatz $W\propto |E|^{-3/2}$ in
Eq.(\ref{felnel}), Eq.(\ref{nelsb}) leads to
\begin{equation}\label{felsb}
f(E,L^2)\propto |E|^{3/2}\exp\big(-\beta(E+b'L^2)\big)~~~.
\end{equation}
The formula (\ref{felsb}) was proposed by Stiavelli and Bertin (1985,
1987) as a candidate to fit the distribution function of spherically
anisotropic systems.
This can also be generalized to axisymmetric systems according to the
formula
\begin{equation}\label{fsbtri}
f(E,L_z,I_3) \propto |E|^{3/2}\exp\big(-\beta(E+b'{L_z^2\over 2}+cI_3)\big)
\end{equation}
where we consider an axisymmetric potential
$$
\Phi(r,\theta) ={\zeta(r)\over r^2} + {\eta\cos\theta\over r^2}
$$
which yields an integrable system third integral
$I_3={(v_\theta^2+v_\phi^2)r^2\over 2} + \eta\cos\theta$.
A more general formula involving axisymmetric St\"{a}ckel
potentials is given in Stiavelli and Bertin (1985).

The Stiavelli-Bertin distribution function can be derived on the
basis of statistical mechanical considerations (Stiavelli and
Bertin 1987). This is done by implementing the microcanonical
approach of statistical mechanics, but assigning {\it unequal} a
priori probabilities of a phase-element to visit one of the
macrocells of the $\mu-$space such as in Fig.10, or partitioning
this space to macrocells of unequal volume. The resulting entropy
can be written in the form of a Boltzman-Gibbs entropy functional
defined in the Lindblad space:
\begin{equation}\label{snsb}
S[N]=-\int N(E,L^2)\log N(E,L^2)dEdL^2
\end{equation}
under the usual constraints of mass and energy conservation, and
one additional constraint regarding a combination of the energy
and angular momentum that is quasi-preserved during the collapse
(Stiavelli and Bertin 1987). The maximization of the entropy
(\ref{snsb}) leads then to an exponential law for the number
density function $N(E,L^2)$ such as in Eq.(\ref{nelsb}). A
partitioning of the phase-space in terms of unequal volumes
($\propto |E|^{-3/2}$) seems quite justified by the fact that, in
an integrable potential, the foliation of invariant tori create a
natural partition in phase space and that, when a system is in
equilibrium, there are no motions of the phase flow transverse to
these tori (see discussion of Fig.15). In the next subsection we
discuss the types of distribution functions found in N-Body
experiments by use of similar arguments.

Independently on whether the Stiavelli - Bertin formula for $N$ is the
most convenient choice or not, the important point in the above analysis
is the shift of emphasis from entropy functionals depending on $f$ to
entropy functionals depending on $N$, which thus becomes the important
quantity to study. This point is emphasized by Tremaine (1987), see
also Merritt et al. (1989). An entropy functional similar to (\ref{snsb})
was proposed by Spergel and Hernquist (1992) in the case of isotropic
spherical systems. The resulting distributions were also found in good
agreement with the results of numerical experiments.

\subsection{The distribution function found in N-Body experiments of
violent relaxation}

The number of particles used in galactic N-Body simulations has
grown from $10^4 - 10^5$ in the 90's to $N=10^6 - 10^7$ today.
Even so, it remains a hardly tractable task to obtain
numerically the distribution function of a relaxed system by the
counting method, i.e., by counting the number of particles in
cells of the six-dimensional $\mu-$space. Even a very coarse
division of the phase space, say by 10 bins per dimension, would
result in $10^6$ cells to consider, implying 1 to 10 particles per
cell on average. Thus the signal would be hidden by the statistical
noise.

On the other hand, with such a number of particles it is possible
to do statistics in the space of integrals, or approximate
integrals of motion, such as the Lindblad space $(E,L^2)$, which
has dimension equal to two, i.e., allowing for a meaningful
statistics. If the system has a spherical symmetry, one can then
pass from $N$ to $f$ according to the formulae of the previous
subsection. Scatter plots of the positions of the particles in the
space $(E,L^2)$ can be found in a number of papers (e.g. May and
van Albada 1984, Aguilar and Merritt 1990). But the first
systematic study of the resulting number density function
$N(E,L^2)$ was made by Voglis (1994a), who proposed fitting
formulae to represent the {\it contours} of $N$ in the space
$(E,L^2)$ (Fig.17). Similar figures were given by Natarajan et al.
(1997) and Trenti et al. (2005).

\begin{figure}[tbp]
\centering{\includegraphics[width=\textwidth]{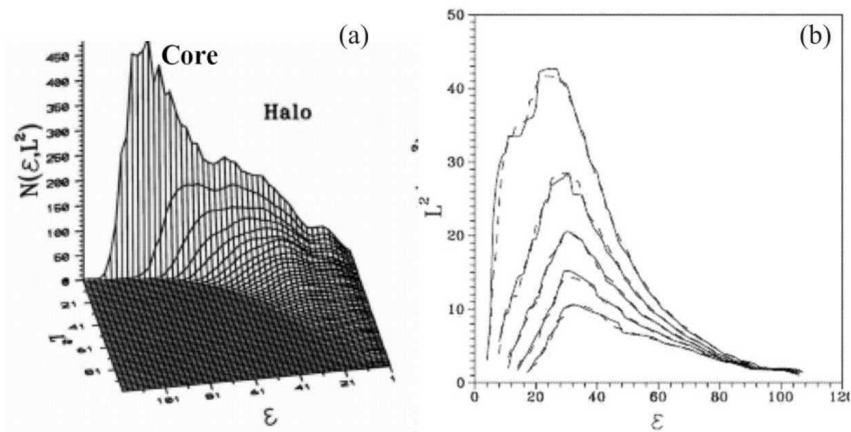}}
\caption{(a) A typical form of the `number density' distribution
$N(E,L^2)$ of an N-Body system after the relaxation. (b) The
contours of the function $N(E,L^2)$ (solid line) together with
the fitting by the model of Voglis 1994a (dashed line).
The quantity in the abscissa is ${\cal E} = -E$ (after Voglis
1994a).} \label{fig17-0}
\end{figure}

Voglis' method gave three main results:

\textbf{a)} A violently relaxed system in equilibrium is
characterized by the existence of a {\it time-invariant} function
$N(E,L^2)$ despite the fact that the arguments $(E,L^2)$ are not
precise integrals of motion. In particular, the energy of
particles has small fluctuations due to numerical fluctuations in
the coefficients of the potential of the N-Body code. On the other
hand, the modulus of the angular momentum $L^2$ is not even
approximately preserved because the final system is not spherical.
Nevertheless, the function $N(E,L^2)$ is found to remain invariant
in time as a result of a `detailed equilibrium' established in the
space $(E,L^2)$, namely the numbers of exchanged particles between
any two elementary cells of the space $(E,L^2)$ are equal in the
course of the N-Body run.

\textbf{b)} The distribution $N(E,L^2)$ is characterized by the
existence of two main loci of maximum of the distribution.

The first locus, called the `core' is given by pairs of values
$(E,L^2)$ which are very close to the locus of the energy of circular
orbits $E_c(L^2)$ if we only consider the monopole term of the multipole
potential expansion of the system. The function $N(E,L^2)$ near
this maximum can be fitted by a modified Lynden-Bell's
formula:
\begin{equation}\label{ncore}
N(E,L^2)\propto {|E|^p\over \exp\big(-\beta(E-E_c(L^2))\big)+1}
\end{equation}
where the function $E_c(L)$ plays the role of `chemical
potential'. The numerator $|E|^p$ represents a {\it polytropic}
function. The polytropic index $p$ can be shown to depend monotonically
on the power-exponent of the initial density perturbation $n$ that
caused the system to collapse (Efthymiopoulos and Voglis 2001).

The second locus, called `halo', is in mild energies but extends to
high values of the angular momentum. The associated function $E_m(L_m^2)$
is given by two formulae relating the energy $E_m$ or angular momentum
$L_m$ of the halo maximum with the value of the number density $N$ at
the maximum, namely
\begin{equation}
\log |E|_m = \log{\cal E}_0 + \nu\log\log N,~~~L_m^2=L_0^2(P-\log\log N)
\end{equation}
with parameters ${\cal E}_0,\nu,L_0,P$ depending again monotonically on
the exponent $n$ of density perturbations.

c) The behavior of the system near both loci indicates a local
change of the sign of the temperature of the system from positive
to negative. In fact, the concept of `negative temperature' was
introduced by Merritt et al. (1989) who had proposed `negative
temperature' Stiavelli-Bertin like models
\begin{equation}\label{negtem}
N(E,L^2)\propto \exp(\beta(E+b'L^2))
\end{equation}
i.e., with a positive factor $\beta$ appearing in the exponential
dependence of $N$ on $E$. Merritt et al. (1989) suggested that
such models better fit the observed surface density profiles as
well as the energy distributions of violently relaxed systems. The
detailed fits to numerical experiments by Aguilar and Merritt
(1990) favored the negative temperature models. However, even
these models failed to reproduce the N-Body distribution of
particles $N(E,L^2)$, or $N(E)$, in the region of energies close
to zero. The authors suggested that this might be attributed to
incomplete relaxation in the outer parts of the systems. On the
other hand, Voglis' study indicated that there is no fundamental
reason to consider a unique sign of the constant $\beta$
throughout the whole available Lindblad space.

Efthymiopoulos and Voglis (2001) presented a more fundamental
understanding of these results in terms of modified Stiavelli-Bertin
number density statistics. In the same time, they showed that the
method is applicable to systems that deviate considerably from the
spherical symmetry, i.e., triaxial systems corresponding to $E5 - E6$
galaxies. The key remark is that if one considers a multipole expansion
of the potential written in spherical coordinates:
\begin{equation}\label{sphehar}
\Phi(r,\theta,\phi) = \Phi_0(r)+\sum_{l=0}^\infty\sum_{m=-l}^l
\Phi_{lm}(r)Y_l^m(\theta,\phi)
\end{equation}
where $Y_l^m(\theta,\phi)$ are spherical harmonics, then the only
part of the potential which is guaranteed to yield an integrable
system is the monopole term $\Phi_0(r)$. One can then use this
term to define tori of constant label values $E,L^2$ under the
flow induced by the Hamiltonian $H_0$ corresponding to $\Phi_0$.
These, of course, are not invariant tori of the full Hamiltonian
of the system. They are, however, well-defined geometrical objects
in the phase space, and, therefore, they can be used in order to
produce a {\it partition} of the phase space in terms of volumes
$d\Omega(E,L)$ given by Eq.(\ref{omededl2}), with $\Phi_0$ in the
place of $\Phi$. This partition is a geometrical structure, not
depending on the dynamics. One can then ask what is the value of
the {\it coarse-grained} distribution function $F(d\Omega)$ within
each elementary volume $d\Omega$. In the case of a spherical
system, this is, by definition, equal also to the value of the
fine-grained distribution function $f(\mathbf{x},\mathbf{v})$ at
any point $\mathbf{x},\mathbf{v}$ of $d\Omega$. In the case of an
axisymmetric or triaxial system, however, $F$ is only an average
value of $f$ throughout the volume $d\Omega$. We find nevertheless
that $F$ can be used in the place of $f$ to reproduce the profile
of the density and the profile of the anisotropy parameter $\beta$
of the system under study with very good accuracy (Efthymiopoulos
and Voglis 2001).

\begin{figure}[tbp]
\centering{\includegraphics[width=7cm]{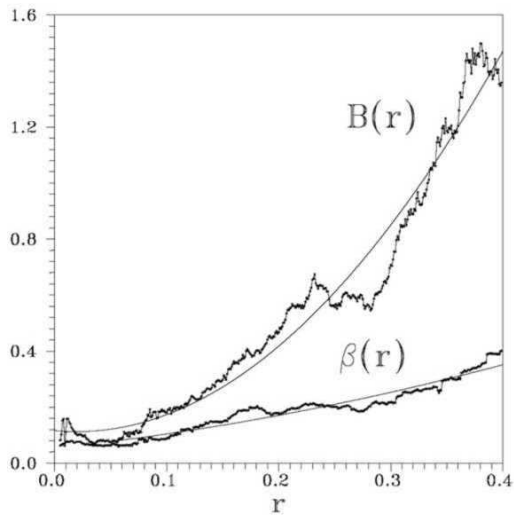}}
\caption{The Lagrange multipliers $\beta(r)$ and $B(r)$ of the
`spherical-shell' modified Stiavelli-Bertin statistics (Eq.(\ref{nuel}))
for a relaxed system as functions of the distance $r$ from the center
(after Efthymiopoulos and Voglis 2001).} \label{fig18-0}
\end{figure}

\begin{figure}[tbp]
\centering{\includegraphics[width=8cm]{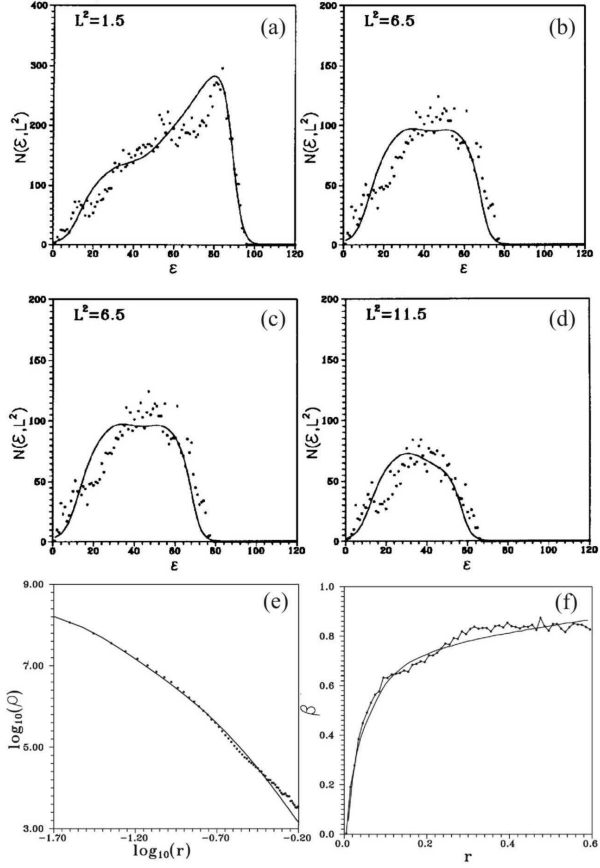}}
\caption{(a-d) Four different slices of the distribution $N({\cal
E},L^2)$ for constant angular momentum values $L^2=1.5,5.5,11.5$
and $29.5$. The fitting by the model of Efthymiopoulos and Voglis
(2001) is shown as a solid line. (e) Reproduction of the N-Body
density profile $\rho(r)$ (points) by the model (solid line)
(f) same as (e) for the anisotropy parameter profile
$\beta_{an}(r)$.} \label{fig19-0}
\end{figure}

As regards the functional form of $F$, this is given by
Eq.(\ref{felnel}) with $F$ in the place of $f$. The problem is
thus again transferred to the determination of the number density
function $N(E,L^2)$. It was found that if the system is divided in
a number of spherical shells of radii $r$, width $dr$, then
locally, within each shell, the number density function
$\nu(E,L^2,r)=\Delta N/\Delta E\Delta L^2\Delta r$ takes the form
of a modified Stiavelli - Bertin's formula
\begin{equation}\label{nuel}
\nu(E,L^2,r) = {\exp\big(-\beta(r) E+B(r)L^2\big)\over
\exp[-\beta_c(E-E_a(r,L^2))]+1}
\end{equation}
This function fits well the numerical function $\nu(E,L^2,r)$ found
in the N-Body experiments. The latter has the same property as (a) above,
i.e., it remains practically invariant in different time snapshots.
The numerator of Eq.(\ref{nuel}) has, precisely, the form of
Stiavelli - Bertin statistics for the number density function
(Eq.(\ref{nelsb})). However, we find that the parameters $\beta$ and $B$,
measuring the temperature and velocity anisotropy within the
shell, are functions of the shell radius $r$ (Fig.18). Since these
parameters enter as Lagrange multipliers in the maximization of an
entropy functional in the Lindblad space, such as the functional
(\ref{snsb}), the authors concluded that the results hint towards
a new type of statistics that incorporates the different degree of
mixing in phase-space during relaxation between the inner and
outer system's shells. We finally note that the denominator in
Eq.(\ref{nuel}) introduces again a cut-off of the shell number
density function $\nu$ for energies lower than $E_a(r,L)$, the
energy of an orbit reaching the shell at its apocenter:
\begin{equation}\label{ear}
E_a(r,L^2) = {L^2\over 2r^2}+\Phi_0(r)~~~.
\end{equation}
In the spherical case, Eq.(\ref{ear}) provides an absolute cut-off, i.e,
no particle with energy $E<E_a$ can reach the shell. But in a triaxial
system there is some tolerance around this cut-off introduced by the
multipole terms of Eq.(\ref{sphehar}), which is measured by the value
of the constant $\beta_c$.

The global number density function $N(E,L^2)$ found by integrating
$\nu(E,L^2,r)$ over the radii of all shells
\begin{equation}\label{nelshell}
N(E,L^2) = \int_0^\infty \nu(E,L^2,r)dr
\end{equation}
fitted quite nicely the numerical data in a series of experiments
collapsing under either spherically symmetric or clumpy initial
conditions (Figs.19a-d). The goodness of the fit was also evident
in the profiles of the density $\rho(r)$ and of the anisotropy
parameter $\beta_{an}(r)$ of the same systems (Fig.19e and
Fig.19f respectively).

\section{The orbital approach. Global dynamics and self-consistent
models of galaxies}

In the previous section, the focus were on studying the distribution
function of galaxies on the basis of statistical mechanical
considerations. However, a different approach to the same problem
lies in studying the {\it orbital content} of stellar systems. An
orbital study should give the main characteristics of the phase space
structure and find which types have the dominant contribution in
the self-consistency of the system. As a rule, a type of orbits
is important if the form of the orbits supports the form of the
galaxy.

In the sequel, we analyze the main types of orbits in spherical,
axisymmetric and triaxial systems (we focus on non-rotating
systems). We then refer to applications of `global dynamics' in
galaxies, based mostly on the frequency analysis of Laskar (see
e.g. Laskar 1990, 1993a,b, Laskar et al. 1992, Dumas and Laskar
1993, Sidlichovsky and Nesvorny 1997, Laskar 1999, Laskar 2003).
Finally, we discuss the method of {\it self-consistent models}
(Schwazschild 1979) which is widely used today in order to explore
the relative contribution of various types of orbits in the
composition of the distribution function of a galaxy.

\subsection{Orbits in spherical systems}
\label{sec:2}

As already discussed in subsection 2.3, the orbits in a spherical
potential $\Phi(r)$ are confined to planes normal to their
(constant) angular momentum vector
$\mathbf{L}=\mathbf{r}\times\mathbf{\dot{r}}$. The modulus of
$\mathbf{L}$ appears as a parameter in the effective one degree of
freedom Hamiltonian
\begin{equation}\label{hameff}
H(r,p_r;L^2) = {p_r^2\over 2}+\Phi_{eff}(r,L) = {p_r^2\over 2}+{L^2\over 2r^2}
+\Phi(r)
\end{equation}
with $p_r=\dot{r}$. The Hamiltonian (\ref{hameff}) yields the radial motion
on the orbital plane. The value $E_c =
L^2/2r_c^2 +\Phi(r_c)$, where
\begin{equation}\label{circ}
{d\Phi(r_c)\over dr}-{L^2\over r_c^3}=0
\end{equation}
yields the energy $E_c$ of the circular orbit with radius $r_c$. In
galactic potentials, the radius $r_c$ corresponds to a minimum of
the effective potential. As a result, the circular orbits are
stable against {\it radial perturbations}. On the other hand, for
any value of the energy $0<E\leq E_c$ the orbits are confined
between a minimum pericentric distance $r_p$ and a maximum
apocentric distance $r_a$. These are the roots of
Eq.(\ref{apside}). The forms of the orbits are rosettes (Fig.20).
The radial period is given by Eq.(\ref{tr}), while the azimuthal
period is (e.g. Binney and Tremaine 1987, p. 107)
\begin{equation}\label{tazi}
T_\theta(E,L^2) = \frac{2\pi T_r(E,L^2)} {\Delta\phi}
\end{equation}
with
$$
\Delta\phi = 2L\int_{r_p(E,L^2)}^{r_a(E,L^2)} {dr\over
r^2\sqrt{2(E-\Phi(r))-L^2/r^2}}
$$
\begin{figure}[tbp]
\centering{\includegraphics[width=7cm]{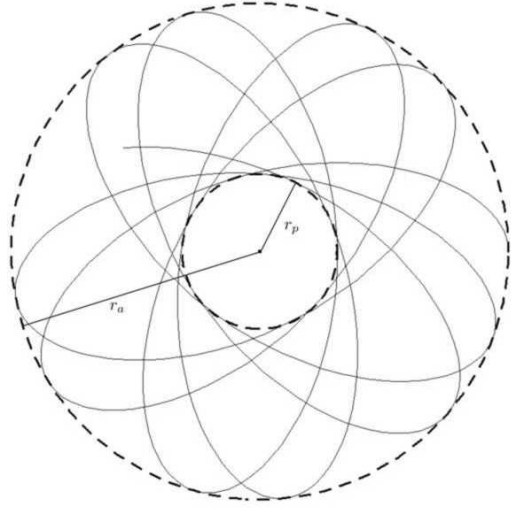}}
\caption{A typical orbit (rosette) in a spherical potential.}
\label{fig20-0}
\end{figure}

If the orbit is close to circular ($r_a-r_p<<r_c$), the radial period
tends to the {\it epicyclic period} $T_\kappa = 2\pi/\kappa$, with
\begin{equation}\label{epic}
\kappa^2 = \frac{\partial^2\Phi_{eff}(r_c)}{\partial r^2} =
{3L^2\over r_c^4} + {\partial^2\Phi\over\partial r_c^2}
\end{equation}
If the density $\rho(r)$ is a decreasing function of $r$, then
$1<T_\theta/T_r<2$, that is, the angle $\Delta\phi$ covered within
one radial period lies between $\pi$ and $2\pi$ (Contopoulos
1954). Limiting cases are the Keplerian $\Phi(r)\propto -1/r,
\rho(r)=\delta(r)$, where $\Delta\phi=2\pi$, and the homogeneous
$\Phi(r)\propto r^2, \rho(r)= const$, where $\Delta\phi=\pi$. In
these potentials, there are no rosettes but only closed (periodic)
orbits. In any other case, we have closed orbits if
\begin{equation}
    \Delta\phi=\frac{m}{n}2\pi~~~.
\end{equation}
with $m, n$ integers, $n\ne 0$.

\begin{figure}[tbp]
\centering{\includegraphics[width=\textwidth]{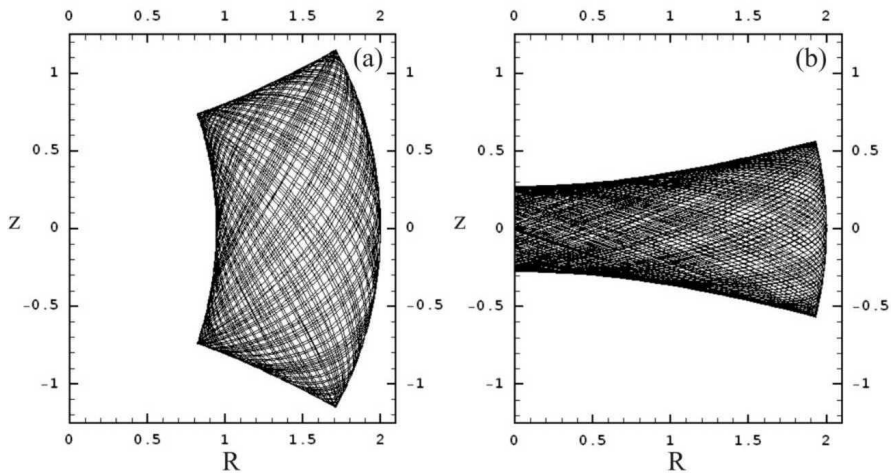}}
\caption{(a) A tube type orbit with $L_z\ne 0$ on the rotating
meridional plane $(R,z)$ in the case of an axisymmetric potential.
(b) When $L_z=0$ we have a box orbit. This orbit is 2D and lies on
the meridional plane $(R,z)$.} \label{fig21-0}
\end{figure}

\subsection{Orbits in axisymmetric systems}
\label{sec:3}

The effective Hamiltonian of motion in the meridional plane of an
axisymmetric galaxy is given by Eq.(\ref{hamaxi}). We consider the
potential symmetric on both sides of the plane $z=0$,
$(\Phi(R,z)=\Phi(R,-z))$. All orbits preserve, besides the energy,
the angular momentum component $L_z=R^2\dot{\varphi}$ which is a
parameter in the two degrees of freedom Hamiltonian
(\ref{hamaxi}). We call
$\Phi_{eff}(R,z)=\Phi(R,z)+\frac{L_z^2}{2R^2}$ the effective
potential. The 3D orbit of a star is the result of the combination
of the motion on the meridional plane $(R,z)$ and of the rotation
about the $z$ axis with angular speed $\dot{\varphi}=L_z/R^2$
(which is not constant). If the orbit obeys a `third integral',
the orbit is called regular, otherwise it is called chaotic.

\begin{figure}[h]
\centering{\includegraphics[width=\textwidth]{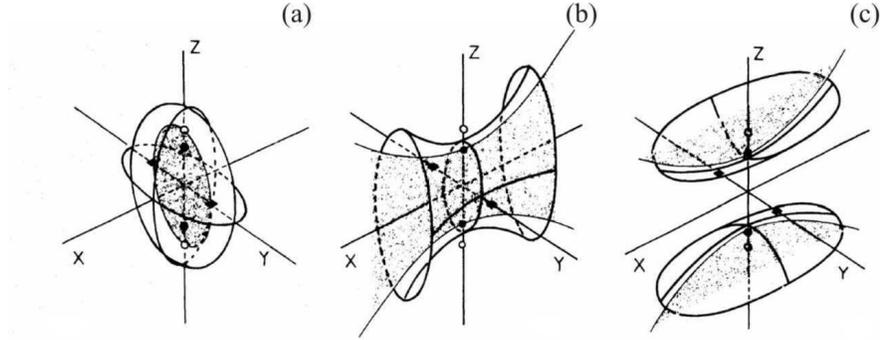}}
\caption{(a), (b), (c) Contour surfaces of constant ellipsoidal
coordinates $\lambda, \mu,  \nu$, respectively. The surfaces of
constant $\lambda$ are ellipsoids. The surfaces of constant $\mu$
are hyperboloids of one sheet and the surfaces of constant $\nu$
are hyperboloids of two sheets (after de Zeeuw 1985).}
\label{fig22-0}
\end{figure}

Fig.21a shows an example of regular orbit on the meridional plane
with $L_z\ne 0$. The model used has a potential function that
corresponds to a flat central density profile. The orbit appears
as a deformed parallelogram in the meridional plane. However, as
the orbit also rotates, there is a cylindrical hole around the
z-axis that is created by the rotation of the left boundary of the
parallelogram. Such regular orbits are called `tubes' (see e.g.
Dehnen and Gerhard 1993). On the other hand, when $L_z=0$ the
orbit's left boundary touches the axis $z=0$, and the hole
disappears (Fig.21b). Furthermore, there is no rotation because
$\dot{\varphi}=0$. Thus, the orbits are two-dimensional, and they
are called `box' orbits, because their shape on the meridional
plane resembles a box with curvilinear sides.

The box orbits are quasi-periodic orbits associated with two
independent oscillations with incommensurable frequencies, on the
$R$ and $z$ axes respectively. The limiting periodic orbits are
stable orbits along the z-axis, or the R-axis. In general, the
z-axis orbit is stable for values of the energy close to the
central potential value. At larger values the z-axis orbit become
unstables and there can be no box orbits around it. At the
transition to instability, a 1:1 stable periodic orbit bifurcates
from the z-axis orbit. The 1:1 orbit forms a loop on the
meridional plane. Such is the orbit of figure 28c below, that
corresponds to the center of the island of stability of the 1:1
resonance marked with (B) in figure 6a. Higher order periodic
orbits can also exist that correspond to various ratios of the
fundamental frequencies in the $z$ and $R$ axes.

In models with flat central profiles most orbits are regular (e.g.
Gerhard and Binney 1985). An exploration of the phase space by
means of Poincar\'{e} surfaces of section yields typically
invariant curves corresponding to boxes or tubes, and only small
secondary resonances with limited chaos. If, however, the galaxy
has a central black hole or, more generally, a `\textbf{C}entral
\textbf{M}ass \textbf{C}oncentration' (CMC), the box orbits or
tube orbits with low values of $L_z$ lose their regular character
and they are converted to chaotic orbits (subsection 4.4 below).

Finally, the orbits laying on the equatorial plane follow the
same rules as the orbits in spherical potentials, since they
feel a 2D axisymmetric potential $\Phi(R,0)$.

\begin{figure}[tbp]
\centering{\includegraphics[width=8cm]{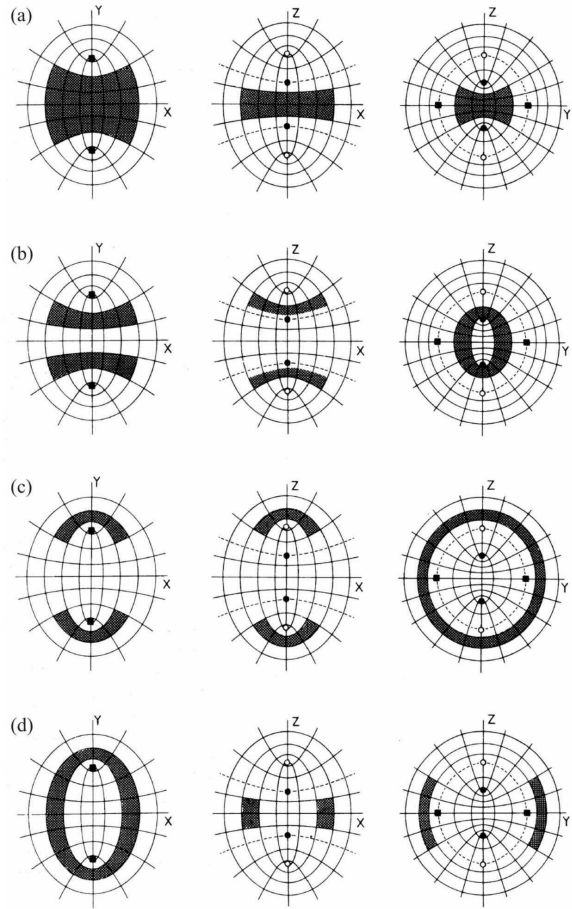}} \caption{The
cross sections of the four types of (regular) orbits with the
three principal planes in the case of perfect ellipsoid
Eq.(\ref{perfell}). The coordinate surfaces confine the various
types of orbits (after de Zeeuw 1985). The are (a) Box, (b) ILAT,
(c) OLAT and (d) SAT orbits.} \label{fig23-0}
\end{figure}

\subsection{Orbits in triaxial systems}
\label{sec:4}

In generic triaxial models of galaxies only the energy is a global
integral of motion. An exception is the perfect ellipsoid
(Eq.(\ref{perfell})) which yields an integrable St\"{a}ckel
potential (de Zeeuw 1985). The regular orbits of this model have
served as a basic guide for the form of regular orbits in generic
triaxial potential models. Figs.22a-c show the contour surfaces of
the ellipsoidal coordinates $(\lambda, \mu, \nu)$, respectively
(subsection 2.3). The surfaces of constant $\lambda$ are
ellipsoids, while the surfaces of constant $\mu$ and $\nu$ are
hyperboloids, of one and two sheets respectively (de Zeeuw 1985).
The orbits can be of four types: \textbf{a)} box, \textbf{b)}
Inner Long Axis Tube - ILAT, \textbf{c)} Outer Long Axis Tube -
OLAT, and \textbf{d)} Short Axis Tube -- SAT. Fig.23 shows the
cross-sections of these orbits with the three principal planes as
well as the limits of these orbits determined by the ellipsoidal
coordinate lines.

\begin{figure}[tbp]
\centering{\includegraphics[width=10cm]{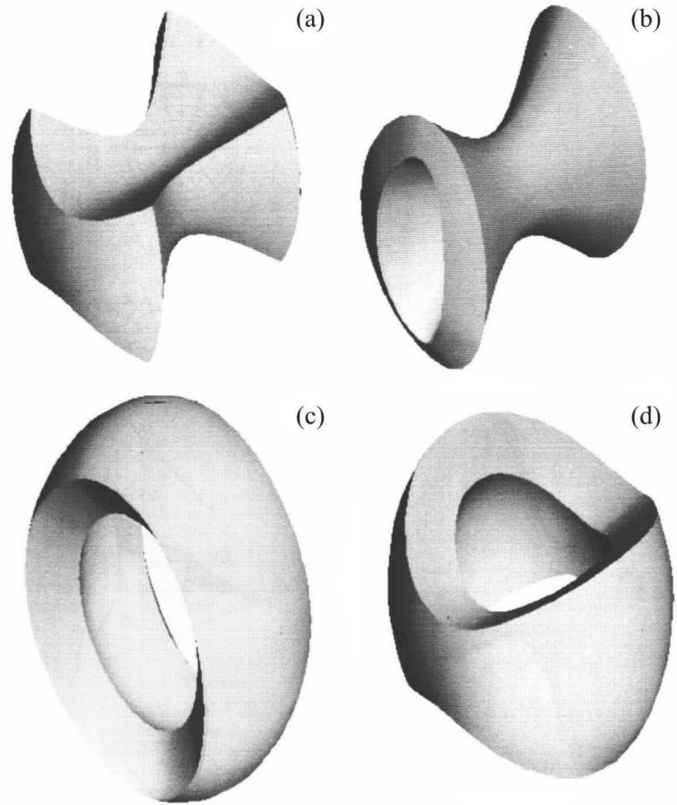}} \caption{(a)
Box, (b) ILAT, (c) OLAT, (d) SAT. The four types of (regular)
orbits in 3D space (for perfect ellipsoid). These orbits are very
good guides for the form of regular orbits that exist in most
galactic models (after Statler 1987).}\label{fig24-0}
\end{figure}

Fig.24 shows the same orbits in 3D configuration space (Statler
1987). Box orbits fill a region that resembles a parallelepiped
with curved surfaces. These orbits pass arbitrarily close to the
system's center. 3D boxes do not exist in an axisymmetric
configuration. On the other hand, ILATs are tube orbits which fill
an elongated region around the long axis, they have a hole along
the same axis and they are compatible with triaxial or prolate
configurations. OLATs are tube orbits with a hole also around the
long axis (like ILATs), which however do not approach close to the
center of the system. OLATs are also compatible with triaxial or
prolate configurations. SATs resemble like OLATs except that their
hole is around the small axis. SATs are compatible with triaxial
or oblate configurations. Orbits such as in Fig.21a are limiting
cases of either a SAT in an oblate configuration, or an ILAT or
OLAT in a prolate configuration.

Besides the above main families of orbits, in generic triaxial potentials
there can be higher order periodic orbits corresponding to different
commensurabilities of the basic frequencies of oscillation in the
three axes (see Merritt 1999 for some examples of such orbits).
When stable, these orbits are surrounded by quasi-periodic orbits
which form thin tubes around the periodic orbits. We call the
these orbits `Higher Order Resonant Tubes' (HORT).

\subsection{Chaotic orbits. The role of chaos in galaxies}

The role of chaos in galaxies is currently a very active field of
research (see the volume of proceedings Contopoulos and Voglis
2003). In the case of elliptical galaxies, the successful
construction of self-consistent models of triaxial galaxies
composed practically only by regular orbits (Schwarzschild 1979,
1982, subsection 4.6) suggested that galactic equilibria favor,
for some reason, nearly-integrable models with mostly regular
orbits. An explanation was provided on the basis of Statler's
(1987) self-consistent models of the perfect ellipsoid. In these
models (which are integrable) there was a clear predominance of
box orbits, and it was naturally expected that such a predominance
should be generic. Besides the usual box orbits, which are symmetric
with respect to the three axes, Levison and Richtone's
work (1987) on self-consistent models of the logarithmic
potential demonstrated that there were many `tilted' box orbits
that were probably not associated with the axial periodic orbits,
but with other higher order periodic orbits. On
the other hand, Schwarzschild (1993) studied triaxial models of
galactic halos of the form $\varrho\propto r^{-2}$ (cuspy density
profiles) and found a significant percentage of chaotic orbits
indicating thereby the substantial role of chaos for systems with
cuspy profiles. The role of chaos in such systems was emphasized
in recent years mostly by Merritt and his collaborators (e.g.
Valluri and Merritt 1998, Merritt and Fridman 1996, Merritt and
Valuri 1996, 1999, see Merritt 1999, 2006 for a review),
supporting the view that the percentage of chaotic orbits in an
elliptical galaxy with a central density cusp may raise up to
60\%.

Central black holes or CMCs are known also to
contribute to the creation of a large percentage of chaotic
orbits. From the early 60s, it was known that black holes possibly
exist at the centers of galaxies (see e.g. Salpeter 1964,
Zel'dovich 1964, Lynden-Bell 1969). The presence of the black
holes was proposed, initially, in order to explain the Active
Galactic Nuclei (AGN). However, during the last 10-15 years, in
view of better quality observations (e.g. with the Hubble Space
Telescope), many researchers (e.g. Kormendy and Richstone 1995,
Kormendy et al. 1997, 1998, van der Marel et al. 1997, van der
Marel and van den Bosch 1998, Magorrian et al. 1998, Cretton and
van den Bosch 1999, Gebhardt et al. 2000) found evidence of the
existence of massive black holes at the centers of galaxies. The
density of matter in many galaxies is not constant at the center
but it appears in a similar `cuspy' form as in the models of
Schwarzschild (1993) (e.g. Crane et al. 1993, Ferrarese et al.
1994, Lauer et al. 1995, Gebhardt et al. 1996, Faber et al. 1997).
Today, the dominant point of view is that practically all galaxies
contain a massive black hole at their center.

The presence of a CMC produces a significant number of chaotic
orbits in galaxies that have triaxial form, by destroying the
regular character of many regular orbits (e.g. Gerhard and Binney
1985, Merritt and Fridman 1996, Merritt and Valluri 1996, Fridman
and Merritt 1997, Valluri and Merritt 1998, Merritt and Quinlan
1998, Siopis 1999, Siopis and Kandrup 2000, Holley-Bockelmann
2001, 2002, Poon and Merritt 2001, 2002, 2004, Kandrup and Sideris
2002, Kandrup and Siopis 2003, Kalapotharakos et al. 2004,
Kalapotharakos and Voglis 2005). In particular, with the inclusion
of a massive CMC, many (previously box) orbits acquire positive
Lyapunov exponents that correspond to Lyapunov times much smaller
than the age of galaxies in which they reside. The reason for this
destabilization of the orbits is that, when approaching
arbitrarily close to the center, the box orbits, are scattered by
the CMC and become chaotic, tending to fill the whole available
space inside the equipotential surface corresponding to the
constant energy condition. As a consequence, the orbits cover a
more spherical domain. The insertion of a CMC in a triaxial galaxy
produces many chaotic orbits, which cannot, in general, support a
triaxial equilibrium state. In reality, after such an insertion,
the chaotic orbits cause a secular evolution of the system towards
a different equilibrium state. We show below (subsection 5.4)
that, while under certain circumstances the final equilibrium can
still be triaxial, more often it is very close to axisymmetric
(oblate spheroid). In any case, the structure of the system in the
final equilibrium state is mainly supported by regular orbits of
the SAT type, which have a large amount of angular momentum,
because the latter condition is required in order that an orbit
avoids the (singular) center.

Another example of the importance of chaos is the case of disk
galaxies. Chaos is known to play an important role mostly near the
{\it corotation} region (Contopoulos 1983, Kauffmann and Contopoulos
1996, Contopoulos et al. 1996). In the case of barred galaxies,
the chaos is prominent near corotation and it is considered as
responsible for the termination of strong bars (see Contopoulos
2004a, section 3.3.8). On the other hand, recent findings from
N-Body experiments (Voglis et al. 2006a) suggest that the spiral
structure beyond corotation is also composed almost entirely by
chaotic orbits. A theoretical mechanism explaining this phenomenon
was proposed by Voglis et al. (2006b).

\subsection{Global dynamics}
\label{sec:2}

In two degrees of freedom (DOF) Hamiltonian systems, an easy way
to visualize the structure of the phase-space is by means of
Poincar\'{e} surfaces of section. In 3DOF cases, however, the
surface of section is four-dimensional and cannot be visualized.
In such systems, an efficient method to study the phase space
structure is by means of the analysis of fundamental frequencies
of orbits. This is usually called the study of `Global Dynamics'
of galaxies. An early example of frequency analysis was given by
Binney and Spergel (1982), who used the Fourier transform to test
the variability of the frequencies of orbits in a logarithmic
potential model of a galaxy. But the most precise treatment of the
same problem can be made by the Frequency Map method of Laskar
(1990, 1993a,b).
The frequency map offers a clear representation of the picture of
the phase space by providing a distinction of regular or
chaotic domains in the space of {\it actions}, or of their
associated frequencies. Thus, one may visualize the Arnold web of
the various resonances and identify which resonances play the
dominant role.

\begin{figure}[tbp]
\centering{\includegraphics[width=\textwidth]{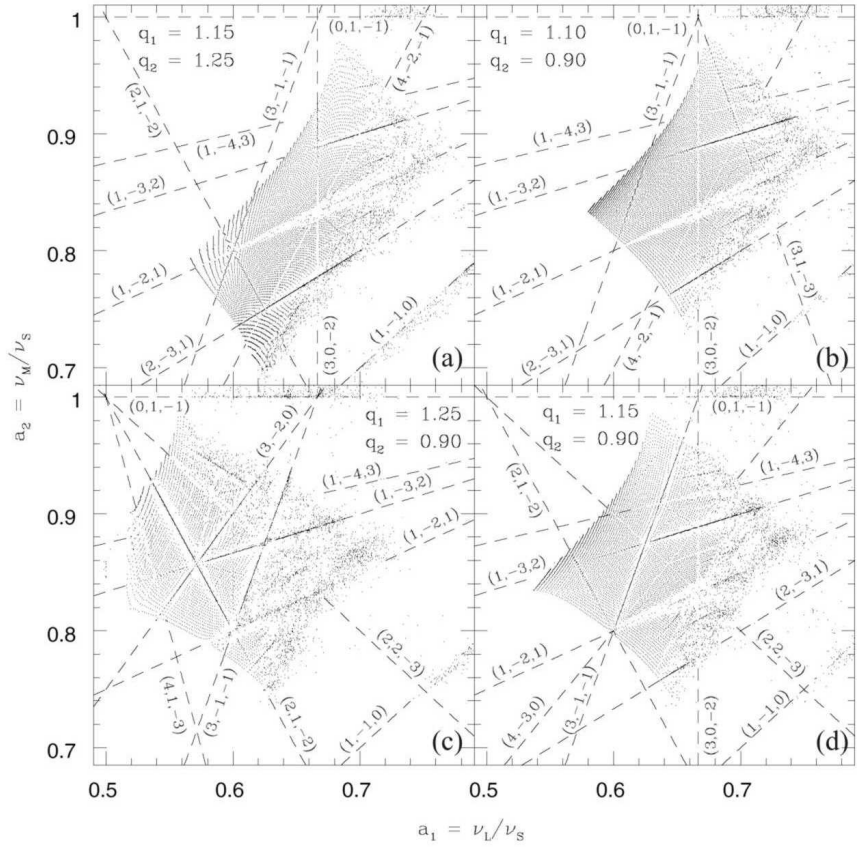}}
\caption{Frequency maps (rotation numbers $(a_1,a_2)$) of the
box orbits in the case of the logarithmic potential
Eq.(\ref{logphi01}) for a fixed energy level and for various
pairs of the parameters $q_1,q_2$. Each point in these diagrams
corresponds to one orbit. We distinguish regions of regular orbits
(well ordered points), regions of chaotic orbits (scattered
points), and various resonance lines (after Papaphilippou and
Laskar 1998).} \label{fig25-0}
\end{figure}

The distinction between the chaotic and regular orbits is based on
the fact that the regular orbits have constant frequencies whereas
chaotic orbits show a variability of the frequencies calculated in
different time windows. The calculation of the frequencies takes
place with an advanced numerical technique that reduces, in
general, the scaling of the error with respect to the width of the
time window $T$ to $O(1/T^4)$, instead of $O(1/T)$ as in the fast
Fourier transform. This method was implemented in the galactic
problem firstly by Papaphilippou and Laskar (1996, 1998). The
potential adopted was the logarithmic potential
\begin{equation}\label{logphi01}
    V(x,y,z)=\ln\left(R_c^2+x^2+\frac{y^2}{q_1^2}+\frac{z^2}{q_2^2}\right)
\end{equation}
that represents elliptical galaxies with flat density profiles at
the center. The parameter $R_c$ is a softening radius, and $q_1,
q_2$ are two parameters that control the ellipticity and the
triaxiality of the system.

Figs.25a-d show a characteristic example of a frequency map of box
orbits for four different sets of the parameters' values of the
potential (\ref{logphi01}) and for one particular value of the
energy (Papaphilippou and Laskar 1998). Every {\it point} in these
diagrams corresponds to one {\it orbit} in phase space. The axes
give the {\it rotation numbers} $a_1, a_2$ of the orbits. The
horizontal axis corresponds to the ratio $a_1=\omega_x/\omega_z$
of the orbital frequency along the long axis $x$ over the
frequency along the short axis $z$. Similarly,
$a_2=\omega_y/\omega_z$ is the ratio of the frequency of
oscillation along the middle axis to the frequency along the short
axis.

\begin{figure}[tbp]
\centering{\includegraphics[width=9cm]{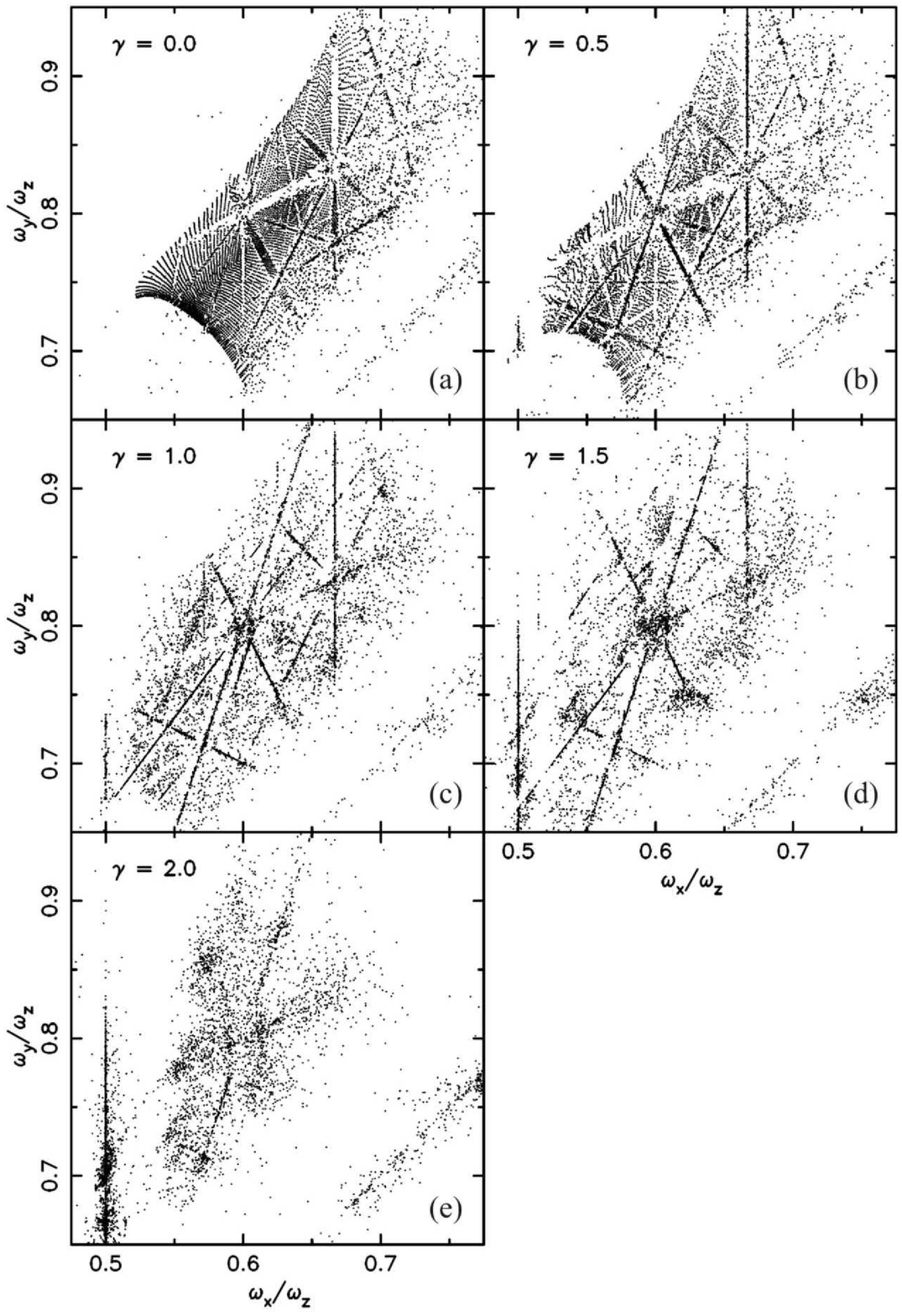}}
\caption{As in Fig.25 for the Dehnen (or $\gamma$) model Eq.(\ref{gamma}),
for various values of the parameter $\gamma$. Chaos becomes more prominent
as the parameter $\gamma$ increases (after Valluri and Merritt 1998).}
\label{fig26-0}
\end{figure}

In these diagrams, areas filled with well ordered points
correspond to regular orbits, whereas areas with scattered points
correspond to chaotic orbits. We also distinguish various resonance
lines and resonance strips with borders covered by chaotic orbits.
A resonance line is specified by a linear combination of the form
$k_1 a_1+k_2 a_2+k_3=0$ with integer $k_1,k_2,k_3$. At the intersection
of two resonance lines there are periodic orbits of various stability
types. Inside each resonance strip, on the other hand, there are
invariant tori of dimensionality lower than three. The orbits near
the central resonance lines are usually on 2D elliptic tori, which
cause a concentration of points along the resonant line. On the other
hand, the orbits in resonance lines devoid of points are usually on
tori which are at least partially hyperbolic. The study of Papaphilippou
and Laskar demonstrated in a clear way the complexity of the phase space
in 3D galactic systems by giving detailed information not only about
the existence of periodic orbits but also about the interaction of
resonances. They also confirmed that triaxial systems with a flat
central density profile contain all the types of regular orbits
found in the simple perfect ellipsoid model (see Fig.24), but also
many chaotic orbits that appear to play an important role in the
system's global dynamics.

\begin{figure}[tbp]
\centering{\includegraphics[width=9cm]{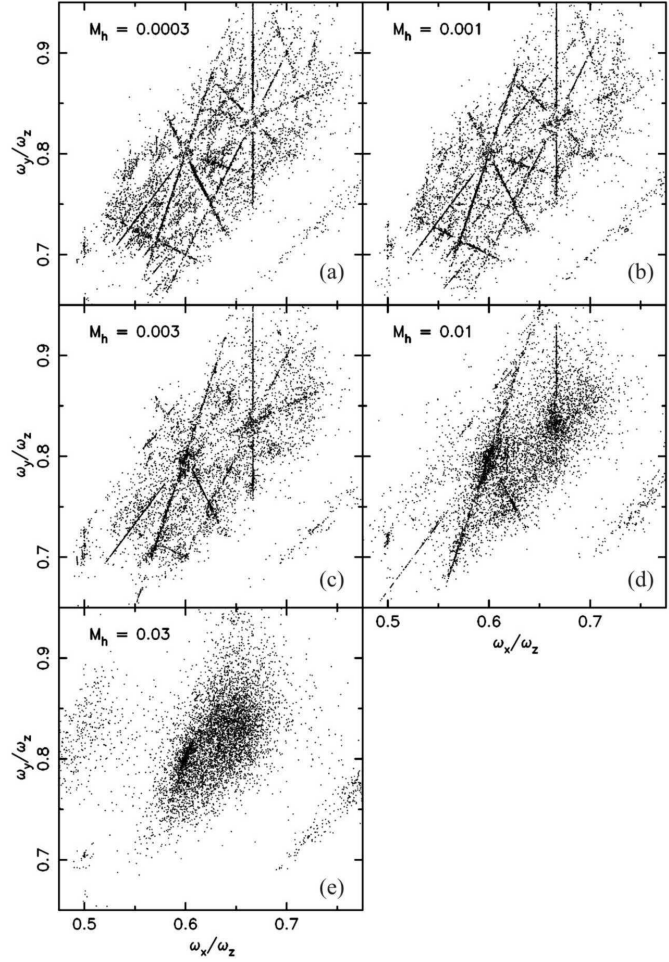}}
\caption{As in Fig.26, for $\gamma=0.5$ and various mass values of
a central black hole. The chaos in these cases is even more prominent
than in the cases of Fig.26 (after Valluri and Merritt 1998).}
\label{fig27-0}
\end{figure}

Wachlin and Ferraz-Mello (1998) and Valluri and Merritt (1998)
used the same technique as Papaphillippou and Laskar (1998) in
order to study the dynamics of triaxial galaxies with cuspy
central density profiles and massive black holes. These studies
use the Dehnen (or $\gamma$) density model (Dehnen 1993, Tremaine
et al. 1994)
\begin{equation}\label{gamma}
   \varrho(m)=\frac{(3-\gamma)M}{4\pi abc}m^{-\gamma}(1+m)^{-(4-\gamma)}
\end{equation}
\noindent where
$m^2=\frac{x^2}{a^2}+\frac{y^2}{b^2}+\frac{z^2}{c^2}$ with
$a>b>c$. The total mass of the system is given by $M$ and the
equidensity surfaces are stratified ellipsoids with axial ratio
$a:b:c$. The parameter $\gamma$ specifies the form of the central
density profile and can take values $0\leq \gamma <3$. For
$\gamma=0$ we have a flat central density profile, while for
$\gamma >0$ we have a cuspy profile, with $\rho\rightarrow\infty$
as $m\rightarrow 0$, and $\gamma$ regulating the logarithmic
slope of the density profile. The studied values of $\gamma$ were
$0\le \gamma \le 2$. Fig.26 shows the frequency maps in the domain
of box orbits for different values of $\gamma$. These diagrams
render obvious that, as the value of $\gamma$ increases, the total
volume occupied by regular orbits decreases while that of chaotic
orbits increases. Furthermore, the density of points near
resonance lines increases. When $\gamma$ is large, the orbits near
resonances are the only surviving regular orbits of the system.

Valluri and Merritt (1998) studied the case with $\gamma=0.5$
combined with massive central black holes. They found that chaos
becomes even more prominent at the presence of the black hole,
leading to full stochasticity when the mass of the black hole
becomes of order $M_h\approx 0.03M_{galaxy}$ (Fig.27). The same
authors noticed that the regular character of tube orbits is not
greatly affected by the presence of a black hole or CMC, since,
by definition, these orbits avoid anyway the center of the system.

\subsection{Self-Consistent Models - Schwarzschild's Method}

The next step after a study of global dynamics is to
examine the relative contribution of various types of orbits in supporting
self-consistently the equilibrium state of the considered galactic model.
The basic method towards such a study was introduced in stellar dynamics
by M. Schwarzschild (1979). The main steps of Schwarzschild's method are
the following:

1)  a spatial density function $\rho(\mathbf{r})$ is initially selected
and we pose the question whether this function can represent the density
of a galaxy in steady-state equilibrium. Via Poisson's equation, the
potential $\Phi(\mathbf{r})$ corresponding to $\rho(\mathbf{r})$ is obtained.

2) A grid of initial conditions is specified in a properly chosen
subset of the phase space, e.g. on equipotential surfaces. The
orbits with these initial conditions are integrated for
sufficiently long time intervals. This creates a `library of
orbits' (typical number is of order $10^4$ orbits).

3) The configuration space is divided into small cells, and the time is
recorded that each orbit spends inside each cell. Let
$N_c$ be the number of cells, $N_o$ the number of orbits $(N_c<<N_o)$ and
$t_{oc}$ the time that the orbit $o$ spends inside the cell $c$. We then
assign statistical weights $w_o$ $(o=1,...,N_o)$ to the orbits that
represent the relative contribution of each orbit, i.e., percentage of
stars that follow the same orbit, in the system. Based on these weights,
it is possible to construct a {\it response} density, that is the density
in ordinary space created by the superposition of the orbits with the above
weights. The problem is then to find for which values of the weights
the response density can be made to match the {\it imposed} density, namely
$\rho(\mathbf{r})$. Mathematically, we look for solutions for $w_o$ of
\begin{equation}\label{selfco}
    \sum_{i=1}^{N_o}w_o t_{oc}=m_c\quad c=1,...,N_c
\end{equation}
where $m_c$ is the total mass in the $c$-th cell determined by the
value of the imposed density $\rho(\mathbf{r_c})$ at the center
$\mathbf{r_c}$ of the cell times the volume of the cell. We
furthermore impose the constraint $w_o\geq 0$, since the density
contributed by each orbit can only be a positive quantity.

In most implementations of the above algorithm, a solution to
Eq.(\ref{selfco}), under the imposed constraints, is seeked by
reformulating the problem as an optimization problem. Namely, we
look for weights that minimize the absolute difference, for all
$c$, of the left and right hand sides of Eq.(\ref{selfco}). This
request takes the form of an objective function that is to be
minimized. The problem is then solved by various algorithmic
techniques such as linear programming solved e.g. by the Simplex
method (see e.g. Schwarzschild 1979, Richstone and Tremaine 1984,
Statler 1987), Non Negative Least Squares NNLS, (e.g. Pfenniger
1984, Wozniak and Pfenniger 1997, Rix et al. 1997), Lucy algorithm
(e.g. Lucy 1974, Newton and Binney 1984, Statler 1987),
maximization of a suitably defined `entropy' functional (e.g.
Richstone 1987) and Quadratic Programming (e.g. Merritt and
Fridman 1996).

The main drawback in the above method is that the algorithm
usually yields non-unique solutions. This is not physically
unacceptable, since it is known that there can be more than one
distribution functions compatible with a particular density
function (Pfenniger 1984). However, the non-uniqueness of
Schwarzschild's solutions is also partly due to numerical reasons
or to the fact that the problem's formulation is not sufficiently
constrained. This means that the number of unknowns in the
equations (i.e. the weights) is larger than the number of
available equations. This is because in Schwarzschild's method we
seek to match the distribution of matter only in configuration
space but we ignore the velocity distribution which is an outcome
of the method for any particular solution ${w_o}$. Thus, different
solutions imply the same distribution in ordinary space but quite
different distributions in velocity space. An alternative version
of Schwarzschild's method has been applied (Cretton et al. 1999,
2000) that requests, besides the spatial density, to reproduce
also the velocity profiles observed along the line of sight
(subsection 2.2). This new request reduces the number of solutions
by increasing the number of equations, yet it still leaves some
possibility for more than one solutions.

Another selection criterion among different solutions is the {\it
stability} properties of each solution. Actually this is also a
drawback of Schwarzschild's method, since the latter cannot decide
on whether a particular solution found by the method is stable or
not. This can only be decided after an N-Body realization of the
system is prepared and runs in the computer. Such
a stability analysis for the original Schwarzschild (1979) model
(Smith and Miller 1982) demonstrated that, despite the authors
characterization of the model as `robust', the model was actually
{\it not} remaining in steady-state (see also Siopis 1999, p.80). This
can be due to two reasons: \textbf{a)} the self-consistent fitting is not
perfect, i.e., we find weights that minimize the difference
between the imposed and response density, but the difference is
not actually equal to zero, and \textbf{b)} the model found is not
stable. Smith and Miller (1982) concluded that while the N-Body system
was evolving in time, the differences in time were seemingly not growing
exponentially.

\begin{figure}[tbp]
\centering{\includegraphics[width=\textwidth]{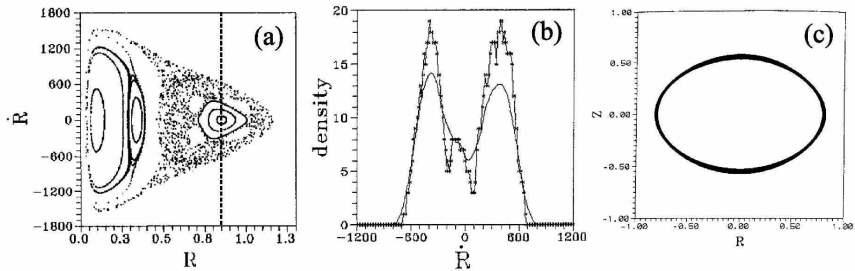}}
\caption{(a) Same as in Fig.7a. (b) The density of real particles
of the N-Body system along a vertical line passing from the center
of the 1:1 resonance (dashed line in (a)) has a {\it minimum} at
the center of the resonance. (c) The form of a very thin tube
orbit in the meridional plane around the stable 1:1 periodic
orbit. The orbit is elongated perpendicularly to the Z-axis, while
the system as a whole is elongated along the Z-axis (Fig.4).}
\label{fig28-0}
\end{figure}

For most other self-consistent models of galaxies presented so far
in the literature, there have been no accompanying N-Body tests of
the stability of the models. We believe, however, that such tests
are indispensable, otherwise the conclusions drawn by the
self-consistent modelling alone may not reflect real properties of
the systems under study.

We finally mention a variant of Schwarzschild's method implemented
in disk galaxies, i.e., spiral (Contopoulos and Grosb{\o}l 1986,
Patsis et al. 1991), or barred (Kaufmann and Contopoulos 1996). In
these authors' approach, one has first to calculate the basic {\it
families of periodic orbits} that (presumably) constitute the
backbone of the galaxy. Then, one builds the library of
orbits by considering initial conditions preferably in the
neighborhood of the periodic orbits (e.g. with a Gaussian
distribution). This method can be advantageous over Schwarzschild's
method in that one starts with some insight as regards which
orbits play the major role in the galaxy, rather than using a
blind grid of initial conditions. On the other hand, this can also
turn to be a disadvantage in case the leading hypothesis about which
orbits are important is wrong.

A further complication is due to the fact that there are cases in
which the distribution of matter has a {\it minimum} rather than
maximum near some particular stable periodic orbits. Such an
example is shown in Fig.28, referring to the same N-Body
experiment as in Figs.(4-7). In Fig.28a we see that, for a
particular value of the energy, there is a large island of
stability near a 1:1 stable periodic orbit of the system. If,
however, we find the number of N-Body particles as a function of
the distance from the central periodic orbit (Fig.28b), this plot
has a {\it minimum} rather than maximum at the position of the
periodic orbit. This is because the form of this orbit is an
elongated ellipse (Fig.28c), but the elongation is at right angle
with the elongation of the galaxy (Fig.4a,b). This means that the
distribution of matter has to have a minimum near the particular
periodic orbit, because the shape of this orbit, and of other nearby
quasi-periodic orbits, cannot support the shape of the galaxy.

\subsection{The importance of chaos through self-consistent models of
galaxies}

Schwarzschild (1979) constructed self-consistent models of
triaxial elliptical galaxies using only regular orbits (box and
SAT). In the decade following his paper, Schwarzschild's method
was implemented for the construction of self-consistent models
using almost exclusively regular orbits (Schwarzschild 1982,
Richstone 1980, 1982, 1984, Richstone and Tremaine 1984, Levison
and Richstone 1987, Stattler 1987). These systems were either
integrable (e.g. spherical or the perfect ellipsoid), or
nearly-integrable (e.g. axisymmetric with a flat central density
profile). Goodman and Schwarzschild (1981) pointed out that a
large number of the orbits that were considered as regular in
their models were in reality weakly chaotic (called
`semistochastic'). However, these orbits exhibited a nearly
regular behavior for times comparable to the galaxy's lifetime.
This means that the orbits obey an approximate integral of motion
(subsection 2.3.2), or that they exhibit the `stickiness
phenomenon' (subsection 2.3) associated with slow Arnold diffusion
in phase space.

On the other hand, as we have seen above, Schwarzschild (1993)
searched for self-consistent solutions of cuspy halo models of the
form $\varrho \propto r^{-2}$ near the center. In these models, he
found a large number of chaotic orbits. For systems with a
moderate ellipticity (smaller than E5), it was in fact possible to
find multiple solutions that contained either a mixed population
of regular and chaotic orbits or only regular orbits. But for
more elongated models (ellipticities larger than E5) it was not
possible to find any solution composed entirely by regular orbits.
It thus became evident that cuspy models of elliptical galaxies
render necessary the presence of chaotic orbits.

Merritt and Fridman (1996) studied the construction of
self-consistent models in systems with the potential function corresponding
to the $\gamma$ model for the values $\gamma = 1$ (weak cusp) and
$\gamma = 2$ (strong cusp). In both models the axial ratios were
$c/a=0.5$ and $b/a=0.79$. These correspond to a
case of a maximally triaxial E5 galaxy, i.e. a galaxy with
triaxiality
\begin{equation}
    T=\frac{a^2-b^2}{a^2-c^2}
    \label{triax}
\end{equation}
equal to $T\simeq 0.5$ (T=0 in oblate systems ($a=b$), and T=1 in
prolate ($b=c$) systems). Furthermore, the same authors quantified
the distinction of orbits into regular or chaotic by means of the
Lyapunov characteristic number. They concluded that efforts to
construct self-consistent models without the presence of chaotic
orbits were unsuccessful in both cases of weak or strong cusps. If
instead chaotic orbits are included in the same manner as regular
ones (one weight assigned to each chaotic orbit),
then it becomes possible to find successful models.
In fact, these models are {\it not} precisely stationary,
because the chaotic orbits exhibit observable diffusion in phase
space, especially at low energy levels (where diffusion times are
much smaller than the Hubble time). As a result, the superposition
of orbits creates a model that changes shape in time. Furthermore,
many chaotic orbits do not manage to fill ergodically the whole
available connected chaotic domain during an integration time
$\approx 100T_{dyn}$, which corresponds to the age of the system.
Nevertheless, by ignoring chaotic orbits at low energy levels
(which are responsible for the fastest diffusion), it was possible
to produce quasi-stationary solutions that provided good
self-consistent models, especially in the case of weak cusp.

The next step in the same study was the search of {\it fully
mixed} solutions. These are solutions in which all the chaotic
orbits of the library that correspond to the same value of the
energy can essentially be considered as different pieces of only
{\it one} orbit. If this can be established, the library
representatives of this one orbit do not cause a macroscopic
change in the shape of the response density field due to chaotic
diffusion. This means that the resulting models are guaranteed to
be stationary. Merritt and Fridman (1996) concentrated on such
solutions applicable to the chaotic orbits of low energy levels.
This was completely successful in the case of weak cusps, but only
partly successful in the case of strong cusps. The so found `fully
mixed' solutions yielded large percentages of chaotic orbits in
the final orbital composition, up to 45\% in the case of weak
cusps, and 60\% in the case of strong cusps.

\section{The N-Body approach}

\subsection{Numerical integration of the N-Body problem}

The simplest method to integrate the N-Body problem, with $N$
large, is the so-called {\it direct} method, namely the direct
numerical solution of the softened equations of motion
\begin{equation}\label{eqmonb}
\mathbf{\ddot{r}_i}=-\sum_{j=1,j\neq i}^N
{Gm_j(\mathbf{r_i}-\mathbf{r_j})\over
\big(|\mathbf{r_i}-\mathbf{r_j}|^2+\epsilon^2)^{3/2} }
\end{equation}
for each particle $i=1,\ldots,N$. The softening parameter
$\epsilon$ is introduced in order to avoid the singular behavior
of the Keplerian force whenever two particles have
a close approach. The value of $\epsilon$ should be such that the
equations (\ref{eqmonb}) `mimic' the collisionless character of
the system under study (subsection 2.1), without, however,
introducing a large error to Newton's law. Typical values are a
fraction of $D/N^{1/3}$, where $D$ is the typical scale length of
the integrated system. Specialized softening techniques are often
combined with numerical implementations of regularization
techniques (in the case of two-body encounters) or approximate
regularization (for triple encounters). A review of these techniques
is made by Aarseth (1994), a leading scientist in this field over
decades.

The direct method is very accurate, but its algorithmic complexity
is $O(N^2)$, hence prohibitive for $N$ large. For this reason,
Aarseth has developed codes that use a direct summation only for
neighboring particles, while they use a multipole expansion of the
potential, or the force, for groups of distant particles. The
algorithm used to split the particles into neighboring or distant
was introduced by Ahmad and Cohen (1973). The introduction of this
scheme reduces the algorithmic complexity from $O(N^2)$ to about
$O(N^{1.5})$.

Another direction followed in order to reduce the computational
cost of N-Body calculations was the construction of special
hardware (GRAPE, e.g. Sugimoto et al. 1990, Makino and Funato
1993, Makino et al. 1997) based on
specially designed chips to perform the sum (\ref{eqmonb}) with
speed exceeding by orders of magnitude any program written in
conventional programming languages.

In simulations of galactic systems we are usually not interested
in having an accuracy comparable to that requested in Celestial
Mechanics. In the latter case the integration must often extend
over billions of periods of the solar system bodies, while in the
former case we are interested in integration times of order $10^2 -
10^3$ periods of the stars. Furthermore, in galactic dynamics we
usually pose questions regarding the collective behavior of the system,
which do not require an accuracy of integration of individual orbits
as high as celestial mechanical calculations.

Such differences have led to the consideration of special
techniques to integrate the N-Body problem with N large. Besides
traditional `particles-in-cell' or `grid' methods (see the review
by Sellwood 1987) that are applicable also to plasma physics,
there are two methods that fit particularly the nature of the
gravitational N-Body problem: the TREE method (Barnes and Hut
1986, Hernquist 1987, McMillan and Aarseth 1993), and the smooth
potential field method (Clutton-Brock 1972, 1973, Allen et al.
1990, Hernquist and Ostriker 1992, Weinberg 1999). The so-called
`spherical harmonics' method (Villumsen 1982, McGlynn 1984,
Merritt and Aguilar 1985) is a hybrid method similar to the smooth
field code but with a `stepwise' numerical calculation of the radial
part of the spherical harmonics expansion of the potential of the
system (Sellwood 1987), that requires sorting of the parcles with
respect to their distances from the center.
There have also been simulations in which
the collisionless Boltzmann equation (\ref{bol}) is solved
directly. This is numerically tractable in systems forced to
retain a particular symmetry (usually spherical). The Boltzmann
equation can be integrated either through calculation of its
moments (e.g. Hoffman et al. 1979), or by its characteristic
system of ordinary differential equations (e.g. H\'{e}non 1964,
Burkert 1990, Henriksen and Widrow 1997).

The main idea of the TREE algorithm (Barnes and Hut 1986) is to
define a criterion by which, considering, say, the i-th particle
of the system, the other particles can be divided in groups of
close or distant particles with respect to the position of the
i-th particle. The forces on i-th particle by close particles are
added by direct summation. However, the forces of distant
particles are added by considering one term for each group rather
than for each particle. This effectively reduces the algorithmic
complexity of the code from $O(N^2)$ to $O(N\log N)$.

In the algorithm of Barnes and Hut (1986) all the particles
are set initially in one cubic cell of volume $s_0$. This cell
is divided by consecutive bisections into subcells of volume
$s_k=s_0/2^{3k}$, where the index $k=1,2,\ldots$ denotes
the order of division. If, for some order $k$, one subcell
contains no more than one particle, this subcell is not further
divided, otherwise the subdivision continues at order $k+1$.
The data structure storing the hierarchy of all subcells is
called the `tree'.

A `tolerance parameter' $\theta_{tol}$ is also defined in to
distinguish close cells from distant cells with respect to the
position of one particle (suggested values are around $\theta_{tol}
\approx 1$, Hernquist 1987). Considering the i-th particle, a
cell is called distant if the following condition holds:
\begin{equation}\label{treetol}
r\geq {s_k^{1/3}\over\theta_{tol}}
\end{equation}
where $r$ is the distance of the cell from the i-th particle.
If condition (\ref{treetol}) is true, the particles in the cell
are viewed as one group of mass $M_c$ (the sum of the particles'
masses). The contribution of this group to the force on the
i-th particle is calculated by a multipole expansion (usually
up to quadrupole terms).

The TREE method is very efficient. While the integration time is
drastically reduced, there are many different cases of N-Body
experiments that can be effectively treated with TREE. Examples
are \textbf{a)} collapsing galaxies (e.g. Cannizzo and Holister
1992, Curir et al. 1993, Voglis 1994a, Voglis et al. 1995)
\textbf{b)} merging galaxies (Barnes 1988, 1992, Hernquist 1992,
Viturro and Carpintero 2000, Burkert and Naab 2003), \textbf{c)}
multiple merger events (Efthymiopoulos and Voglis 2001), and
\textbf{d)} Cosmological simulations (where TREE is often combined
with a particle-mesh (PM) algorithm, Bouchet and Hernquist 1988,
Kravtsov et al. 1997). This flexibility is due to the fact that
the TREE code can follow simultaneously the evolution of different
parts of a system that may have large density contrasts or a
rapidly varying spatial distribution. For these reasons, TREE
codes or hybrid TREE - PM codes have been developed continually
over the years, resulting in drastic improvements of the $O(N\log
N)$ scaling (e.g. Dehnen 2000) and in parallel implementations of
the algorithm for either galactic or cosmological simulations
(e.g. Warren and Salmon 1993, Dubinski 1996, Kravtsov et al. 1997,
Viturro and Carpintero 2000, Springel et al. 2001, Becciani and
Antonuccio-Delogu 2001, Miocchi and Capuzzo-Dolcetta 2002, Bode
and Ostriker 2003). The TREE code can also be combined with the
special hardware GRAPE (e.g. Fukushige et al. 1991, Athanassoula
et al. 1998).

On the other hand, the main disadvantage of the TREE code is that
it does not allow one to have the potential function of the system
$\Phi(\mathbf{x},t)$ in a closed analytical form. This means that
one cannot make global dynamical studies with TREE, such as, e.g.,
the calculation of orbits, variational equations, Poincar\'{e}
sections, frequency maps etc. (section 4).

The class of {\it self-consistent field codes} is particularly suited
to global dynamical studies of galaxies. The main idea in such
codes is that a spatial distribution of particles represents a
Monte Carlo realization of an ideally smooth density field.
The smooth density $\rho$ is given by Eq.(\ref{rho}), i.e., in terms
of a smooth distribution function $f$. If the system's geometry is
not very peculiar, the smooth function $\rho(\mathbf{x})$ can be
expanded in a truncated series of basis functions. Different basis
functions can be chosen taylored to the particular properties of the
system under study.

We shall follow the formalism of Weinberg (1999) in order to show the
method to obtain suitable sets of basis functions for triaxial systems.
If we anticipate that the average density profile of the system that
is to be simulated will not be very different from a model function
$\rho_{00}(r)$, we express the monopole term of the density as a
truncated series of the form:
\begin{equation}
\rho_{monopole}(r) =
\rho_{00}(r)\sum_{n=0}^{n_{max}}b_{n00}u_{n00}(r)
\end{equation}
The sum in the r.h.s. represents the residuals of the fit of the
monopole term of the real density of the system by the model density
$\rho_{00}(r)$. The coefficients $b_{n00}$ are unknown and the
main task of the N-body code is to find their values. The functions
$u_{n00}(r)$, on the other hand, are known functions which are
eigenfunctions of a Sturm-Liouville problem specified below.
We can similarly express all multipole contributions to the
density, i.e., we fix some model functions $\rho_{ml}(r)$
and express the density as:
\begin{equation}\label{rhoscf}
\rho(r,\theta,\phi) = \sum_{l=0}^{l_{max}}
\sum_{m=-l}^l\sum_{n=0}^{n_{max}}b_{nml}\rho_{ml}(r)u_{nml}(r)
Y_l^m(\theta,\phi)
\end{equation}
with functions $u_{nml}(r)$ specified by a Sturm-Liouville problem
and coefficients $b_{nml}$ calculated by the N-Body code.

The Sturm-Liouville problem for $u_{nml}(r)$ is formulated as
follows: writing the potential in a form similar to (\ref{rhoscf})
\begin{equation}\label{phiscf}
\Phi(r,\theta,\phi) = \sum_{l=0}^{l_{max}}
\sum_{m=-l}^l\sum_{n=0}^{n_{max}}c_{nml}\Phi_{ml}(r)u_{nml}(r)
Y_l^m(\theta,\phi)
\end{equation}
we couple equations (\ref{rhoscf}) and (\ref{phiscf}) via Poisson
equation (\ref{pot}) in spherical coordinates. After the separation
of variables, this leads to:
\begin{multline}\label{stlv1}
-\dfrac{d}{dr}\left(r^2 \Phi_{ml}^2
\dfrac{du_{nml}}{dr}\right)+\left[l(l+1)\Phi_{ml}^2-\Phi_{ml}\dfrac{d}{dr}\left(r^2\dfrac{d\Phi_{ml}}{dr}\right)\right]u_{nml}=\\
-(4\pi G \lambda_{nml} r^2 \Phi_{ml}\rho_{ml})u_{nml}
\end{multline}
with $\lambda_{nml} = b_{nml}/c_{nml}$. This equation, supplemented with
appropriate boundary conditions at two particular radii $r_a$ and $r_b$
is a case of the Sturm-Liouville eigenvalue problem
\begin{equation}\label{stlv2}
-\dfrac{d}{dx}\left[p(x)\dfrac{dy}{dx}\right]+q(x)y=\lambda w(x)y
\end{equation}
with
\begin{equation}\label{pqwsl}
\begin{split}
p(x)&=x^2 \Phi_{ml}^2(x)\\
q(x)&=l(l+1)\Phi_{ml}^2(x)-\Phi_{ml}(x)\dfrac{d}{dx}\left(x^2\dfrac{d\Phi_{ml}(x)}{dx}\right)\\
w(x)&=-4\pi G x^2 \Phi_{ml}(x)\rho_{ml}(x)
\end{split}
\end{equation}
The functions $u_{nml}(r)$ are eigenfunctions of a differential
operator acting on $u_{nml}$ in the l.h.s. of Eq.(\ref{stlv1}).
Since this operator does not depend on $n$, the index $n$ can be
identified to the serial index of successive eigenvalues and
eigenvectors, starting from the ground state value $n=0$. The
problem is well defined if boundary conditions are given in the
form
\begin{equation}\label{bcstlv}
\begin{split}
a_1 u-a_2\left(p(r) \dfrac{du}{dr}\right)&=\lambda \left(a_1'
u-a_2'
\dfrac{du}{dr}\right)\qquad \text{at}\quad r=r_a\\
b_1 u+b_2\left(p(r)
\dfrac{du}{dr}\right)&=0\qquad\qquad\qquad\quad\quad\quad
\text{at}\quad r=r_b
\end{split}
\end{equation}
In galactic problems, the radii $r_a, r_b$ are set equal to
$r_a=0$ (center of the system), and $r_b=R_p$ or
$r_b\rightarrow\infty$. If $r_b=R_p$, the radius $R_p$ is set to
represent the size of the system, and the boundary conditions at
$R_p$ are obtained by the request of continuity, and continuous
derivative, of the potential function at the point $R_p$ where we
pass from Poisson to Laplace equation.

If the model functions $\Phi_{ml}(r)$, $\rho_{ml}(r)$
satisfy Poisson's equation, they are called {\it potential
- density pair} functions. The eigenfunctions of the Sturm-Liouville
problem (\ref{stlv1}) are mutually orthogonal with respect to the
inner product definition:
\begin{equation}\label{inpro}
<f|g> = \int_{r_a}^{r_b} f(r)g(r)w(r)dr
\end{equation}

The above equations provide the general framework of the self-consistent
field method. In order to have a concrete N-body implementation we proceed
by the following steps:

a) Specify a set of potential - density model functions $\Phi_{ml}(r)$,
$\rho_{ml}(r)$. These are arbitrary functions which may, or may not
really depend on the indices $l$, or $m$.

b) Substitute the functions $\Phi_{ml}(r)$ and $\rho_{ml}(r)$ in
Eq.(\ref{stlv1}) and solve the Sturm-Liouville problem. This will
specify the eigenfunctions $u_{nml}$ and eigenvalues $\lambda_{nml}$.
Although a numerical solution of
Eq.(\ref{stlv1}) is in principle possible for any choice of
$\Phi_{ml}(r),\rho_{ml}(r)$, we prefer to use sets of functions
for which the solution of Eq.(\ref{stlv1}) is reduced to
known functions from the literature. Examples are:

- the Clutton-Brock (1973) set:
\begin{equation}
\begin{split}
\rho_{ml}(r)&=\sqrt{4\pi}\dfrac{r^l}{(1+r^2)^{l+5/2}}\\
\Phi_{ml}(r)&=-\sqrt{4\pi}\dfrac{r^l}{(1+r^2)^{l+1/2}}
\end{split}
\end{equation}
in which the eigenfunctions $u_{nlm}(r)$ are Gegenbauer
polynomials of the form $C_n^{l+1}(\xi)$ where
$\xi=\frac{r^2-1}{r^2+1}$,

- the set of Allen et al. (1990)
\begin{equation}
\begin{split}
\rho_{ml}(r)&=-1\\
\Phi_{ml}(r)&=1
\end{split}
\end{equation}
where the eigenfunctions $u_{nml}(r)$ are spherical Bessel
functions, and

- the Hernquist - Ostriker (1992) set
\begin{equation}
\begin{split}
\rho_{ml}(r)&=\sqrt{4\pi}\dfrac{1}{2\pi}\dfrac{(2l+1)(l+1)}{r}\dfrac{r^l}{(1+r)^{2l+3}}\\
\Phi_{ml}(r)&=-\sqrt{4\pi}\dfrac{r^l}{(1+r)^{2l+1}}
\end{split}
\end{equation}
where the eigenfunctions $u_{nml}$ are Gegenbauer polynomials of
the form $C_n^{2l+\frac{3}{2}}(\xi)$ where $\xi =\frac{r-1}{r+1}$.

In the case of galactic disks, potential-density pairs were
proposed by Clutton-Brock (1972), Kalnajs (1976), Aoki and Iye (1978)
and Earn (1996). We note that such pairs are also extremely useful in
the study of the stability properties of galaxies (e.g. Palmer 1995).

c) Given the positions of the $N$ particles, we calculate the coefficients
$b_{nml}$ and $c_{nml}$ of the full density and potential expansions
(\ref{rhoscf}) and (\ref{phiscf}). We determine first the coefficients
$b_{nml}$ by exploiting the orthogonality of the functions $u_{nml}$,
i.e. $<u_{nml}|u_{n'ml}>=\delta_{n,n'}$, as well as the orthogonality
of the spherical harmonic functions. If we multiply both sides of
Eq.(\ref{rhoscf}) by $w(r)u_{nml}(r)Y_l^m(\theta,\phi)$
and take the integral over all positions we find:
\begin{equation}\label{bnml1}
b_{nml} = \int_{r_a}^{r_b}\int_0^\pi\int_0^{2\pi}
w(r)\rho(r,\theta,\phi)u_{nml}(r)Y_l^{m*}(\theta,\phi)dr d\theta
d\phi
\end{equation}
Assuming that the positions of the N particles provide a Monte
Carlo sampling of the function $\rho(r,\theta,\phi)$, the triple
integral in (\ref{bnml1}) can be approximated by a sum over particles:
\begin{equation}\label{bnml2}
b_{nml} \simeq \sum_{i=1}^N
w(r_i)u_{nml}(r_i)Y_l^{m*}(\theta_i,\phi_i)
\end{equation}
The potential coefficients $c_{nml}$ are finally determined via
the relation $c_{nml} = b_{nml}/\lambda_{nml}$.

The use of the Monte Carlo integration method implies that, contrary to
what the term `smooth field' might signify, there is some sort of noise
in the system introduced by its discreteness. This noise
appears in a quite different way than in the direct or TREE
algorithm. Namely, the uncertainties, or inevitable small
fluctuations of the coefficients $b_{nml}$ during the simulation
of even an `equilibrium' system result in a relaxation of this
system which is essentially due to discreteness effects (see
Weinberg 1998 for a detailed discussion). However, this noise is
reduced, in general, as the number of particles $N$ increases.
As $N$ increases we may also use a larger number of basis
functions to fit the density, or the potential.

Since the calculation of the sum (\ref{bnml2}) has an $O(N)$ algorithmic
complexity, the overall complexity of a smooth field code scales
linearly with $N$. This, in combination with the fact the parallelization
of the sum (\ref{bnml2}) is straightforward, renders such codes very
powerful even in small computers or computer clusters,
with applications reaching $N=10^7 - 10^8$.

Besides its linear algorithmic complexity, the main power of a smooth
field code lies in that the outcome of the potential evaluation can
be expressed analytically in terms of a series of basis functions.
This allows one to have the Hamiltonian of the system in closed form,
a fact which greatly facilitates the orbital or global dynamical study
of the system. The use of smooth field codes has been very fruitful
in galactic dynamics so far, and we can anticipate only better prospects
for the future.

\subsection{The Global Dynamics of N-Body systems}

We have already discussed Schwarzschild's method for the construction of
self-consistent models of galaxies as well as the limitations of this
method (subsection 4.5). Such limitations are not present if one uses
N-body simulations. The equilibria of such simulations are by definition
self-consistent and stable.
Thus, questions like what is the relative importance of ordered or chaotic
orbits in a galaxy are better answered within the framework of global dynamical
studies of N-Body systems. The N-Body method allows one to deal also with
systems exhibiting secular evolution such as, e.g., systems with a
CMC or central black hole.

An early example of orbital analysis in triaxial systems resulting from N-Body
simulations was given by Udry and Martinet (1994). These authors presented
histograms yielding the distribution of particles with respect to the
ratios of the fundamental frequencies of their orbits. They subsequently
discuss the link between the orbital structure and the shape of the
N-Body systems.

A extended study of global dynamics in N-Body systems was made by
Voglis et al. (2002), Kalapotharakos et al. (2004) and
Kalapotharakos and Voglis (2005). This is a study of N-Body
systems in equilibrium resulting from the cosmological simulations
described in subsection 2.5. We have seen that spherically
symmetric (or `quiet') initial conditions lead to very elongated
galaxies as a result of the radial orbit instability. We call
such an experiment the `Q-system'. On the other hand, clumpy
initial conditions lead to a less elongated final state (the
`C-system'). These systems remain in a steady state for long time
periods ($>>1t_{Hubble}$). The smooth field code of Allen et
al. (1990) is used to calculate a smooth analytic potential
and the corresponding density. These are given by:
    \begin{equation}\label{phiin}
        \Phi(r,\vartheta,\varphi)=-\frac{G}{R_0}\sum_{l=0}^{\infty}\sum_{n=0}^{\infty}\sum_{m=-n}^{n}b_{lmn}j_n\left(a_{ln}\frac{r}{R_0}\right)P_n^{|m|}(\cos\vartheta)e^{im\varphi}
        ~~~\mbox{if}~~~r\leq R_0
    \end{equation}
    \begin{equation}\label{phiout}
        \Phi(r,\vartheta,\varphi)=-\frac{G}{R_0}\sum_{l=0}^{\infty}\sum_{n=0}^{\infty}\sum_{m=-n}^{n}b_{lmn}j_n\left(a_{ln}\right)\left(\frac{R_0}{r}\right)^{n+1}P_n^{|m|}(\cos\vartheta)e^{im\varphi}
        ~~~\mbox{if}~~~ r>R_0
    \end{equation}
    \begin{equation}\label{rhoin}
       \rho(r,\vartheta,\varphi)=\frac{1}{4\pi R_0^3}\sum_{l=0}^{\infty}\sum_{n=0}^{\infty}\sum_{m=-n}^{n}a^2_{ln}b_{lmn}j_n\left(a_{ln}\frac{r}{R_0}\right)P_n^{|m|}(\cos\vartheta)e^{im\varphi}
        ~~~\mbox{if}~~~ r\leq R_0
    \end{equation}
where $b_{lmn}$ are the coefficients of the expansion and $R_0$ is a
parameter of the code fixing the outermost radius inside which Poisson's
equation is solved. The functions $j_n(r)$ are spherical Bessel functions
and $P_n^m$ are Legendre polynomials.

\begin{figure}[tbp]
\centering{\includegraphics[width=8cm]{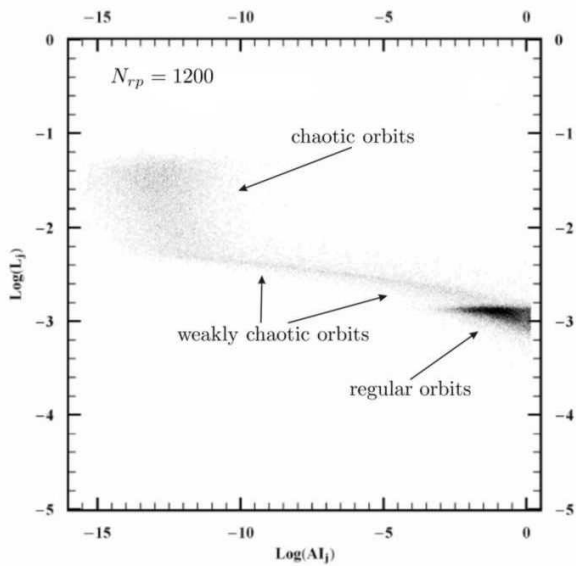}}
\caption{The distinction of regular and chaotic orbits on the
$(\log L_j-\log AI_j)$ plane. The orbits with small values of
$AI_j$ and stabilized values of $L_j$ are identified as chaotic
orbits. The orbits with large values of $AI_j$ ($>10^{-3}$) and
decreasing values of $L_j$ (as $t^{-1}$) are identified as
regular orbits. The orbits on the lane joining the two
regions are weakly chaotic orbits (after Voglis et al. 2002).}
\label{fig29-0}
\end{figure}

When calculating orbits, care is needed as regards the characterization
of an orbit as regular or chaotic. The usual criterion of the Lyapunov
characteristic number
\begin{equation}\label{lcn}
LCN = \lim_{t\rightarrow\infty}{1\over
t}\ln\left|{\xi(t)\over\xi(0)}\right|
\end{equation}
cannot be applied in a straightforward manner in the
case of galaxies. The reason has to do with the (inevitably) finite time
of numerical integration of orbits (in order to obtain an estimate of the LCN),
that has to be compared with the lifetime of the system. In fact, the periods
of stars in the inner and outer parts of a galaxy differ by about three
orders of magnitude. This suggests that the Lyapunov times of orbits
(inverse of the LCNs) be normalized with respect to the periods of orbits.
On the other hand,
the lifetime of the galaxy defines a second relevant timescale, i.e.,
an orbit is effectively chaotic only if its Lyapunov time is of the order
of the galaxy's lifetime or smaller. This second timescale is uniform for
all orbits. The question then is what is a proper normalization of
Lyapunov times (or Lyapunov exponents) that
provides a fair measure of the `chaoticity' of an orbit.

In order to address this question, Voglis et al. (2002) introduced
a new type of calculation based on the combination of two methods:
\textbf{a)} the `specific time Lyapunov number' $L_j$, normalized
with respect to the orbit's inverse of the period, and \textbf{b)}
the Alignment Index (AI). The Alignment Index is a numerical
method based on certain properties of the time evolution of
deviation vectors (Voglis et al., 1998, 1999, Skokos, 2001). Along
regular orbits the index AI has a value close to unity while along
chaotic orbits it tends exponentially to zero.

\begin{figure}[tbp]
\centering{\includegraphics[width=\textwidth]{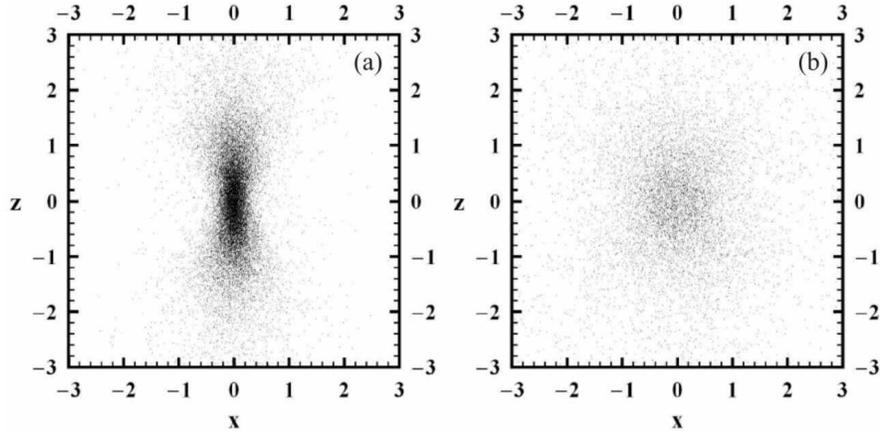}}
\caption{, Projection of the particles in (a) regular and
(b) chaotic orbits on the plane of short - long axis.
The distribution of chaotic orbits is almost spherical while
the distribution of regular orbits is strongly elongated along
the long axis of the system. (after Voglis et al. 2002).}
\label{fig30-0}
\end{figure}

In order to make the distinction of the orbits, one considers a
particular snapshot of the system and uses the potential expansion
(\ref{phiin}-103) as a time-independent potential in which orbits can
be calculated. The orbits with initial conditions given by the
positions and velocities of the N-Body particles are integrated.
The distinction of the orbits in the case of the Q system, after
an integration for $N_{rp}=1200$ radial periods of each orbit, is
shown in Fig.29. The triangular group of points in the down-right
part of the diagram corresponds to regular orbits with
$\log(AI)\gtrsim -3$ and Lyapunov numbers decreasing in time as
$t^{-1}$. The group of points in the up-left part ($\log(AI)<-12$)
are chaotic orbits. The index $AI$ of such orbits reaches, after a
fast decrease, the accuracy limit of the computer. Furthermore,
their calculation of the LCN stabilizes to a positive limit.
We also distinguish a lane of points connecting the two groups.
These are particles in weakly chaotic orbits. In this case
the time evolution of the index $AI$ shows a relatively slow
decrease, compared to that of strongly chaotic orbits.

Based on the above method, the bodies found in chaotic orbits are
23\% and 32\% of the total mass in the C and Q system
respectively. When the whole procedure was repeated after 100
half-mass crossing times ($T_{hmct}$), the above percentages
remained essentially unaltered. Keeping track of the identities of
particles that where characterized in regular or chaotic orbits,
there was no change of character except for a 3\% of the particles.

It should be stressed that only a fraction of the particles
characterized chaotic can, in fact, develop appreciable chaotic
diffusion within a Hubble time. The estimated percentage of such
particles is less than 8\%. Although most orbits are only
weakly chaotic, the spatial distribution of the chaotic mass
component is very different from that of the regular component
(Fig.30). Namely, the chaotic component is distributed rather
spherically, i.e., isotropically, while the regular component has
a spatial distribution elongated in the direction of the long axis
of the system. Moreover, contrary to the regular component, the
chaotic component has a flat central surface density profile
(Fig.31). The superposition of the two profiles creates a hump
in the total surface density profile roughly at the point where the
two profiles cross each other. In the case of the Q-system, this
transition is manifested also in the ellipticity profile (Fig.32)
which has an abrupt decrease by two units from the inner region,
where regular orbits dominate, to the outer region, where chaotic
orbits dominate.

The conclusion is that there is an upper limit to the ellipticity
of galaxies containing chaotic orbits, while systems with only regular
orbits can reach much larger values of the ellipticity.

\begin{figure}[tbp]
\centering{\includegraphics[width=7cm]{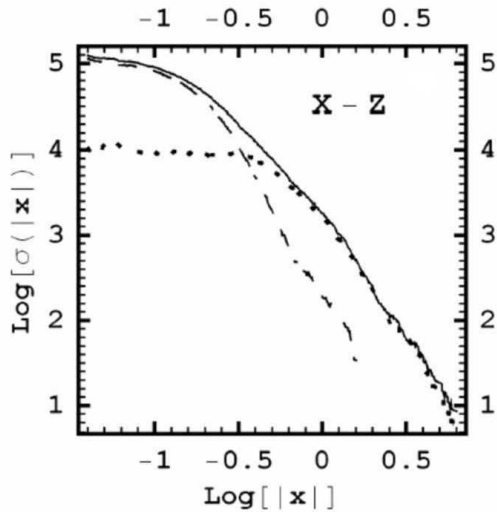}}
\caption{Profiles of the surface density on the projected plane of
short - long axis for the regular component (dashed line), chaotic
component (dotted line) and the overall system (solid line). The
superposition of the two profiles (regular and chaotic) creates a
hump in the overall surface density profile roughly at the point
where the two partial profiles cross each other (after Voglis et
al. 2002).} \label{fig31-0}
\end{figure}

In a similar study, Muzzio et al. (2005) found that the
fraction of mass in chaotic motion in their system (similar to the
Q-system above) is about $53\%$. This larger fraction is due to two
reasons: \textbf{a)} the use of a smaller threshold in the Lyapunov
number for the characterization of an orbit as chaotic, and \textbf{b)}
a less flat density profile near the center. However, the different
spatial distribution of the particles in regular or chaotic orbits
(as in Fig.30), was also found in the experiments of Muzzio et al.

There is a variety of different orbital structures that are able
to support systems with smooth centers (Contopoulos et al. 2002,
Kalapotharakos and Voglis 2005, Kalapotharakos 2005, see also
Jesseit et al. 2005). This seems to apply both to regular and
chaotic orbits. The self-consistency condition imposes some
restrictions on the permissible orbital distributions. For
example, the Q-system, being very elongated, is formed by
particles moving almost exclusively in box orbits, and it has only
a small fraction of particles in tube types (SAT or LAT). This,
despite the fact that the SAT and LAT types of orbits are very
stable and occupy an extended domain in phase space. On the other
hand, in the C system, which is more spherical, the particles move
preferentially in tube orbits, especially SAT. This fact clearly
shows that even if a global dynamical analysis establishes the
existence of large domains of stability in phase space, this does
not imply that the real particles of the N-Body system will fill
these domains. The preferential domains in phase-space are only
partly determined by the regular or chaotic character of the
orbits. The other determining factor is the request for
self-consistency.

\begin{figure}[tbp]
\centering{\includegraphics[width=7cm]{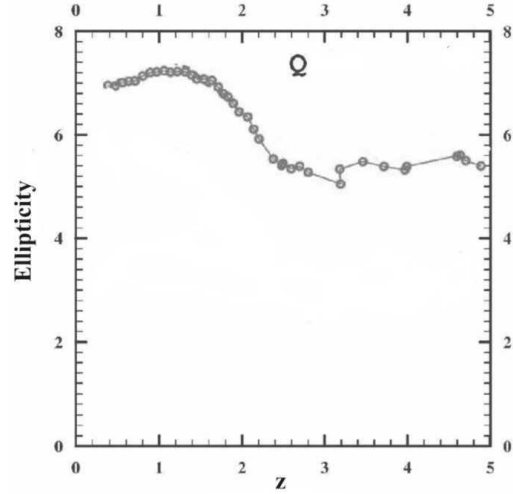}}
\caption{The ellipticity of the Q-system on the short-long axis
plane as a function of the distance from the center along the long
axis $z$. The abrupt decrease of the ellipticity at about $r_{rc}=2$
marks the transition from a region where regular orbits dominate
(inside $r_{rc}$) to a region where chaotic orbit dominate (outside
$r_{rc}$) (after Voglis et al. 2002).}
\label{fig32-0}
\end{figure}

Fig.33 shows the orbital content of the Q-system in frequency
space (the fundamental frequencies are calculated by the algorithm
of Sidlichovsky and Nesvorny 1997). The points correspond to
particles moving on regular orbits (Fig.33a), and chaotic orbits
(Fig.33b). In Fig.33a we can distinguish the distribution of
particles in different domains of the frequency space, according
to whether an orbit is of the box type or one of the tube subtypes.
The points in Fig.33b show a scatter due to their chaotic character,
which implies variability of the frequencies. The orbits with large
variability of frequencies have also large Lyapunov characteristic
numbers. Nevertheless, there are also chaotic orbits that remain
localized along the resonance lines of other, regular, orbits.
These are weakly chaotic orbits which are temporarily trapped in
particular resonances, diffusing mostly along the resonance lines
and only marginally across these lines.

\begin{figure}[tbp]
\centering{\includegraphics[width=\textwidth]{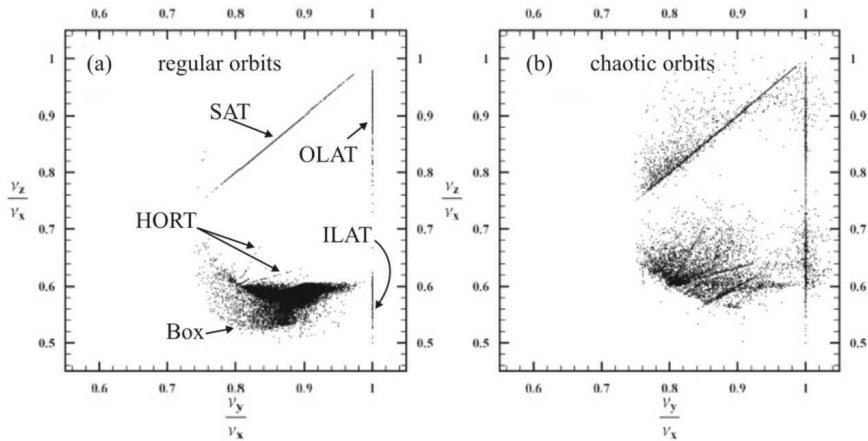}}
\caption{Frequency maps (rotation numbers) for (a) the
regular and (b) the chaotic orbits of the Q-system. The regions
of various types of regular orbits are marked in (a). In (b) many
chaotic orbits are concentrated in particular resonance lines and
they diffuse mainly along these lines (after Kalapotharakos and
Voglis 2005).} \label{fig33-0}
\end{figure}

\subsection{Secular evolution under the presence of a CMC.
Self-organization}

Secularly evolving models can be created by inserting a black
hole, or CMC, to a Q or C system (Kalapotharakos et al. 2004). The
evolution and the properties of these systems depend on the value
of the relative mass parameter $m=\frac{M_{cmc}}{M_{galaxy}}$. We
consider values of $m$ in a range [0.0005, 0.01]. Just after the
insertion of the CMC, the fraction of mass in chaotic motion
increases suddenly to the level of 80\% in systems generated by
the Q-system, and 50\% in systems generated by the C-system. The
sudden rise of the chaotic component causes a secular evolution in
these systems (subsection 4.6). This is mainly due to the
anisotropic, i.e. {\it non-mixed} distribution of the chaotic
orbits caused by the fact that, before the insertion of the CMC,
these were mostly regular orbits (boxes) of the original system.
Voglis and Kalapotharakos (2006) found that the mean rate of
exponential divergence (or mean level of LCN) of this chaotic
component has a narrow correlation with $m$, scaling as $m^{1/2}$.
Furthermore, in order to measure the effectiveness of chaotic
diffusion, these authors defined a parameter called `effective
diffusion momentum' ${\cal L}$ as the product of the
anisotropically distributed chaotic mass times the mean
logarithmic divergence of the orbits of this mass. Numerically, it
is found that the parameter ${\cal L}$ measures the ability of
secular evolution of the system. Namely, if ${\cal L}\lesssim
0.0045$ there is negligible secular evolution due to chaotic
diffusion even for times longer than a Hubble time. On the other
hand, if ${\cal L}\gtrsim 0.0045$ the models evolve following a
process of self-organization that converts chaotic orbits to
regular. The resulting reduction of entropy is partly balanced by
the increase of the mean level of exponential divergence of the
remaining chaotic orbits. During the whole process, the fraction
of chaotic mass distributed anisotropically decreases in time,
resulting in smaller values of ${\cal L}$. The evolution ceases
when ${\cal L}$ goes below the value 0.0045.

\begin{figure}[tbp]
\centering{\includegraphics[width=\textwidth]{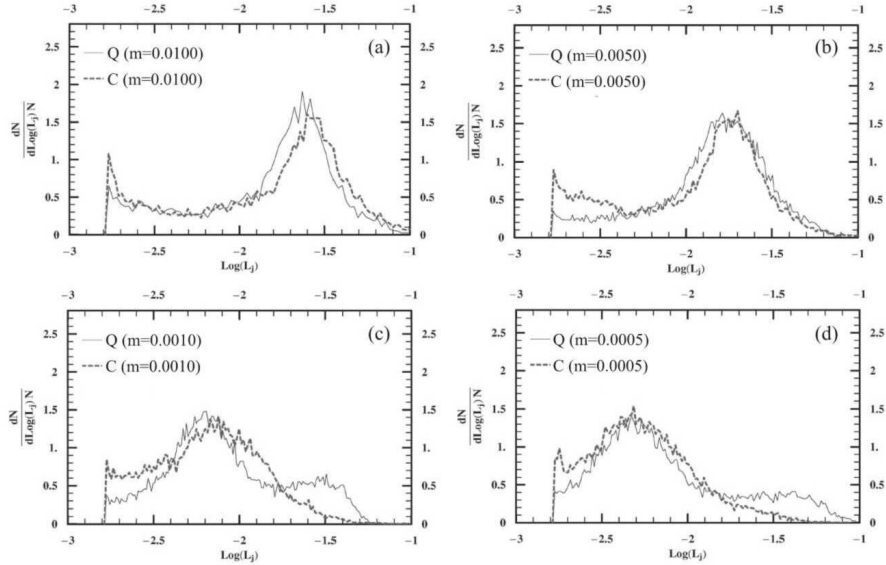}}
\caption{The distribution of the real particles with respect to
values of $\log L_j$ in four models with CMCs. For larger values
of the mass $m$, the maximum of the distributions is shifted at
larger values of $\log L_j$ (after Kalapotharakos et al. 2004).}
\label{fig34-0}
\end{figure}

Fig.34 shows the distributions of the Lyapunov numbers for all the
systems after half a Hubble time from the CMC insertion.
Smaller CMCs produce in general smaller Lyapunov numbers (the
Lyapunov number at the peak of the distribution scales with $m$ as
$m^{1/2}$). This also implies a slower rate of secular evolution. In
fact, the morphology of systems with small CMCs ($m\le 0.001$)
remains close to the morphology of the original Q or C systems for
at least a Hubble time, as indicated by a plot of the time evolution
of the triaxiality index $T$ (Fig.35). The fact that regular orbits of
the original systems are now characterized as weakly chaotic does
not have serious consequences in the resulting morphology of the
systems.

In systems with $m\ge 0.005$ the secular evolution is faster,
and it leads from a prolate, or maximally triaxial shape to a
final equilibrium which is characterized either by almost zero
triaxiality (oblate), or moderate triaxiality, depending on the
size of $m$ and on the initial orbital distribution of the system
at the time when the CMC is inserted. Larger CMCs and small
initial percentages of tube orbits (like in the Q-system) favor
oblate final equilibria.

During the secular evolution of the systems, the fraction of mass
in chaotic motion decreases in time, and in the final equilibrium
it reaches a range 12\% to 25\%. As already mentioned, the systems
present strong indications of {\it self-organization}. This means
that in the course of secular evolution, many chaotic orbits are
gradually converted into regular orbits of the SAT type. This
process can be understood with the help of Figs.36 and 37.
Figs.36a-d show projections of the 4D Poincar\'{e} sections at
successive snapshots of the secular evolution of a Q-system with a
CMC $m=0.01$. The phase portraits in the background are obtained
by integrating many reference orbits in a potential frozen at the
time corresponding to each snapshot. On the other hand, the orbits
of the real particles of the N-Body system are integrated in
an evolving N-Body potential and superposed on the phase portraits
at different snapshots. Fig.36 shows the successive Poincar\'{e}
consequents (stars or dots) of one orbit of a real particle.
Initially, before the insertion of the CMC, this is box orbit.
Thus, immediately after the insertion this orbit becomes a
chaotic orbit yielding Poincar\'{e} consequents on the chaotic
domain of the surface of section (stars). However, as the volume
of the regular domain progressively increases, the character of
the orbit is converted at a particular moment from chaotic to
regular (open circles). The new regular orbit is of the SAT type.
As a result of many such orbits, the triaxiality parameter $T$ of
the system decreases.

\begin{figure}[tbp]
\centering{\includegraphics[width=7cm]{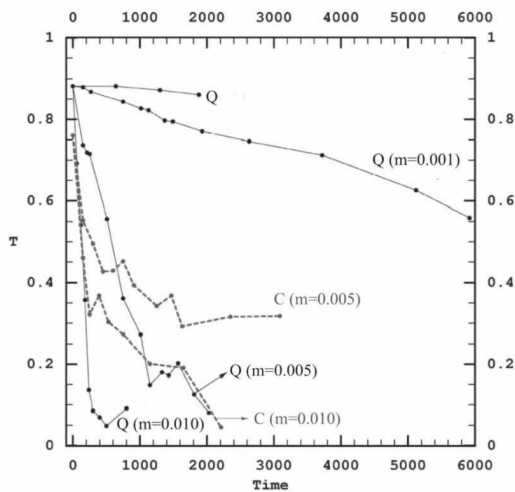}}
\caption{The time evolution of the triaxiality parameter $T$ in
various systems. One Hubble time corresponds to $t_{Hub}\approx 300$.
The systems (Q, $m=0.01$), (Q, $m=0.005$) and (C, $m=0.01$) reach
an oblate ($T=0$) final equilibrium state. Only the system (Q, $m=0.01$)
achieves this equilibrium within a Hubble time. The (C, $m=0.005$) system
reaches an equilibrium with modest triaxiality. Systems with
smaller CMCs $(m\leq 0.001)$ do not appear to evolve significantly
within a Hubble time (e.g. system (Q, $m=0.001$) (after Kalapotharakos
et al. 2004).} \label{fig35-0}
\end{figure}

\begin{figure}[h]
\centering{\includegraphics[width=\textwidth]{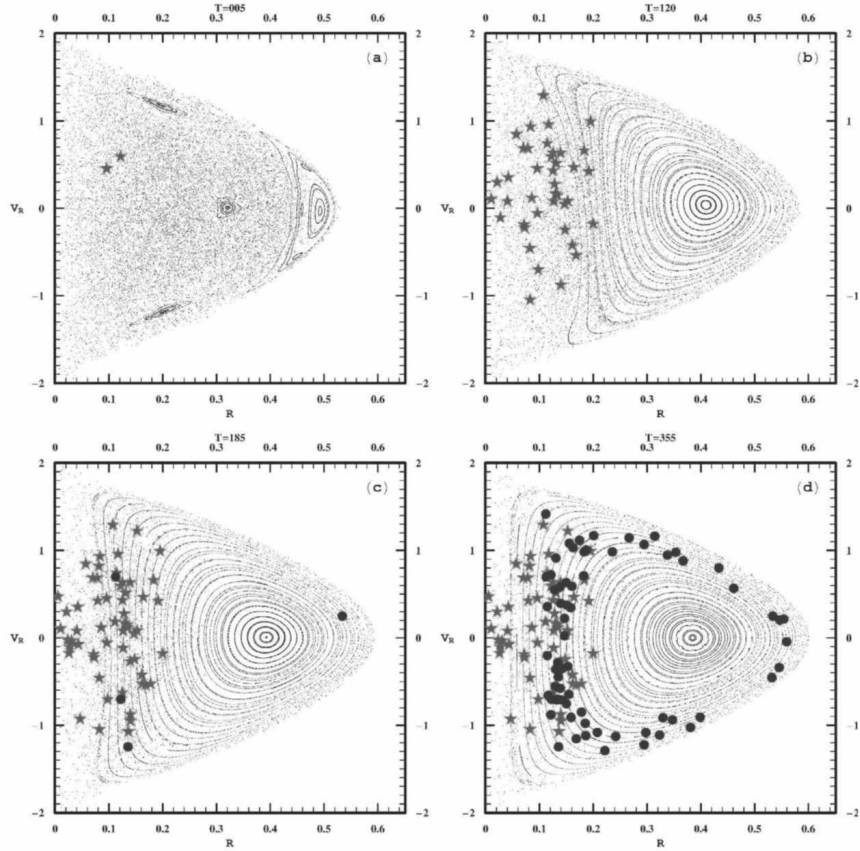}}
\caption{Projections on the ($R,V_R$) plane of the 4D Poincar\'{e}
section $z=0, \dot{z}>0$ at four different snapshots of the
evolution of the (Q, $m=0.01$) system, at times a) $t = 5$, b) $t
= 120$, c) $t = 185$ and d) $t = 355$. The island of stability to
the right corresponds to SAT orbits and its size increases as a
result of changes in the self-consistent potential. The stars or
dots give the successive Poincar\'{e} consequents of an orbit
which was a box in the original Q-system (before the insertion of
the CMC), up to the time corresponding to each panel. A star is
plotted as long as the orbit falls in the chaotic domain of the
surface of section. part of the portrait. A dot is plotted after
the moment when the orbit is captured in the regular domain (after
Kalapotharakos et al. 2004).} \label{fig36-0}
\end{figure}

Figs.37a-d show the particles of the Q-system (with $m=0.01$) on
the plane of rotation numbers at four different snapshots
($t=30,90,150,210$, a Hubble time corresponds to 300 time units).
Initially the box orbits of the original Q system are converted to
chaotic orbits or they are trapped along the various resonance
lines of HORT orbits. As the chaotic orbits diffuse in the phase
space the system's geometry changes. Namely, the system becomes
less elongated and its triaxiality parameter $T$ decreases. Due to
this evolution, some particles are trapped in orbits confined on
SAT tori. During this self-consistent evolution the areas of HORT
and ILAT orbits move upwards, approaching the line of SAT orbits.
The number of SAT orbits increases while the number of all other
orbital types decreases. At the equilibrium position there are
only regular orbits of SAT type and chaotic orbits.

\begin{figure}[tbp]
\centering{\includegraphics[width=\textwidth]{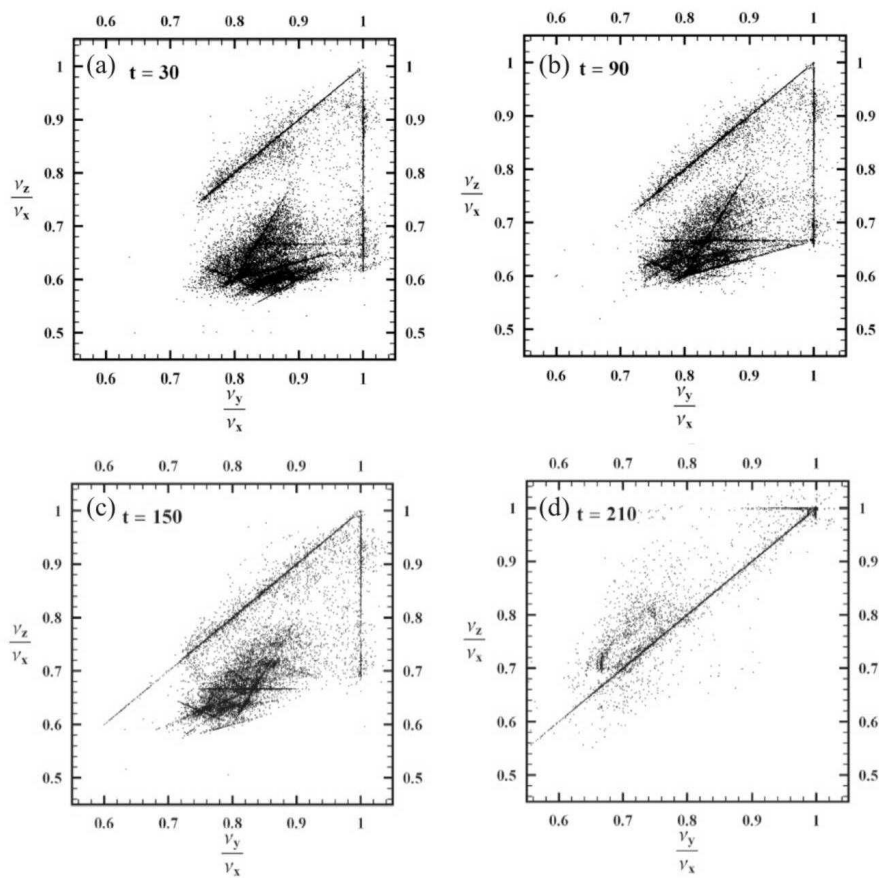}}
\caption{Frequency maps for the orbits of all the particles of the
(Q, $m=0.01$) system at the snapshots a) $t=30$, b) $t=90$, c)
$t=150$, and d) $t=210$. Initially there are many (previously box)
chaotic orbits, trapped in various resonance lines (a). During the
self-consistent evolution of the system,  the population of SAT
orbits increases while the population of all other types of orbits
decreases (b,c). At equilibrium all orbits are either regular (of
SAT type) or chaotic (d) (after Kalapotharakos and Voglis 2005).}
\label{fig37-0}
\end{figure}

Kalapotharakos et al. (2004) concluded that a system must have a
CMC of at least $m\approx 0.01$, in order to complete its
evolution and reach a new equilibrium state within a Hubble time.
For smaller mass parameters the evolution to equilibrium is
prolonged over many Hubble times. The exact evolution rate of the
systems with CMCs depends mainly on two factors. \textbf{a)} the
fraction of non-mixed chaotic orbits and \textbf{b)} the mean
Lyapunov number of these orbits (Voglis and Kalapotharakos 2006).
For example, the C-system with $m=0.01$ needs longer time than the
Q-system with the same $m$ in order to evolve towards the final
oblate configuration (Fig.35). This is because the C-System, as we
have seen above, had initially a smaller fraction of box orbits
than the Q-system. Therefore, after the insertion of the CMC, the
C-system has a smaller fraction of chaotic orbits than the
Q-system.

These results are in agreement with previous studies of Merritt
and Quinlan (1998) and Holley-Bockelmann et al. (2002). Merritt
and Quinlan (1998) studied the evolution of systems for various
values of the mass of the central black hole. The black hole is
inserted at the center of a maximally triaxial E5 elliptical
galaxy in equilibrium. They found that black holes with $m=0.01$
are capable to make the system evolve towards an oblate
axisymmetric configuration within a Hubble time. On the other
hand, Holley-Bockelmann et al. (2002) found that the insertion of
a black hole with mass $m=0.01$ leaved unchanged their system
(especially in the external parts). However, the original system
that they used was a triaxial E2 elliptical galaxy, while the
original system of Merritt and Quinlan (1998) is similar to the
Q-system considered above (many box orbits). As explained above,
given that box orbits become chaotic after the insertion of the
black hole, such a system evolves rapidly towards a new
equilibrium. On the contrary, the system of Holley-Bockelmann et
al. (2002) (similar to the C-system above) has a smaller initial
fraction of box orbits and it evolves at a slower rate.

\bibliographystyle{astron}
\bibliography{evk}
\nocite{cvk02,alletal1990,dz85,fabea97,ferea94,fm97,gebea96,gebea00,gb85,hb01,hb02,kvc04,ksd02,ksp03}
\nocite{kr95,korea97,korea98,las90,las93a,las93b,lfc92,lauea95,magea98,mfr96,mq98}
\nocite{mv96,muzea05,plas96,plas98,pm02,pm04,siopphd,spk00,sn97,sc01,st87,vm98,vdmea97,vdmvdb98}
\nocite{vce98,vks02,wfm98,lyn69,vog94,cmuz95,ev01,her1987,aar94,bt87,pm01}
\nocite{schw79,las99,crdz00,cdzmr99,dlas93,schw93,las03,hg80,schw82,gschw81}
\nocite{contg88,kcont96,pcg91,stack1890,stack1893,lb67,bh86,bish87,vce99,nek1977,spitz87}
\nocite{salp64,zel64,con2004,arn1978,ber2000,bocpuc1996,con1966,con2004b,gio2002,hag1970,ogo1965}
\nocite{eft1999,con2004,arn1978,ber2000,bocpuc1996,cha1942,con1966,con2004b,gio2002}
\nocite{hag1970,ogo1965,fripol1984,pal1995,sze1967,bir1927,con1994,varetal2003,conetal1995,con1967}
\nocite{leccoh1972,mer2005,kan2003,tre1987,agumer1990,bercap1975,bin1976,bin1978,bin1982a,bin1982b}
\nocite{con1960,davetal1983,eggetal1962,gursav1986,jea1915,lyn1962a,lyn1962b,pfe1986,spihar1971}
\nocite{udrpfe1988,ber1978,eftetal2004,eft2005,fasetal1998,froetal2000,gio1988,giosko1997}
\nocite{guzetal1998,loc1992,morguz1997,nek1977,nie1998,pos1993,seretal1983,baretal1986}
\nocite{bur1990,canhol1992,curetal1993,devau1948,dubcal1991,edd1916,got1973}
\nocite{got1975,her1987,hjomad1991,hofetal1979,kat1991,kin1962,kuletal1997}
\nocite{lonetal1991,mad1987,mat1988,mayalb1984,mcg1984,mer1985,mer1999,meretal1989,mic1963}
\nocite{minetal1990,pakpap1987,palvog1983,polshu1984}
\nocite{shu1978,shu1987,stiber1985,stiber1987,treetal1986,alb1982,vil1984}
\nocite{vogetal1995,whi1976,whi1978,whinar1987,zel1970,henhei1964}
\nocite{che1924a,che1924b,whi1916,sie1941,gus1966,con1963,conmou1965,kalrob1992,takina1007,conetal2000}
\nocite{gio1979,giogal1978,conetal2003,dep1969,hor1966,bazmar1991,eftetal2004,benetal1985,eft2005}
\nocite{denger1993,gersah1991,matger1999,eftsan2005,kuz1956,dezlyn1985,frosch1973,conetal1978,milnob1985}
\nocite{milnob1992,mcgbin1990,ratetal1984,pet1983,beletal2006,arn1964,las1993,giocin2004,guzetal2005}
\nocite{con1971,eftetal1997,kar1983,rosetal1966,chi1979,legetal2003,guzetal2002,vaucap1979,mer1996}
\nocite{craetal1993,hen1959,hen1964,arsbin1978,minetal1990,henwid1997,henwid1999,merhen2003}
\nocite{bertre2004,cupetal1969,goletal1969,sevluw1986,kanetal2003,kanmah1994,milsmi1994,louger1988}
\nocite{ant1960,lynwoo1968,hen1973,dehmer1988,pol1981,lyn1979,meragu1985,too1964,mersel1994}
\nocite{hoh1971,ostpee1973,hofetal1979,ipshor1979,ips1974,kan1987,sri1987,dej1987,sok1996,speher1992}
\nocite{filgol1984,nak2000,treetal2005,arajoh2005,aralyn2005,ziewie1989,wieetal1988,kan1998,tsa1988,plapla1993}
\nocite{tarsak2002,tarsak2003,cha1998,cha2006,cha2002,mer1985,con1954,levric1987,crebos1999,merval1999}
\nocite{con1983,vogetal2006a,vogetal2006b,binspeetal1982,deh1993,treetal1994,rictre1984,pfe1984,wozpfe1997}
\nocite{rixetal1997,newbin1984,ric1987,luc1974,smimil1982,ric1980,ric1982,ric1984,kal2005}
\nocite{conetal1996,udrmar1994,kalvog2005,jesetal2005,vogkal2006}
\nocite{mer2006,dejbru2003,capetal2002,veretal2002,dejetal1996,craetal1993}
\nocite{tootoo1972,ger1981,negwhi1983,bar1988,bar1992,her1992,naa1999,burnaa2003,smicon1996}
\nocite{meretal1989,natetal1997,ahmcoh1973,athetal1998,sugetal1990,maketal1997}
\nocite{makfun1993,mcmaar1993,wei1999,clu1972,clu1973,sel1987,bodost2003,vitcar2000}
\nocite{kraetal1997,bouher1988,dub1996,warsal1993,spretal2001,miocap2002,becant2001}
\nocite{fuketal1991,wei1998,ear1996,deh2000,too1966,alletal1992,athsel1986}
\nocite{vog1994b,herost1992,valb1987,vogetal1991,vil1982,gebetal2001,aokiye1978}
\nocite{ill1977,ver1979,kendez1991,edd1915,dezmer1983,kol1954,arn1963,mos1962,ser1963,ser1968,kal1976}

\end{document}